\PassOptionsToPackage{unicode}{hyperref}
\PassOptionsToPackage{hyphens}{url}
\PassOptionsToPackage{dvipsnames,svgnames,x11names}{xcolor}
\documentclass[
  11pt,
  a4paper,
]{article}

\usepackage{amsmath,amssymb}
\usepackage[]{libertinus}
\usepackage{iftex}
\ifPDFTeX
  \usepackage[T1]{fontenc}
  \usepackage[utf8]{inputenc}
  \usepackage{textcomp} 
\else 
  \usepackage{unicode-math}
  \defaultfontfeatures{Scale=MatchLowercase}
  \defaultfontfeatures[\rmfamily]{Ligatures=TeX,Scale=1}
\fi
\IfFileExists{upquote.sty}{\usepackage{upquote}}{}
\IfFileExists{microtype.sty}{
  \usepackage[]{microtype}
  \UseMicrotypeSet[protrusion]{basicmath} 
}{}
\makeatletter
\@ifundefined{KOMAClassName}{
  \IfFileExists{parskip.sty}{%
    \usepackage{parskip}
  }{
    \setlength{\parindent}{0pt}
    \setlength{\parskip}{6pt plus 2pt minus 1pt}}
}{
  \KOMAoptions{parskip=half}}
\makeatother
\usepackage{xcolor}
\usepackage[a4paper]{geometry}
\ifx\paragraph\undefined\else
  \let\oldparagraph\paragraph
  \renewcommand{\paragraph}[1]{\oldparagraph{#1}\mbox{}}
\fi
\ifx\subparagraph\undefined\else
  \let\oldsubparagraph\subparagraph
  \renewcommand{\subparagraph}[1]{\oldsubparagraph{#1}\mbox{}}
\fi

\usepackage{color}
\usepackage{fancyvrb}

\DefineVerbatimEnvironment{Highlighting}{Verbatim}{commandchars=\\\{\}}
\newenvironment{Shaded}{}{}

\newcommand{\AttributeTok}[1]{\textcolor[rgb]{0.84,0.23,0.29}{#1}}

\newcommand{\BuiltInTok}[1]{\textcolor[rgb]{0.84,0.23,0.29}{#1}}
\newcommand{\CharTok}[1]{\textcolor[rgb]{0.01,0.18,0.38}{#1}}
\newcommand{\CommentTok}[1]{\textcolor[rgb]{0.42,0.45,0.49}{#1}}

\newcommand{\ControlFlowTok}[1]{\textcolor[rgb]{0.84,0.23,0.29}{#1}}

\newcommand{\DecValTok}[1]{\textcolor[rgb]{0.00,0.36,0.77}{#1}}

\newcommand{\ExtensionTok}[1]{\textcolor[rgb]{0.84,0.23,0.29}{\textbf{#1}}}
\newcommand{\FloatTok}[1]{\textcolor[rgb]{0.00,0.36,0.77}{#1}}
\newcommand{\FunctionTok}[1]{\textcolor[rgb]{0.44,0.26,0.76}{#1}}
\newcommand{\ImportTok}[1]{\textcolor[rgb]{0.01,0.18,0.38}{#1}}

\newcommand{\KeywordTok}[1]{\textcolor[rgb]{0.84,0.23,0.29}{#1}}
\newcommand{\NormalTok}[1]{\textcolor[rgb]{0.14,0.16,0.18}{#1}}
\newcommand{\OperatorTok}[1]{\textcolor[rgb]{0.14,0.16,0.18}{#1}}

\newcommand{\SpecialCharTok}[1]{\textcolor[rgb]{0.00,0.36,0.77}{#1}}
\newcommand{\SpecialStringTok}[1]{\textcolor[rgb]{0.01,0.18,0.38}{#1}}
\newcommand{\StringTok}[1]{\textcolor[rgb]{0.01,0.18,0.38}{#1}}
\newcommand{\VariableTok}[1]{\textcolor[rgb]{0.89,0.38,0.04}{#1}}

\providecommand{\tightlist}{%
  \setlength{\itemsep}{0pt}\setlength{\parskip}{0pt}}\usepackage{longtable,booktabs,array}
\usepackage{calc} 
\usepackage{etoolbox}
\makeatletter
\patchcmd\longtable{\par}{\if@noskipsec\mbox{}\fi\par}{}{}
\makeatother
\IfFileExists{footnotehyper.sty}{\usepackage{footnotehyper}}{\usepackage{footnote}}
\makesavenoteenv{longtable}
\usepackage{graphicx}
\makeatletter
\def\maxwidth{\ifdim\Gin@nat@width>\linewidth\linewidth\else\Gin@nat@width\fi}
\def\maxheight{\ifdim\Gin@nat@height>\textheight\textheight\else\Gin@nat@height\fi}
\makeatother
\setkeys{Gin}{width=\maxwidth,height=\maxheight,keepaspectratio}
\makeatletter
\def\fps@figure{htbp}
\makeatother
\newlength{\cslhangindent}
\newlength{\csllabelwidth}
\newlength{\cslentryspacingunit} 
\newenvironment{CSLReferences}[2] 
 {
  \setlength{\parindent}{0pt}
  \ifodd #1
  \let\oldpar\par
  \def\par{\hangindent=\cslhangindent\oldpar}
  \fi
  \setlength{\parskip}{#2\cslentryspacingunit}
 }%
 {}
\usepackage{calc}

\usepackage{tabularx}
\makeatletter
\@ifpackageloaded{caption}{}{\usepackage{caption}}
\AtBeginDocument{%
\ifdefined\contentsname
  \renewcommand*\contentsname{Table of contents}
\else
  \newcommand\contentsname{Table of contents}
\fi
\ifdefined\listfigurename
  \renewcommand*\listfigurename{List of Figures}
\else
  \newcommand\listfigurename{List of Figures}
\fi
\ifdefined\listtablename
  \renewcommand*\listtablename{List of Tables}
\else
  \newcommand\listtablename{List of Tables}
\fi
\ifdefined\figurename
  \renewcommand*\figurename{Figure}
\else
  \newcommand\figurename{Figure}
\fi
\ifdefined\tablename
  \renewcommand*\tablename{Table}
\else
  \newcommand\tablename{Table}
\fi
}
\@ifpackageloaded{float}{}{\usepackage{float}}
\floatstyle{ruled}
\@ifundefined{c@chapter}{\newfloat{codelisting}{h}{lop}}{\newfloat{codelisting}{h}{lop}[chapter]}
\floatname{codelisting}{Listing}

\makeatother
\makeatletter
\@ifpackageloaded{caption}{}{\usepackage{caption}}
\@ifpackageloaded{subcaption}{}{\usepackage{subcaption}}
\makeatother
\makeatletter
\@ifpackageloaded{tcolorbox}{}{\usepackage[many]{tcolorbox}}
\makeatother
\makeatletter
\@ifundefined{shadecolor}{\definecolor{shadecolor}{rgb}{.97, .97, .97}}
\makeatother
\ifLuaTeX
  \usepackage{selnolig}  
\fi
\IfFileExists{bookmark.sty}{\usepackage{bookmark}}{\usepackage{hyperref}}
\IfFileExists{xurl.sty}{\usepackage{xurl}}{} 
\hypersetup{
  pdftitle={Parametrization Cookbook},
  pdfauthor={Jean-Benoist Leger},
  pdfkeywords={computational statistics; parametrizations},
  colorlinks=true,
  linkcolor={blue},
  filecolor={Maroon},
  citecolor={Blue},
  urlcolor={Blue},
  pdfcreator={LaTeX via pandoc}}

\title{Parametrization Cookbook}
\usepackage{etoolbox}
\makeatletter
\providecommand{\subtitle}[1]{
  \apptocmd{\@title}{\par {\large #1 \par}}{}{}
}
\makeatother
\subtitle{A set of Bijective Parametrizations for using Machine Learning
methods in Statistical Inference}
\author{Jean-Benoist Leger}
\date{}

\begin{document}
\begin{center}
    \fontsize{25}{30}\selectfont
    Parametrization Cookbook
    \vspace{.5cm}
    \\
    \fontsize{15}{18}\selectfont
    A set of Bijective Parametrizations for using Machine Learning
    methods in Statistical Inference
\end{center}

\vspace*{2cm}
\begin{center}
  \begin{tabularx}{\textwidth}{lp{1em}X}
          Jean-Benoist Leger &&
            \begin{tabular}[t]{p{.6\textwidth}}
            \setlength{\parindent}{-1em}
      Université de technologie de Compiègne, CNRS, Heudiasyc,
Compiègne, France\\
            \setlength{\parindent}{-1em}
      Université Paris-Saclay, AgroParisTech, INRAE, UMR MIA
Paris-Saclay, Palaiseau, France\\
            \end{tabular}\\
            \end{tabularx}
\end{center}

\bigskip
\bigskip
\begin{abstract}
We present in this paper a way to transform a constrained statistical
inference problem into an unconstrained one in order to be able to use
modern computational methods, such as those based on automatic
differentiation, GPU computing, stochastic gradients with mini-batch.

Unlike the parametrizations classically used in Machine Learning, the
parametrizations introduced here are all bijective and are even
diffeomorphisms, thus allowing to keep the important properties from a
statistical inference point of view, first of all identifiability.

This cookbook presents a set of recipes to use to transform a
constrained problem into a unconstrained one.

For an easy use of parametrizations, this paper is at the same time a
cookbook, and a Python package allowing the use of parametrizations with
numpy, but also JAX and PyTorch, as well as a high level and expressive
interface allowing to easily describe a parametrization to transform a
difficult problem of statistical inference into an easier problem
addressable with modern optimization tools.
\end{abstract}

\bigskip
\noindent%
{\it Keywords:} computational statistics; parametrizations
\vfill

\clearpage\ifdefined\Shaded\renewenvironment{Shaded}{\begin{tcolorbox}[frame hidden, enhanced, breakable, borderline west={3pt}{0pt}{shadecolor}, sharp corners, boxrule=0pt, interior hidden]}{\end{tcolorbox}}\fi

\definecolor{code-block-code}{HTML}{000080}

\definecolor{code-block-stdout}{HTML}{808000}

\definecolor{code-block-stderr}{HTML}{800000}

\colorlet{shadecolor}{code-block-code!30!white}

\def\shadecolor{\color{shadecolor}}

\colorlet{code-block-stdout-light}{code-block-stdout!40!white}

\colorlet{code-block-stderr-light}{code-block-stderr!40!white}

\renewcommand*\contentsname{Table of contents}
{
\hypersetup{linkcolor=}
\setcounter{tocdepth}{3}
\tableofcontents
}
\[
\gdef\p#1{{\left(#1\right)}}
\gdef\abs#1{{\left|#1\right|}}
\gdef\brackets#1{{\left[#1\right]}}
\gdef\braces#1{{\left\{#1\right\}}}
\gdef\ouv#1#2{\p{#1,#2}}
\gdef\repar#1#2{r_{#1\to#2}}
\gdef\reparp#1#2#3{r_{#1;#2\to#3}}
\gdef\reels{{\mathbb R}}
\gdef\reelsp{{\mathbb R_+^*}}
\gdef\natup{\mathbb N^*}
\gdef\simplex#1{\mathcal S_{#1}}
\gdef\osimplex#1{\mathring{\mathcal S}_{#1}}
\gdef\concat{\operatorname{concat}}
\gdef\cumsum{\operatorname{cumsum}}
\gdef\cumprod{\operatorname{cumprod}}
\gdef\flip{\operatorname{flip}}
\gdef\logit{\operatorname{logit}}
\gdef\expit{\operatorname{expit}}
\gdef\logupexp{\operatorname{log1pexp}}
\gdef\logexpmu{\operatorname{logexpm1}}
\gdef\range{\operatorname{range}}
\gdef\Logistic{\operatorname{Logistic}}
\gdef\erf{\operatorname{erf}}
\gdef\erfinv{\operatorname{erfinv}}
\gdef\esp{\mathbb{E}}
\gdef\sphere#1{\mathbf{S}_{#1}}
\gdef\msphere#1{\tilde{\mathbf{S}}_{#1}}
\gdef\hsphere#1{\mathbf{HS}_{#1}}
\gdef\ball#1{\mathbf{B}_{#1}}
\gdef\oball#1{\mathring{\mathbf{B}}_{#1}}
\gdef\appsim{\underset{\text{approx}}\sim}
\gdef\owedge{\bigtriangleup}
\]

\hypertarget{introduction}{%
\section{Introduction}\label{introduction}}

When working with a model for inference or optimization, there are many
cases where the parameter space is different from \(\reels^n\), for
example a space of matrices or symmetric matrices, where a simple
rewriting is enough to get back to \(\reels^n\). But, in other cases,
the parameter space is constrained, as often in statistics where the
parameters can be bounded values, strictly positive variances,
covariance matrices (symmetric positive definite matrices), mixture
proportions (simplex element), correlation matrices, etc.

There are several ways to take these constraints into account. For
example, a constrained optimization can be performed, but this often
requires writing a different method for each parameter space. In many
machine learning applications, a surjective parametrization is often
used, allowing to describe the whole parameter space from \(\reels^n\),
and to use unconstrained optimization methods (such as variants of the
stochastic gradient very used in deep learning).

The use of these non-injective parametrisations in statistics is however
a major problem since the non-injectivity induces a loss of
identifiability for the parametrized model. It is therefore important to
have surjective and injective parametrizations, in order to describe the
whole space of parameters, while keeping the identifiability of the
parameters.

Moreover, the goal of parametrization is often to be able to use
unconstrained gradient-based optimization methods, so it is necessary to
have parametrizations that are differentiable from \(\reels^n\) to the
target space. Moreover the transfer of properties obtained on
\(\reels^n\) to the target space (for example the transfer of an
asymptotic distribution of a maximum likelihood estimator using a delta
method) also requires the parametrizations to be differentiable, and in
some cases to be differentiable in both directions, i.e.
diffeomorphisms, in order to be able to arbitrarily compose the
parametrizations.

From a computational point of view, the use of modern computational
methods leads to compute gradients in an exact way by means of automatic
differentiation, often in backward mode, but also in forward mode, in
particular in the computation of Hessian matrices (for example in a
likelihood objective function framework to obtain Fisher information).
It is therefore important to use parametrizations that are automatically
differentiable in backward and forward modes.

Finally, the use of parametrizations in a computational context must be
done with caution regarding numerical stability. Some naive
parametrizations are numerically unstable or lead to unstable
algorithms, and should be avoided.

The objective of this cookbook is to introduce, in a single medium, a
set of classically useful parametrizations, in particular for
statistical model inference, which are diffeomorphisms, automatically
differentiable in forward and backward modes, and numerically stable.

As it is impossible to be completely exhaustive, it is always possible
to need different parameterizations in particular cases. The
parameterizations here are not intended to be prescriptive, but only to
present a use case that works well in most cases. In order to allow the
user to adapt and define the parametrizations, these will be explained
with the necessary details.

Each elementary parametrization is provided with example Python code
(using numpy). In order to allow for a easy use of parametrizations,
this cookbook is provided together with an optimized Python package
compatible with numpy (\protect\hyperlink{ref-numpy}{Harris et al.
2020}), JAX (\protect\hyperlink{ref-jax}{Bradbury et al. 2018}), and
PyTorch (\protect\hyperlink{ref-pytorch}{Paszke et al. 2019}) and is
specially designed to be used conjointly with optimizers provided in
JAXopt (\protect\hyperlink{ref-jaxopt}{Blondel et al. 2021}) and
PyTorch. The package can be used in two ways. Firstly, a high-level
interface is defined allowing to build complex parametrizations, and to
use them in validated computational methods. Secondly, a low-level
interface allowing to access elementary parametrizations. These two ways
of using the package are detailed in this article.

\hypertarget{sec-repars}{%
\section{Definitions of parametrizations}\label{sec-repars}}

\hypertarget{notations}{%
\subsection{Notations}\label{notations}}

\hypertarget{general-notations}{%
\subsubsection{General notations}\label{general-notations}}

\begin{itemize}
\item
  \(\repar EF\) denotes the proposed parametrization of \(E\) to \(F\).
  Since the parametrization is bijective, the following properties are
  satisfied:

  \begin{itemize}
  \tightlist
  \item
    \(\forall y\in F,\;\exists x\in E, \; \repar EF(x)=y\).
  \item
    \(\forall \p{x_1,x_2}\in E^2, \;  \brackets{\repar EF\p{x_1}=\repar EF\p{x_2}}\Rightarrow  \brackets{x_1=x_2}\).
  \end{itemize}

  We note \(\repar FE=\repar EF^{-1}\).
\item
  When the parametrization depends on a hyper-parameter (such as a scale
  parameter), we will note \(\reparp sEF\) where \(s\) is a
  hyper-parameter to choose.
\item
  We note \(\Logistic\) the probability logistic distribution on
  \(\reels\). This probability distribution has the cumulative
  distribution function \(x\mapsto\expit\p x=\frac1{1+\exp\p{-x}}\).
\item
  We note \(\Logistic_n\) the multivariate distribution on \(\reels^n\)
  with independent components and with each marginal following a
  \(\Logistic\) on \(\reels\).
\end{itemize}

\hypertarget{vectors-notations}{%
\subsubsection{Vectors notations}\label{vectors-notations}}

\begin{itemize}
\item
  Vector/matrix indices: in general, the convention used in Python will
  be used. The indices start at zero and end at \(n-1\). Thus for
  \(x\in\reels^n\) we have \(x=\p{x_0,\ldots,x_{n-1}}\).
\item
  \(\concat\) denotes the concatenation of vectors. For
  \(x\in\reels^n\), \(y\in\reels^m\), we have
  \(\concat(x,y)\in\reels^{n+m}\) with \(\concat(x,y)\) =
  \(\p{x_0,\ldots,x_{n-1},y_0,\ldots,y_{m-1}}\).
\item
  Vector/matrix slices. For \(x\in\reels^n\) we note:

  \begin{itemize}
  \tightlist
  \item
    \(x_{i:j} = \p{x_k}_{k: i\le k<j} = \p{x_i,\ldots,x_{j-1}}\).
    Warning, as in Python, the upper index is not included.
  \item
    \(x_{:i} = x_{0:i} = \p{x_k}_{k: k<i} = \p{x_0,\ldots,x_{i-1}}\).
    Warning, as in Python, the upper index is not included.
  \item
    \(x_{i:} = x_{i:n} = \p{x_k}_{k: k\ge i} = \p{x_i,\ldots,x_{n-1}}\).
  \item
    Same notation is used for matrices.
  \end{itemize}
\item
  Cumulated sum. For \(x\in\reels^n\), we note
  \(\cumsum\p x\in\reels^n\) the vector defined by: \[
    \cumsum\p x = \p{\sum_{i=0}^k x_i}_{k: 0\le k<n}
          = \p{x_0, x_0+x_1, x_0+x_1+x_2, \ldots, \sum_i x_i}
  \]
\item
  Cumulated product. For \(x\in\reels^n\), we note
  \(\cumprod\p x\in\reels^n\) the vector defined by: \[
    \cumprod\p x = \p{\prod_{i=0}^k x_i}_{k: 0\le k<n}
          = \p{x_0, x_0x_1, x_0x_1x_2, \ldots, \prod_i x_i}
  \]
\item
  Reversed vector. For \(x\in\reels^n\), we note \(\flip(x)\in\reels^n\)
  the vector defined by: \[
    \flip\p x
        = \p{x_{n-1},\ldots,x_0}
  \]
\item
  We note respectively \(\odot\), \(\oslash\), and \(\owedge\), the
  element-wise product, the element-wise division and the element-wise
  power. For \(x,y\in\reels^n\):

  \begin{itemize}
  \tightlist
  \item
    \(x\odot y = \p{x_iy_i}_{i: 0\le i<n}\)
  \item
    \(x\oslash y = \p{\frac{x_i}{y_i}}_{i: 0\le i<n}\)
  \item
    \(x\owedge y = \p{x_i^{y_i}}_{i: 0\le i<n}\)
  \end{itemize}
\item
  For \(n\in\natup\), we note \(\range(n)\) the vector of \(\reels^n\)
  defined by: \[
  \range(n) = \p{k}_{k: 0\le k<n} = \p{0, 1, \ldots, n}
  \]
\end{itemize}

\hypertarget{matrix-notations}{%
\subsubsection{Matrix notations}\label{matrix-notations}}

\begin{itemize}
\item
  Notation for diagonal and triangular matrices. All element missing
  will be zero. For example, we note: \[\begin{bmatrix}
            1&\\
            2&3\\
            4&5&6\\
            7&8&9&10\\
          \end{bmatrix}
          =
          \begin{bmatrix}
            1&0&0&0\\
            2&3&0&0\\
            4&5&6&0\\
            7&8&9&10\\
          \end{bmatrix}
  \] \[\begin{bmatrix}
            1&\\
            &2\\
            &&3\\
            &&&4\\
          \end{bmatrix}
          =
          \begin{bmatrix}
            1&0&0&0\\
            0&2&0&0\\
            0&0&3&0\\
            0&0&0&4\\
          \end{bmatrix}
  \]
\item
  Notation for symmetric matrices. We note \((\text{sym})\) to indicate
  the symmetry of the matrix, all missing elements must be filled by
  symmetry. For example, we note: \[\begin{bmatrix}
            1&&&\!\!\!\!\!\!\!\!\text{(sym)}\\
            2&3\\
            4&5&6\\
            7&8&9&10\\
          \end{bmatrix}
          =
          \begin{bmatrix}
            1&2&4&7\\
            2&3&5&8\\
            4&5&6&9\\
            7&8&9&10\\
          \end{bmatrix}\]
\item
  For \(n\in\natup\), and \(M\in\reels^{n\times n}\), we note
  \(\operatorname{diag}(M)\) the diagonal vector of matrix \(M\). \[
    \operatorname{diag}(M)=\p{M_{ii}}_{i:0\le i<n}
  \]
\item
  For \(n\in\natup\), and \(x\in\reels^n\), we note
  \(\operatorname{undiag}(x)\) the diagonal matrix \(M\) which has \(x\)
  as diagonal vector. \[\operatorname{undiag}(x)=
            \begin{bmatrix}
          x_0  \\
          & x_1 \\
          && \ddots \\
          &&& x_{n-1}
        \end{bmatrix}\]
\end{itemize}

\hypertarget{parametrization-of-scalars}{%
\subsection{Parametrization of
scalars}\label{parametrization-of-scalars}}

\hypertarget{sec-softplus}{%
\subsubsection{\texorpdfstring{\(\mathbb R_+^*\) and forms like
\((-\infty,a)\) or
\((a,{+\infty})\)}{\textbackslash mathbb R\_+\^{}* and forms like (-\textbackslash infty,a) or (a,\{+\textbackslash infty\})}}\label{sec-softplus}}

The proposed parametrization with a scale hyper-parameter
\(s\in\reelsp\):

\begin{equation}\protect\hypertarget{eq-softplus}{}{\begin{array}{ccrcl}
  \reparp s\reels\reelsp &\colon& \reels &\longrightarrow& \reelsp \\
    && x &\longmapsto& s\log\p{1+\exp x} \\
  \end{array}}\label{eq-softplus}\end{equation}

\begin{figure}

{\centering \includegraphics{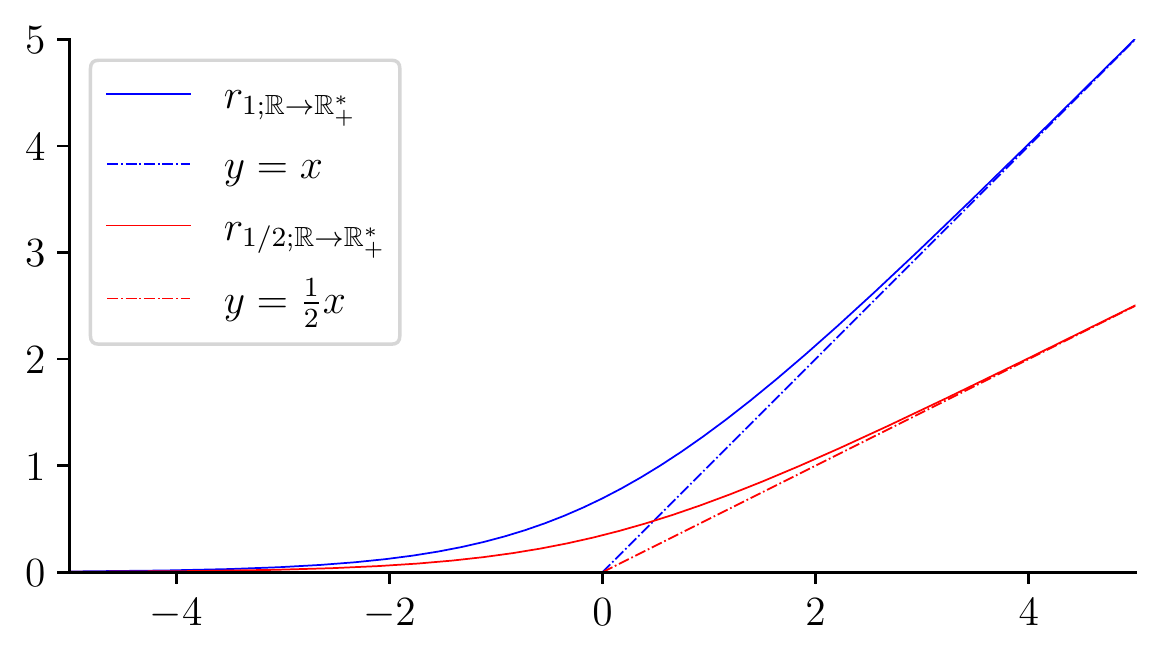}

}

\caption{\label{fig-softplus}Representation of
\(\reparp s\reels\reelsp\)}

\end{figure}

This parametrization is known as the \(\operatorname{softplus}\)
function\footnote{definitions of \(\operatorname{softplus}\) differ
  between sources. These three definitions are used in computing
  softwares \(x\mapsto \log\p{1+\exp x}\),
  \(x\mapsto  s\log\p{1+\exp x}\), and
  \(x\mapsto s\log\p{1+\exp\p{\frac xs}}\).}, and is shown
Figure~\ref{fig-softplus}.

The reciprocal function is
\begin{equation}\protect\hypertarget{eq-softplusinv}{}{\begin{array}{ccrcl}
  \reparp s\reelsp\reels&\colon& \reelsp &\longrightarrow& \reels \\
    && x &\longmapsto& \log\p{-1+\exp\p{\frac xs}} \\
  \end{array}
}\label{eq-softplusinv}\end{equation}

\hypertarget{properties}{%
\paragraph{Properties}\label{properties}}

\begin{itemize}
\item
  If \(x\sim\Logistic\), then \(\esp\p{\reparp s\reels\reelsp(x)}=s\).
\item
  \(\forall x\in\reels,\; \reparp s\reels\reelsp(-x) = \reparp s\reels\reelsp(x)-xs\).
\item
  For \(x\to-\infty\), we have
  \(\reparp s\reels\reelsp(x)=s\exp x + o\p{s\exp x}\).
\item
  For \(x\to+\infty\), we have
  \(\reparp s\reels\reelsp(x)=sx+s\exp\p{-x}+o\p{s\exp-x}\).
\end{itemize}

\hypertarget{alternatives}{%
\paragraph{Alternatives}\label{alternatives}}

The naive parametrization to go from \(\reels\) to \(\reelsp\) is to use
the exponential function. This is the approach often used naively in
statistics to parameterize the variance. This approach should be avoided
in computational approaches using gradient-based algorithms because the
exponential tends to increase the instability of the method, leading to
divergent algorithms.

\hypertarget{choice-of-the-scale-hyper-parameter}{%
\paragraph{Choice of the scale
hyper-parameter}\label{choice-of-the-scale-hyper-parameter}}

We will often choose the scale hyper-parameter \(s\) as the expected
order of magnitude of the parameter. The choice of a too small \(s\) in
an iterative algorithm using the gradient can lead to \(x\) being very
negative, thus of very weak gradient without the optimum being reached.
The choice of a too large \(s\) leads to be locally an exponential and
implies the same problems as an exponential parametrization.

\hypertarget{implementation-details}{%
\paragraph{Implementation details}\label{implementation-details}}

For \(\reparp s\reels\reelsp\), the naive use of the expression
\(x\mapsto s\log\p{1+\exp x}\) in Equation~\ref{eq-softplus} leads to:

\begin{itemize}
\item
  an impossible calculation for slightly high values of \(x\) leading to
  not being able to manipulate \(\exp x\),
\item
  an inexact calculation for very negative \(x\) values, where the
  numerical rounding leads to \(1+\exp x\approx 1\).
\end{itemize}

For the first point, we can notice that
\(\forall y\in\reels,\; \log\p{1+\exp y}=\log\p{1+\exp{-\abs y}}+y^+\)
with \(y^+=\max(x,0)=\frac{\abs{x}+x}2\). Within this expression, the
manipulated exponential cannot become large.

For the second point, there exists in the majority of calculation
libraries the function \(\operatorname{log1p}\) defined as
\(y\mapsto \log\p{1+y}\). This function is designed to calculate the
value precisely, even for argument values close to zero. We therefore
introduce the function \(\logupexp: x\mapsto\log\p{1+\exp\p x}\)
implemented as follows:

\begin{equation}\protect\hypertarget{eq-logupexp}{}{\begin{array}{ccrcl}
  \logupexp &\colon& \reels &\longrightarrow& \reelsp \\
    && x &\longmapsto&
    \operatorname{log1p}\p{\exp\p{-\abs{x}}} + x^+\\
  \end{array}}\label{eq-logupexp}\end{equation}

For the implementation of \(\reparp s\reels\reelsp\), we should use:

\begin{equation}\protect\hypertarget{eq-softplus-impl}{}{\begin{array}{ccrcl}
  \reparp s\reels\reelsp &\colon& \reels &\longrightarrow& \reelsp \\
    && x &\longmapsto&
    s\logupexp\p x\\
  \end{array}}\label{eq-softplus-impl}\end{equation}

For \(\reparp s\reelsp\reels\), the naive use of the expression
\(x\mapsto \log\p{-1+\exp\p{\frac xs}}\) in
Equation~\ref{eq-softplusinv}, leads to:

\begin{itemize}
\item
  an impossible computation for slightly high values of \(\frac xs\)
  leading to not being able to manipulate \(\exp\p{\frac xs}\),
\item
  an impossible calculation or calculation errors when \(\frac xs\) is
  close to \(0\) and \(\exp\p{\frac xs}\) is thus close to \(1\),
  because errors will be made on the calculation
  \(-1+\exp\p{\frac xs}\).
\end{itemize}

To answer the first point, we can notice that
\(\forall y\in\reelsp,\;\log\p{-1+\exp y}=y+\log\p{-\exp\p{-y}+1}\).
Within this last expression the manipulated exponential cannot become
large.

For the second point, there exists in the majority of calculation
libraries the function \(\operatorname{expm1}\) defined as
\(x\mapsto \exp(y)-1\). This function is designed to calculate its value
precisely, including when \(y\) approaches \(0\) (and therefore
\(\exp y\) approaches \(1\)). We therefore introduce the function
\(\logexpmu: x\mapsto\log\p{\exp\p x -1}\) defined as follows:

\[\begin{array}{ccrcl}
  \logexpmu &\colon& \reelsp &\longrightarrow& \reels \\
    && x &\longmapsto&
    x+\log\p{-\operatorname{expm1}\p{-x}}
  \end{array}\]

For the implementation of \(\reparp s\reelsp\reels\), we should use:

\begin{equation}\protect\hypertarget{eq-softplusinv-impl}{}{\begin{array}{ccrcl}
  \repar\reelsp\reels &\colon& \reelsp &\longrightarrow& \reels \\
    && x &\longmapsto&
    \logexpmu\p{\frac xs}
  \end{array}}\label{eq-softplusinv-impl}\end{equation}

\hypertarget{example-of-implementation}{%
\paragraph{Example of implementation}\label{example-of-implementation}}

We provide vectorized form of implementations of
\(\reparp s{\reels^n}{\reelsp^n}\) and
\(\reparp s{\reelsp^n}{\reels^n}\) for \(n\in\natup\).

\begin{Shaded}
\begin{Highlighting}[]
\ImportTok{import}\NormalTok{ numpy }\ImportTok{as}\NormalTok{ np}

\KeywordTok{def}\NormalTok{ log1pexp(x):}
    \ControlFlowTok{return}\NormalTok{ np.log1p(np.exp(}\OperatorTok{{-}}\NormalTok{np.}\BuiltInTok{abs}\NormalTok{(x))) }\OperatorTok{+}\NormalTok{ np.maximum(x, }\DecValTok{0}\NormalTok{)}

\KeywordTok{def}\NormalTok{ softplus(x, scale}\OperatorTok{=}\DecValTok{1}\NormalTok{):}
    \ControlFlowTok{return}\NormalTok{ scale }\OperatorTok{*}\NormalTok{ log1pexp(x)}

\KeywordTok{def}\NormalTok{ logexpm1(x):}
    \ControlFlowTok{return}\NormalTok{ x }\OperatorTok{+}\NormalTok{ np.log(}\OperatorTok{{-}}\NormalTok{np.expm1(}\OperatorTok{{-}}\NormalTok{x))}

\KeywordTok{def}\NormalTok{ softplusinv(x, scale}\OperatorTok{=}\DecValTok{1}\NormalTok{):}
    \ControlFlowTok{return}\NormalTok{ logexpm1(x}\OperatorTok{/}\NormalTok{scale)}
\end{Highlighting}
\end{Shaded}

\hypertarget{forms-like--inftya-or-ainfty}{%
\paragraph{\texorpdfstring{Forms like \((-\infty,a)\) or
\((a,{+\infty})\)}{Forms like (-\textbackslash infty,a) or (a,\{+\textbackslash infty\})}}\label{forms-like--inftya-or-ainfty}}

Definitions of these parametrization involves
\(\reparp s\reels\reelsp\), translation, and optionally a symmetry.

We define, for \(a\in\reels\):

\begin{equation}\protect\hypertarget{eq-forms-reelsp}{}{\begin{array}{ccrcl}
    \reparp s\reels{\ouv{-\infty}a} &\colon& \reels &\longrightarrow& \ouv{-\infty}a \\
    && x &\longmapsto&
    a - \reparp s\reels\reelsp\p{x} \\
    \\
    \reparp s{\ouv{-\infty}a}\reels &\colon& \ouv{-\infty}a &\longrightarrow& \reels \\
    && x &\longmapsto&
    \reparp s\reelsp\reels\p{a-x} \\
    \\
    \reparp s\reels{\ouv a{+\infty}} &\colon& \reels &\longrightarrow& \ouv a{+\infty} \\
    && x &\longmapsto&
    a + \reparp s\reels\reelsp\p{x} \\
    \\
    \reparp s{\ouv a{+\infty}}\reels &\colon& \ouv a{+\infty} &\longrightarrow& \reels \\
    && x &\longmapsto&
    \reparp s\reelsp\reels\p{x-a}
  \end{array}}\label{eq-forms-reelsp}\end{equation}

\hypertarget{sec-expit}{%
\subsubsection{\texorpdfstring{\((0,1)\) and forms like
\((a,b)\)}{(0,1) and forms like (a,b)}}\label{sec-expit}}

The proposed parametrization is:

\begin{equation}\protect\hypertarget{eq-expit}{}{\begin{array}{ccrcl}
    \repar\reels{\ouv01} &\colon& \reels &\longrightarrow& \ouv01 \\
    && x &\longmapsto& \expit\p x = \frac1{1+\exp\p{-x}} \\
  \end{array}
  }\label{eq-expit}\end{equation}

\begin{figure}

{\centering \includegraphics{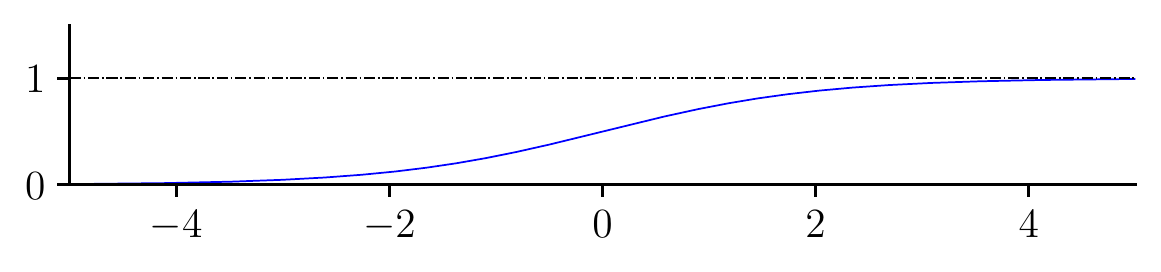}

}

\caption{\label{fig-expit}Representation of \(\reparp s\reels{\ouv01}\)}

\end{figure}

This function is known as \(\operatorname{expit}\) in some libraries.
This parametrization is represented in Figure~\ref{fig-expit}.

The reciprocal parametrization is:

\begin{equation}\protect\hypertarget{eq-logit}{}{\begin{array}{ccrcl}
    \repar{\ouv01}\reels &\colon& \ouv01 &\longrightarrow& \reels \\
    && x &\longmapsto& \logit\p x = \log\frac x{1-x} \\
  \end{array}
  }\label{eq-logit}\end{equation}

This parametrization is the one allowing to describe the Bernoulli
distribution within the exponential family. Thus, this parametrization
is the one used in logistic regression in the framework of generalized
linear models.

\hypertarget{properties-1}{%
\paragraph{Properties}\label{properties-1}}

\begin{itemize}
\item
  If \(x\sim\Logistic\), then
  \(\repar\reels{\ouv01}(x)\sim\mathcal U_{\ouv01}\).
\item
  \(\forall x\in\reels,\; \repar\reels{\ouv01}(-x)=1-\repar\reels{\ouv01}(x)\).
\item
  For \(x\to-\infty\), we have
  \(\repar\reels{\ouv01}(x)=\exp x + o\p{\exp x}\).
\item
  For \(x\to+\infty\), we have
  \(\repar\reels{\ouv01}(x)=1-\exp\p{-x}+o\p{\exp\p{-x}}\).
\end{itemize}

\hypertarget{alternatives-1}{%
\paragraph{Alternatives}\label{alternatives-1}}

Any bijective and differentiable function of \(\reels\) in \(\ouv01\)
can be a parameterization, in particular, any cumulative distribution
function of an absolutely continuous probability law with support on
\(\reels\). For example, the cumulative distribution function of the law
\(\mathcal N(0,1)\) can be theoretically used, but turns out to be a bad
choice in most practical cases because of the lightness of the
distribution tails which leads asymptotes in \(0\) and \(1\) being
reached very quickly, and thus to very weak derivatives leading to
stalled iterative algorithms and numerical errors.

\hypertarget{implementation-details-1}{%
\paragraph{Implementation details}\label{implementation-details-1}}

In many scientific computing libraries, the functions
\(\operatorname{expit}\) and \(\logit\) are available within the special
functions. These implementations should be used when available. This
section is provided only for users of scientific libraries which do not
contains these functions.

The formula presented in Equation~\ref{eq-expit} is a fairly good
choice, but suffers from a problem: when the argument \(x\) is very
negative, \(\exp\p{-x}\) is very large, and an overflow can appear in
the calculation. This overflow is well handled in most cases, but it
often leads to an exception mechanism that can slow down the calculation
and cause undesirable effects in some cases.

A second option is to use the hyperbolic tangent with the following
formula:

\[\forall x\in\reels,\quad \repar\reels{\ouv01}=\frac{1+\tanh\p{\frac x2}}2\]

However, when \(x\) is very negative, then \(\tanh\p{\frac x2}\) on
floats is computed as being exactly \(-1\), this leads to an exactly
zero result, while the result obtained is encodable on floats close to
zero but not zero. The same behavior is observed when the arguments
\(x\) is positive, but as the result obtained is not encodable on floats
close to \(1\), it is not possible to hope for better.

In some works, the function \(\operatorname{expit}\) is introduced as
\(x\mapsto \frac{\exp x}{1+\exp x}\). This expression behaves very well
for negative arguments, but is to be avoided for positive arguments,
since a large value of the argument \(x\) leads to \texttt{NaN}.

It is possible to merge the good behavior of the latter formula which
behaves very well on negative arguments and the formula introduced in
Equation~\ref{eq-expit}, using the following property:

\[\forall x\in\reels,\quad \repar\reels{\ouv01}(x) = \frac{\exp x^-}{\exp
  x^-+\exp\p{-x^+}}\]

Where \(x^-=\min(x,0)\) and \(x^+=\max(x,0)\).

We can therefore use:

\begin{equation}\protect\hypertarget{eq-expit-impl}{}{\begin{array}{ccrcl}
    \repar\reels{\ouv01} &\colon& \reels &\longrightarrow& \ouv01 \\
    && x &\longmapsto& \frac{\exp\p{x^-}}{\exp\p{x^-}+\exp\p{-x^+}} \\
  \end{array}
  }\label{eq-expit-impl}\end{equation}

Regarding the reciprocal, using a ratio in Equation~\ref{eq-logit} poses
a rounding problem for values of \(x\) close to \(1\), and it is better
to use the expanded form of
\(\forall x\in\ouv01,\; \log\frac x{1-x} = \log x-\log\p{1-x}\). One can
use the function \(\operatorname{log1p}\colon y\mapsto \log(1+y)\), but
this is not of major interest here, since
\(\abs{\log(1-y)}\ll\abs{\log y}\) for \(y\) close to \(0\). We can
therefore use:

\begin{equation}\protect\hypertarget{eq-logit-impl}{}{\begin{array}{ccrcl}
    \repar{\ouv01}\reels &\colon& \ouv01 &\longrightarrow& \reels \\
    && x &\longmapsto& \log x - \operatorname{log1p}\p{-x} \\
  \end{array}
  }\label{eq-logit-impl}\end{equation}

\hypertarget{implementation-example}{%
\paragraph{Implementation example}\label{implementation-example}}

We will use the implementations of \texttt{scipy.special}:

\begin{Shaded}
\begin{Highlighting}[]
\ImportTok{import}\NormalTok{ scipy.special}

\NormalTok{expit }\OperatorTok{=}\NormalTok{ scipy.special.expit}
\NormalTok{logit }\OperatorTok{=}\NormalTok{ scipy.special.logit}
\end{Highlighting}
\end{Shaded}

\hypertarget{forms-like-ouv-ab-with-ab}{%
\paragraph{\texorpdfstring{Forms like \(\ouv ab\) with
\(a<b\):}{Forms like \textbackslash ouv ab with a\textless b:}}\label{forms-like-ouv-ab-with-ab}}

We will use a simple affine transformation to go from \(\ouv01\) to
\(\ouv ab\). Thus, the introduced parametrization is:

\begin{equation}\protect\hypertarget{eq-forms-01}{}{\begin{array}{ccrcl}
    \repar\reels{\ouv ab} &\colon& \reels &\longrightarrow& \ouv ab \\
    && x &\longmapsto&
    a + (b-a) \repar\reels{\ouv01}\p{x} \\
    \\
    \repar{\ouv ab}\reels &\colon& \ouv ab &\longrightarrow& \reels \\
    && x &\longmapsto&
    \repar{\ouv01}\reels\p{\frac{x-a}{b-a}} \\
  \end{array}}\label{eq-forms-01}\end{equation}

\hypertarget{detail-of-the-implementation-of-the-special-cases-of-forms-ouv-11-and-ouv-aa-with-ainreelsp}{%
\paragraph{\texorpdfstring{Detail of the implementation of the special
cases of forms \(\ouv{-1}1\) and \(\ouv{-a}a\) with
\(a\in\reelsp\)}{Detail of the implementation of the special cases of forms \textbackslash ouv\{-1\}1 and \textbackslash ouv\{-a\}a with a\textbackslash in\textbackslash reelsp}}\label{detail-of-the-implementation-of-the-special-cases-of-forms-ouv-11-and-ouv-aa-with-ainreelsp}}

When the set is symmetric, we have

\begin{itemize}
\item
  \(\repar\reels{\ouv{-a}a}(0)=0\),
\item
  \({\left.\frac{\operatorname d\repar{\ouv{-a}a}(x)}{\operatorname dx}\right|}_{x=0}=\frac a2\).
\end{itemize}

It can be judicious to be precise near zero. If we take
\(\abs{\varepsilon}\ll1\), we should obtain
\(\repar\reels{\ouv{-a}a}\p\varepsilon\approx\frac\varepsilon2\). But
this is not computable for very small values using a transformation of
the parametrization \(\repar{\reels}{\ouv01}\). In this particular case,
it is interesting to use the function \(\tanh\), noting that
\(\forall x\in\reels,\; \tanh\p{\frac x2} = 2\expit\p x-1\). Thus, we
will use for \(\ouv{-a}a\):

\begin{equation}\protect\hypertarget{eq-forms-01-impl}{}{\begin{array}{ccrcl}
    \repar\reels{\ouv{-a}a} &\colon& \reels &\longrightarrow& \ouv{-a}a \\
    && x &\longmapsto& a\tanh\p{\frac x2} \\
    \\
    \repar{\ouv{-a}a}\reels &\colon& \ouv{-a}a &\longrightarrow& \reels \\
    && x &\longmapsto& 2\operatorname{arctanh}\p{\frac xa} \\
  \end{array}}\label{eq-forms-01-impl}\end{equation}

\hypertarget{parametrization-of-vectors}{%
\subsection{Parametrization of
vectors}\label{parametrization-of-vectors}}

\hypertarget{sec-simplex}{%
\subsubsection{\texorpdfstring{Simplex
\(\mathcal S_n\)}{Simplex \textbackslash mathcal S\_n}}\label{sec-simplex}}

We define the unit simplex of dimension \(n\) as a part of a hyperplane
of \(\reels^{n+1}\):

\[\simplex n = \braces{x\in\reels_+^{n+1} : \sum_i x_i=1} \subset \reels^{n+1}\]

However, \(\simplex n\) is not an open set (w.r.t. the topology on
hyperplane \(\braces{x\in\reels^{n+1} : \sum_ix_i=1}\)), which prevents
the definition of the differentiable bijection with an open set. We will
therefore work with the simplex deprived of its boundary, \emph{i.e.}:

\[\osimplex n = \braces{x\in\reelsp^{n+1} :  \sum_i x_i=1}\]

\hypertarget{construction-of-the-parametrization-between-osimplex-n-and-reelsn}{%
\paragraph{\texorpdfstring{Construction of the parametrization between
\(\osimplex n\) and
\(\reels^n\)}{Construction of the parametrization between \textbackslash osimplex n and \textbackslash reels\^{}n}}\label{construction-of-the-parametrization-between-osimplex-n-and-reelsn}}

The parametrization is expressed naturally at first between
\(\osimplex n\) and \(\ouv01^n\). It will then suffice to compose with
the parametrization of \(\ouv01^n\) in \(\reels^n\) to obtain a
parametrization of \(\osimplex n\) in \(\reels^n\).

The initial idea is to consider that for \(n\in\natup\) we can easily
define a bijection between \(\ouv01\times\osimplex n\) and
\(\osimplex{n+1}\) by: \[\begin{array}{ccrcl}
    && \ouv01\times\osimplex n&\longrightarrow& \osimplex{n+1} \\
    && (x,y) &\longmapsto& \concat\p{1-x, xy} \\
  \end{array}\]

We can then introduce the parametrization defined in a recursive way:

\begin{equation}\protect\hypertarget{eq-simpr}{}{\begin{array}{crcrcl}
    &h_1 &\colon& \ouv01 &\longrightarrow& \osimplex1 \\
    &&& x &\longmapsto& (1-x, x) \\ \\
    \forall n\in\natup,& h_{n+1} &\colon& \ouv01^{n+1}&\longrightarrow& \osimplex{n+1} \\
    &&& x &\longmapsto& \concat\p{1-x_0, x_0h_n\p{x_{1:}}} \\
  \end{array}
  }\label{eq-simpr}\end{equation}

So we can deduce for Equation~\ref{eq-simpr} a un-recursived form of
\(h_n\):
\begin{equation}\protect\hypertarget{eq-simph}{}{\begin{array}{ccrcl}
  h_n &\colon& \ouv01^n &\longrightarrow& \osimplex n\\
  && x &\longmapsto& \concat\p{1-x,1}\odot\concat\p{1,\cumprod\p x}
      \end{array}
  }\label{eq-simph}\end{equation}

And we can also write the reciprocal:
\begin{equation}\protect\hypertarget{eq-simphm1}{}{\begin{array}{ccrcl}
        h_n^{-1} &\colon& \osimplex n &\longrightarrow& \ouv01^n\\
        && x &\longmapsto&
        \flip\p{\cumsum\p{\flip\p{x_{1:}}}}\oslash\p{x_{:n-1}+\flip\p{\cumsum\p{\flip\p{x_{1:}}}}}\\
      \end{array}
  }\label{eq-simphm1}\end{equation}

One can notice that the components of the argument \(x\) in \(h_n(x)\)
have a very different importance. Thus, a small variation of \(x_0\)
will induce a very large variation of \(h_n(x)\), while a small
variation of \(x_{n-1}\) will induce a very small variation of
\(h_n(x)\).

Another way to formulate the problem is to study the distribution of
\(x\) for a \(h_n(x)\) taken uniformly in \(\osimplex n\), the objective
is to obtain a uniform distribution on \(\ouv01\) implying a similar
importance of all components. Let denotes \(\mathcal U_{\osimplex n}\)
the uniform probability distribution on \(\osimplex n\) w.r.t. the
Lebesgue measure of the hyperplane
\(\braces{x\in\reels^{n+1} : \sum_i x_i = 1}\). It can be shown that if
\(y\sim\mathcal U_{\osimplex n}\), and \(x = h_n^{-1}(y)\), then
\(\forall k: 0\le k<n,\;x_k\sim B(n-k,1)\), with \(B(\alpha, \beta)\)
the Beta probability distribution of parameter \((\alpha, \beta)\). The
reciprocal being true if we add the independence between the \(x_k\).
Knowing that the distribution function of \(B(n-k,1)\) is the function
defined on \(\ouv01\), \(x\mapsto x^{n-k}\), then we have
\(\forall k: 0\le k<n, \;x_k^{n-k}\sim \mathcal U_{\ouv01}\).

Therefore, we define the function

\[\begin{array}{ccrcl}
    g_n^{-1} &\colon&\ouv01^n&\longrightarrow&\ouv01^n\\
    &&x&\longmapsto& \p{x_k^{n-k}}_{k:0\le k<n} =
    \p{x_0^{n},\ldots,x_{n-1}^1}\\ \\
    g_n &\colon&\ouv01^n&\longrightarrow&\ouv01^n\\
    &&x&\longmapsto& \p{x_k^{\frac1{n-k}}}_{k:0\le k<n} =
    \p{x_0^{\frac1n},\ldots,x_{n-1}^1}\\
  \end{array}\]

We can now prove\footnote{the ideas are present here, the demonstration
  is not provided, it requires the introduction of many notations and is
  of little interest in this paper.} that
\(y\sim\mathcal U_{\osimplex n}\) if and only if
\(g_n^{-1}(h_n^{-1}(y))\sim\mathcal U_{\ouv01^n}\).

Finally, it only remains to apply a transformation defined between
\(\reels^n\) and \(\ouv01^n\). We choose to use a version of
\(x\mapsto\expit\p{-x}\), similar to the one used in
Section~\ref{sec-expit}. We choose to put a minus on the argument so
that \(\repar{\reels^n}{\osimplex n}(x)_k\) is increasing with respect
to \(x_k\) for \(0\le k<n\). We obtain:

\begin{equation}\protect\hypertarget{eq-simp}{}{\begin{array}{ccrcl}
    \repar{\reels^n}{\osimplex n} &\colon& \reels^n &\longrightarrow&
    \osimplex n\\
    &&x&\longmapsto&h_n\p{g_n\p{\expit\p{-x}}}
  \end{array}
  }\label{eq-simp}\end{equation}

\begin{figure}

{\centering \includegraphics{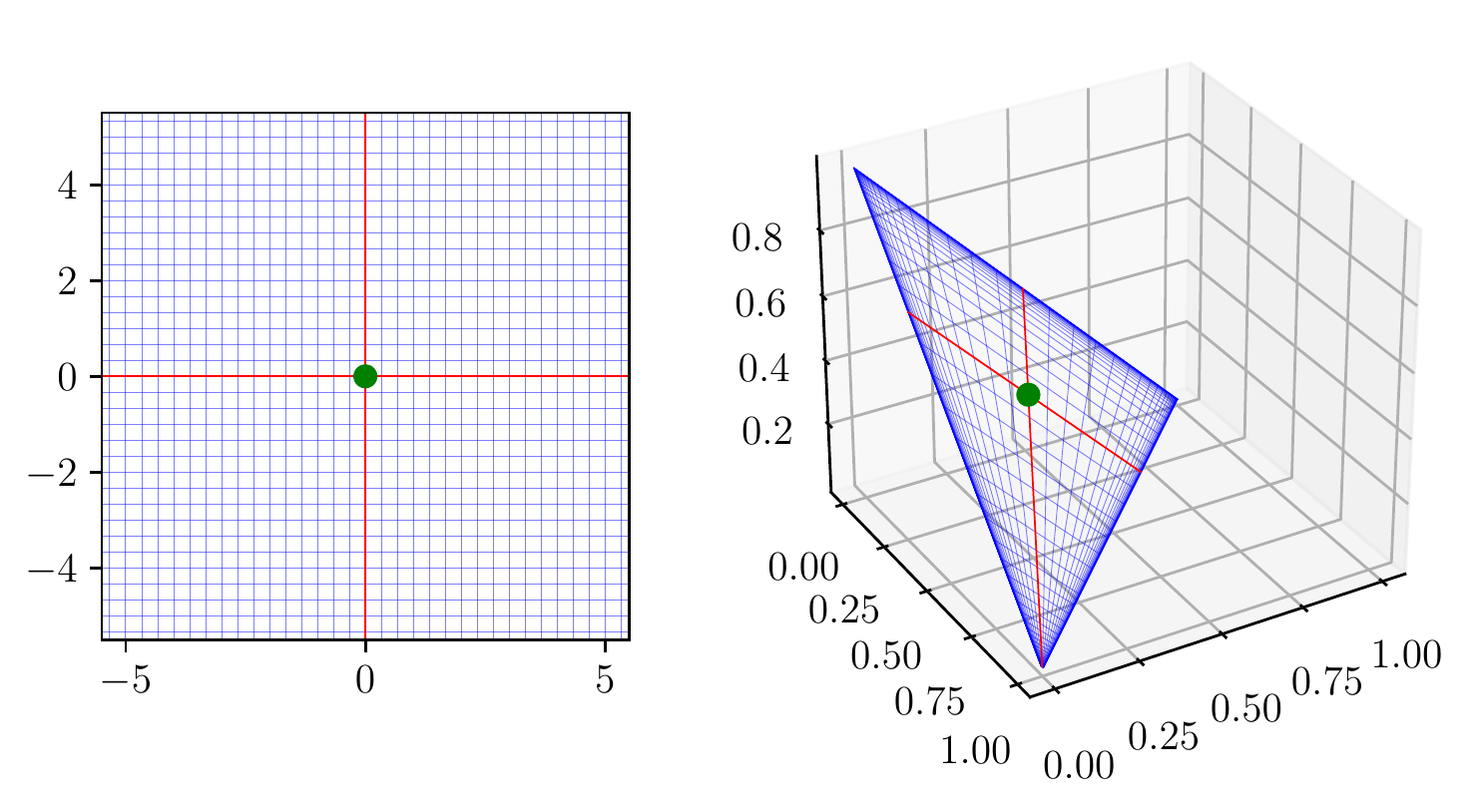}

}

\caption{\label{fig-simp}Representation of
\(\repar{\reels^2}{\osimplex 2}\). Left: a grid in \(\reels^2\). Right:
image of the grid by the transformation
\(\repar{\reels^2}{\osimplex 2}\).}

\end{figure}

For \(n=2\), a representation of \(\repar{\reels^n}{\osimplex n}\) is
shown Figure~\ref{fig-simp}.

We also obtain the reciprocal:

\begin{equation}\protect\hypertarget{eq-simpinv}{}{\begin{array}{ccrcl}
    \repar{\osimplex n}{\reels^n} &\colon& \osimplex n &\longrightarrow&
    \reels^n\\
    &&x&\longmapsto&-\logit\p{g_n^{-1}\p{h_n^{-1}\p{x}}}
  \end{array}}\label{eq-simpinv}\end{equation}

\hypertarget{properties-2}{%
\paragraph{Properties}\label{properties-2}}

\begin{itemize}
\item
  With \(x\) a random variable on \(\reels^n\), the following two
  propositions are equivalent:

  \begin{itemize}
  \tightlist
  \item
    \(x\sim\Logistic_n\)
  \item
    \(\repar{\reels^n}{\osimplex n}(x)\sim\mathcal U_{\osimplex n}\).
  \end{itemize}
\item
  Let \(\p{x^{(t)}}_{t\in\mathbb N}\) be a sequence with value in
  \(\reels^n\). Let \(\p{y^{(t)}}_{t\in\mathbb N}\) be a sequence with
  value in \(\osimplex n\) such that
  \(\forall y, \; y^{(t)} = \repar{\reels^n}{\osimplex n}\p{x^{(t)}}\),
  then for any \(k:0\le k\le n\) the following two propositions are
  equivalent:

  \begin{itemize}
  \item
    \(\lim_{t\to+\infty}y_k^{(t)} = 1\)
  \item
    \(\forall k'<k,\; \lim_{t\to+\infty} x_{k'}^{(t)} = -\infty\) and if
    \(k<n\) \(\lim_{t\to+\infty} x_{k}^{(t)} = +\infty\)
  \end{itemize}
\end{itemize}

\hypertarget{duxe9tails-dimpluxe9mentation-2}{%
\paragraph{Implementation
details}\label{duxe9tails-dimpluxe9mentation-2}}

We can note that the formula presented in Equation~\ref{eq-simp} should
not be used naively, it leads to a power and a product of values between
\(0\) and \(1\). A first step will therefore be to do the calculation in
logarithmic space. For a \(x\in\reels^n\) we have:

\[\begin{aligned}
  \log\p{h_n\p{g_n\p{\expit\p{-x}}}} =&
  \log\big[\concat\p{1-g_n\p{\expit\p{-x}},1} \\
  &\odot\concat\p{1,\cumprod\p{g_n\p{\expit\p{-x}}}}\big]
  \\
  =&
  \concat\p{\log\p{1-g_n\p{\expit\p{-x}}},0}\\
  &+\concat\p{0,\cumsum\p{\log\p{g_n\p{\expit\p{-x}}}}}\end{aligned}\]

We isolate the last element to be calculated:
\(\log\p{g_n\p{\expit\p{-x}}}\).

\[\begin{aligned}
  \log\p{g_n\p{\expit\p{-x}}}
  &= \log\p{\expit\p{-x}}\oslash\p{n-\range(n)} \\
  &= -\log\p{1+\exp\p{x}}\oslash\p{n-\range(n)} \\
  &= -\logupexp\p{x}\oslash\p{n-\range(n)} \\\end{aligned}\]

where the function \(\logupexp\) is the one defined in
Equation~\ref{eq-logupexp}.

Then we return to the first element, using the fact that it is better to
calculate a power setting as a product in logarithmic space:

\[\begin{aligned}
  \log\p{1-g_n\p{\expit\p{-x}}}
  &= \log\p{1-\exp\p{\log\p{g_n\p{\expit\p{-x}}}}} \\\end{aligned}\]

Knowing that the values of \(\log\p{g_n\p{\expit\p{-x}}}\) are negative
and can be close to \(0\), it will be better to use the function
\(\operatorname{expm1}\):

\[\begin{aligned}
  \log\p{1-g_n\p{\expit\p{-x}}}
  &= \log\p{-\operatorname{expm1}\p{\log\p{g_n\p{\expit\p{-x}}}}} \\
  &= \log\p{-\operatorname{expm1}\p{-\logupexp\p{x}\oslash\p{n-\range(n)}}} \\\end{aligned}\]

Thus we are able to calculate \(\log\p{h_n\p{g_n\p{\expit\p{-x}}}}\). It
is therefore theoretically sufficient to apply the function \(\exp\) to
obtain the value of the parametrization. However, the accumulation of
successive calculation errors can lead to the sum of the coordinates
being numerically different from \(1\). To mitigate this error, it is
preferable to use the function \(\operatorname{softmax}\) which is
strictly equivalent to the function \(\exp\) in this case. This one is
defined as:

\begin{equation}\protect\hypertarget{eq-softmax}{}{\begin{array}{ccrcl}
    \operatorname{softmax}&\colon&\reels^{n+1}&\longrightarrow&\osimplex n\\
    &&x&\longmapsto&\frac{\exp\p x}{\sum_{k=0}^n \exp x_k}\\
  \end{array}}\label{eq-softmax}\end{equation}

For the sake of numerical stability, we prefer to use the equivalent
implementation:

\begin{equation}\protect\hypertarget{eq-softmax-impl}{}{\begin{array}{ccrcl}
    \operatorname{softmax}&\colon&\reels^{n+1}&\longrightarrow&\osimplex n\\
    &&x&\longmapsto&\frac{\exp\p{x-\max x}}{\sum_{k=0}^n \exp\p{x_k-\max x}}\\
  \end{array}}\label{eq-softmax-impl}\end{equation}

We thus obtain the implementation of the parametrization:

\begin{equation}\protect\hypertarget{eq-simp-impl}{}{\begin{array}{ccrcl}
    \repar{\reels^n}{\osimplex n} &\colon& \reels^n &\longrightarrow&
    \osimplex n\\
    &&x&\longmapsto&\operatorname{softmax}\big(\\
    &&&&\quad\concat\p{\log\p{-\operatorname{expm1}\p{\xi(x)}},0}\\
    &&&&\quad+\concat\p{0,\cumsum\p{\xi(x)}}\\
    &&&&\big)\\
    \\
    \text{with }\xi&\colon&x&\longmapsto&-\logupexp\p{x}\oslash\p{n-\range(n)}\\
  \end{array}
  }\label{eq-simp-impl}\end{equation}

For the reciprocal, in Equation~\ref{eq-simpinv} the first step is to
express the calculation as a function of
\(\log\p{g_n^{-1}\p{h_n^{-1}\p{x}}}\), for which the passage to the
power of \(g_n^{-1}\) is less problematic.

\[\begin{aligned}
  -\logit\p{g_n^{-1}\p{h_n^{-1}\p{x}}} =&
  \log\p{1-g_n^{-1}\p{h_n^{-1}\p{x}}}-\log\p{g_n^{-1}\p{h_n^{-1}\p{x}}} \\
  =&
  \log\p{1-\exp\p{\log\p{g_n^{-1}\p{h_n^{-1}\p{x}}}}}-\log\p{g_n^{-1}\p{h_n^{-1}\p{x}}} \\
  =&
  \log\p{-\operatorname{expm1}\p{\log\p{g_n^{-1}\p{h_n^{-1}\p{x}}}}}-\log\p{g_n^{-1}\p{h_n^{-1}\p{x}}}
  \\\end{aligned}\]

It is now sufficient to calculate \(\log\p{g_n^{-1}\p{h_n^{-1}\p{x}}}\):

\[\begin{aligned}
  \log\p{g_n^{-1}\p{h_n^{-1}\p{x}}}
  =& \p{n-\range(n)}\odot\log\p{h_n^{-1}\p x} \\
  =& \p{n-\range(n)}\odot\log\big(\\
  &\quad\flip\p{\cumsum\p{\flip\p{x_{1:}}}}\oslash\p{x_{:n-1}+\flip\p{\cumsum\p{\flip\p{x_{1:}}}}}\big) \\
  =& -\p{n-\range(n)}\odot\log\p{1+x_{:n-1}\oslash\flip\p{\cumsum\p{\flip\p{x_{1:}}}}} \\
  =& -\p{n-\range(n)}\odot\operatorname{log1p}\p{x_{:n-1}\oslash\flip\p{\cumsum\p{\flip\p{x_{1:}}}}} \\\end{aligned}\]

We thus obtain the implementation of the reciprocal parametrization:

\begin{equation}\protect\hypertarget{eq-simpinv-implem}{}{\begin{array}{rcrcl}
    \repar{\osimplex n}{\reels^n} &\colon& \osimplex n &\longrightarrow&
    \reels^n\\
    &&x&\longmapsto&\log\p{-\operatorname{expm1}\p{\xi(x)}}-\xi(x)\\
    \\
    \text{with } \xi&\colon&x&\longmapsto&
    -\p{n-\range(n)}\\
    &&&&\quad\odot\operatorname{log1p}\p{x_{:n-1}\oslash\flip\p{\cumsum\p{\flip\p{x_{1:}}}}} \\
  \end{array}
  }\label{eq-simpinv-implem}\end{equation}

\hypertarget{implementation-example-1}{%
\paragraph{Implementation example}\label{implementation-example-1}}

This implementation uses the \texttt{log1pexp} function defined in the
example implementation in Section~\ref{sec-softplus}.

\begin{Shaded}
\begin{Highlighting}[]
\ImportTok{import}\NormalTok{ numpy }\ImportTok{as}\NormalTok{ np}

\KeywordTok{def}\NormalTok{ reals\_to\_simplex(x):}
\NormalTok{    n }\OperatorTok{=}\NormalTok{ x.size}
\NormalTok{    ksi }\OperatorTok{=} \OperatorTok{{-}}\NormalTok{log1pexp(x) }\OperatorTok{/}\NormalTok{ np.arange(n, }\DecValTok{0}\NormalTok{, }\OperatorTok{{-}}\DecValTok{1}\NormalTok{)}
\NormalTok{    logvalues }\OperatorTok{=}\NormalTok{ (}
\NormalTok{        np.concatenate((np.log(}\OperatorTok{{-}}\NormalTok{np.expm1(ksi)), (}\DecValTok{0}\NormalTok{,)))}
        \OperatorTok{+}\NormalTok{ np.concatenate(((}\DecValTok{0}\NormalTok{,), np.cumsum(ksi)))}
\NormalTok{    )}
\NormalTok{    values }\OperatorTok{=}\NormalTok{ np.exp(logvalues }\OperatorTok{{-}}\NormalTok{ logvalues.}\BuiltInTok{max}\NormalTok{())}
    \ControlFlowTok{return}\NormalTok{ values }\OperatorTok{/}\NormalTok{ values.}\BuiltInTok{sum}\NormalTok{()}

\KeywordTok{def}\NormalTok{ simplex\_to\_reals(x):}
\NormalTok{    n }\OperatorTok{=}\NormalTok{ x.size }\OperatorTok{{-}} \DecValTok{1}
\NormalTok{    ksi }\OperatorTok{=}\NormalTok{ (}
        \OperatorTok{{-}}\NormalTok{np.arange(n, }\DecValTok{0}\NormalTok{, }\OperatorTok{{-}}\DecValTok{1}\NormalTok{) }
        \OperatorTok{*}\NormalTok{ np.log1p(x[:}\OperatorTok{{-}}\DecValTok{1}\NormalTok{] }
        \OperatorTok{/}\NormalTok{ np.cumsum(x[}\OperatorTok{{-}}\DecValTok{1}\NormalTok{:}\DecValTok{0}\NormalTok{:}\OperatorTok{{-}}\DecValTok{1}\NormalTok{])[::}\OperatorTok{{-}}\DecValTok{1}\NormalTok{])}
\NormalTok{    )}
    \ControlFlowTok{return}\NormalTok{ np.log(}\OperatorTok{{-}}\NormalTok{np.expm1(ksi)) }\OperatorTok{{-}}\NormalTok{ ksi}
\end{Highlighting}
\end{Shaded}

\hypertarget{sec-sphere}{%
\subsubsection{\texorpdfstring{Sphere
\(\mathbf S_n\)}{Sphere \textbackslash mathbf S\_n}}\label{sec-sphere}}

Let the unit \(n\)-sphere of dimension \(n\) (as part of
\(\reels^{n+1}\)):

\[\sphere n = \braces{x\in\reels^{n+1}: \sum_i x_i^2=1} \subset \reels^{n+1}\]

It is not possible to build a bijection of \(\reels^n\) in
\(\sphere n\), so we will consider a subpart of the sphere in which it
is possible to build a bijection:

\[\msphere n = \braces{x\in\sphere n: \sum_{i=0}^{n-2} x_i^2<1 \wedge
  \p{x_{n-1},x_n}\neq\p{0,-1}}\]

In the topology of \(\sphere n\), the adherence of \(\msphere n\) is
\(\sphere n\), and \(\sphere n\setminus\msphere n\) is of zero Lebesgue
measure w.r.t. Lebesgue measure on \(\sphere n\).

\hypertarget{construction-of-the-parametrization-between-msphere-n-and-reelsn}{%
\paragraph{\texorpdfstring{Construction of the parametrization between
\(\msphere n\) and
\(\reels^n\)}{Construction of the parametrization between \textbackslash msphere n and \textbackslash reels\^{}n}}\label{construction-of-the-parametrization-between-msphere-n-and-reelsn}}

The parametrization is expressed naturally in a first step between
\(\msphere n\) and
\(\ouv{-\frac\pi2}{\frac\pi2}^{n-1}\times\ouv{-\pi}\pi\). It will then
suffice to compose with the parametrization of
\(\ouv{-\frac\pi2}{\frac\pi2}^{n-1}\times\ouv{-\pi}\pi\) in \(\reels^n\)
to obtain a parametrization of \(\msphere n\) in \(\reels^n\).

As in the case of the simplex, the initial idea is to consider that for
\(n\in\natup\), by means of the polyspherical coordinates, we can easily
define a bijection between
\(\ouv{-\frac\pi2}{\frac\pi2}\times\msphere n\) and \(\msphere{n+1}\):

\[\begin{array}{ccrcl}
    && \ouv{-\frac\pi2}{\frac\pi2}\times\msphere n&\longrightarrow& \msphere{n+1} \\
    && (x,y) &\longmapsto& \concat\p{\sin(x), \cos(x)y} \\
  \end{array}\]

We can then introduce the parametrization defined in a recursive way:

\begin{equation}\protect\hypertarget{eq-sphr}{}{\begin{array}{crcrcl}
    &h_1 &\colon& \ouv{-\pi}\pi &\longrightarrow& \msphere1 \\
    &&& x &\longmapsto& (\sin(x), \cos(x)) \\ \\
    \forall n\in\natup,& h_{n+1} &\colon&
    \ouv{-\frac\pi2}{\frac\pi2}^n\times\ouv{-\pi}\pi&\longrightarrow& \msphere{n+1} \\
    &&& x &\longmapsto& \concat\p{\sin\p{x_0}, \cos\p{x_0}h_n\p{x_{1:}}} \\
  \end{array}
  }\label{eq-sphr}\end{equation}

So we can deduce for Equation~\ref{eq-sphr} a un-recursived form of
\(h_n\):

\begin{equation}\protect\hypertarget{eq-sphrh}{}{\begin{array}{ccrcl}
  h_n &\colon& \ouv{-\frac\pi2}{\frac\pi2}^{n-1}\times\ouv{-\pi}\pi
        &\longrightarrow& \msphere n\\
        && x &\longmapsto&
        \concat\p{\sin\p x,1}\\
        &&&&\quad\odot\concat\p{1,\cumprod\p{\cos\p x}}
      \end{array}
  }\label{eq-sphrh}\end{equation}

And we can also write the reciprocal:
\begin{equation}\protect\hypertarget{eq-sphrhm1}{}{\begin{array}{ccrcl}
        h_n^{-1} &\colon& \msphere n &\longrightarrow& \ouv{-\frac\pi2}{\frac\pi2}^{n-1}\times\ouv{-\pi}\pi\\
        && x &\longmapsto&
        \operatorname{arctan2}\big(x_{:n-1},\\
        &&&&\quad\concat\p{\flip\p{\cumsum\p{\flip\p{x_{1:}}\owedge2}_{1:}\owedge\frac12},x_{n}}\big) \\
      \end{array}
  }\label{eq-sphrhm1}\end{equation}

Similar to the simplex, one can notice that the components of the
argument \(x\) in \(h_n(x)\) have very different importance. Thus, a
small variation of \(x_0\) will induce a very large variation of
\(h_n(x)\), while a small variation of \(x_{n-1}\) will induce a very
small variation of \(h_n(x)\).

To rephrase the problem, let us consider the uniform probability
distribution on \(\msphere n\) noted \(\mathcal U_{\msphere n}\). After
the transformation, the objective is to obtain a uniform distribution on
\(\ouv01\) implying a similar importance of all components. We can show
that if \(y\sim\mathcal U_{\msphere n}\), and \(x = h_n^{-1}(y)\), then
for \(k: 0\le k<n-1\), \(x_k\) follows a distribution on
\(\ouv{-\frac\pi2}{\frac\pi2}\) of density
\(x\mapsto \frac{\cos(x)^{n-1-k}}{B\p{\frac12, \frac{n-k}2}}\) where
\(B\) is the Beta function.

Unfortunately, it is not possible to express the cumulative distribution
function of this distribution other than by using a generalized
hypergeometric function, which does not make it possible to implement a
differentiable transformation to transform the uniform law of
\(\msphere n\) into the uniform distribution on the orthotope
\(\ouv{-\frac\pi2}{\frac\pi2}^{n-1}\times\ouv{-\pi}\pi\). Therefore it
is not possible to apply a transformation to obtain exactly a logistic
probability distribution on \(\reels^n\).

The proposed approach is then to transform
\(\ouv{-\frac\pi2}{\frac\pi2}^{n-1}\times\ouv{-\pi}\pi\) into
\(\reels^n\), and apply a scaling factor on the dimensions to obtain the
variance of a logistic probability distribution.

The transformation from
\(\ouv{-\frac\pi2}{\frac\pi2}^{n-1}\times\ouv{-\pi}\pi\) to \(\reels^n\)
is simply written as:

\begin{equation}\protect\hypertarget{eq-sph-trsf-reals}{}{\begin{array}{ccrcl}
    g_n & \colon & \reels^n & \longrightarrow &
    \ouv{-\frac\pi2}{\frac\pi2}^{n-1}\times\ouv{-\pi}\pi \\
    &&x&\longmapsto&\concat\big(\repar{\reels^{n-1}}{\ouv{-\frac\pi2}{\frac\pi2}^{n-1}}\p{x_{:n-1}},\\
    &&&&\quad\repar\reels{\ouv{-\pi}\pi}\p{x_{n-1}}\big)\\
    \\
    g_n^{-1} & \colon & \ouv{-\frac\pi2}{\frac\pi2}^{n-1}\times\ouv{-\pi}\pi &
    \rightarrow& \reels^n \\
    &&x&\longmapsto&\concat\big(\repar{\ouv{-\frac\pi2}{\frac\pi2}^{n-1}}{\reels^{n-1}}\p{x_{:n-1}},\\
    &&&&\quad\repar{\ouv{-\pi}\pi}\reels\p{x_{n-1}}\big)\\
  \end{array}}\label{eq-sph-trsf-reals}\end{equation}

Thus, if \(y\sim\mathcal U_{\msphere n}\), and
\(x = g_n^{-1}\p{h_n^{-1}(y)}\), for \(k: 0\le k<n\), we have
\(\mathbb{E}\p{x_k}=0\), and
\(\operatorname{Var}\p{x_k} \approx \frac{\pi^2}{3\p{2(n-k)-1}}\). By
introducing \(z = \sqrt{2\p{n-\range(n)}-1}\odot x\), we have
\(\mathbb{E}\p{z_k}=0\) and
\(\operatorname{Var}\p{z_k}\approx\frac{\pi^2}3\), which is the variance
of a logistic distribution.

We will therefore consider the transformation:
\begin{equation}\protect\hypertarget{eq-sph2}{}{\begin{array}{ccrcl}
        \repar{\reels^n}{\msphere n} &\colon& \reels^n& \longrightarrow & \msphere n\\
        && x & \longmapsto & h_n\p{g_n\p{x\oslash\sqrt{2\p{n-\range(n)}-1}}} \\
        \\
        \repar{\msphere n}{\reels^n} &\colon& \msphere n & \longrightarrow & \reels^n\\
        && x & \longmapsto & \sqrt{2\p{n-\range(n)}-1}\odot g_n^{-1}\p{h_n^{-1}(x)} \\
      \end{array}}\label{eq-sph2}\end{equation}

\begin{figure}

{\centering \includegraphics{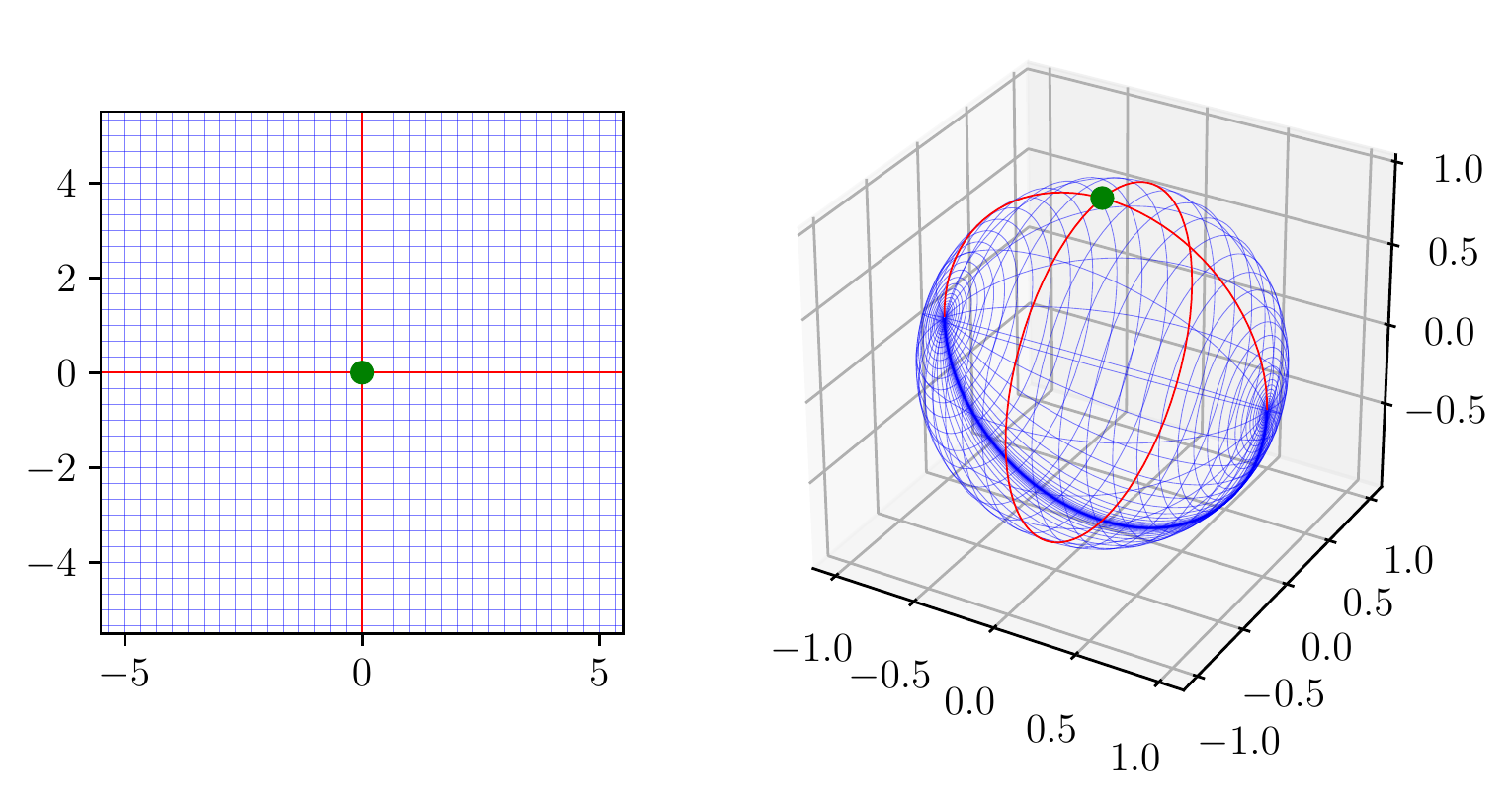}

}

\caption{\label{fig-sphr}Representation of
\(\repar{\reels^2}{\msphere 2}\). Left: a grid in \(\reels^2\). Right:
image of the grid by the transformation
\(\repar{\reels^2}{\msphere 2}\).}

\end{figure}

For \(n=2\), a representation of \(\repar{\reels^n}{\msphere n}\) is
shown Figure~\ref{fig-sphr}.

\hypertarget{properties-3}{%
\paragraph{Properties}\label{properties-3}}

\begin{itemize}
\item
  \(\repar{\reels^n}{\msphere n}(0) = \p{0,\cdots,0,1}\),
\item
  If \(x\sim\Logistic_n\), then \(\repar{\reels^n}{\msphere n}(x)\)
  follows approximately \(\mathcal U_{\msphere n}\),
\item
  If \(x\sim\mathcal U_{\msphere n}\), then
  \(\repar{\msphere n}{\reels^n}(x)\) follows approximately
  \(\Logistic_n\).
\end{itemize}

\hypertarget{implementation-details-2}{%
\paragraph{Implementation details}\label{implementation-details-2}}

The parametrization of \(\reels^n\) in
\(\ouv{-\frac\pi2}{\frac\pi2}^{n-1}\times\ouv{-\pi}\pi\) described in
Equation~\ref{eq-sph-trsf-reals} must be done with the hyperbolic
tangent to ensure good stability around \(0\) (see implementation
details for \(\ouv{-a}a\) in Section~\ref{sec-expit}).

Then for \(x\in\ouv{-\frac\pi2}{\frac\pi2}^{n-1}\times\ouv{-\pi}\pi\),
the computation of \(\cumprod\p{\cos x}\) in \(h_n(x)\) must be done
carefully to be numerically stable and differentiable. We can notice
that \(\forall k: 0\le k<n-1,\,\cos x_k>0\) (which is not true for
\(\cos x_{n-1}\)). Thus, we will compute \(\cumprod\p{\cos x}\) in two
steps:

\[\begin{aligned}
  \cumprod\p{\cos x} =
  & \concat\p{\zeta(x_{:n-1}), {\zeta(x_{:n-1})}_{n-2}\cos x_{n-1}} \\
  \\
  \text{with }\zeta(x_{:n-1}) =& \exp\p{\cumsum\p{\log\p{\cos x_{:n-1}}}}\\\end{aligned}\]

We will use the following formulation:

\begin{equation}\protect\hypertarget{eq-sphr-impl}{}{\begin{array}{ccrcl}
        \repar{\reels^n}{\msphere n} &\colon& \reels^n
        &\longrightarrow& \msphere n\\
        && x &\longmapsto& \psi\p{\xi(x)} \\
        \\
        \text{with }\psi &\colon& y &\longmapsto&
        \concat\p{\sin\p{y},1}\\
        &&&&\quad\odot\concat\big(1,
        \concat\big(\zeta\p{y_{:n-1}},\\
        &&&&\qquad
        {\zeta\p{y_{:n-1}}}_{n-2}\cos\p{y_{n-1}}\big)
        \big) \\
        \\
        \text{with }\zeta &\colon& u &\longmapsto&
        \exp\p{\cumsum\p{\log\p{\cos\p u}}} \\
        \\
        \text{with }\xi &\colon& x &\longmapsto&
        \frac\pi2\concat\p{1_{n-1},2}\\
        &&&&\quad\odot\tanh\p{\frac x2\oslash\sqrt{2\p{n-\range(n)}-1}}
      \end{array}}\label{eq-sphr-impl}\end{equation}

And for the reciprocal, we will use:

\begin{equation}\protect\hypertarget{eq-sphrinv-impl}{}{\begin{array}{ccrcl}
    \repar{\msphere n}{\reels^n} &\colon& \msphere n& \longrightarrow& \reels^n
    \\
    &&x&\longmapsto& 2\operatorname{arctanh}\p{\xi\p
    x\oslash\p{\frac\pi2\concat\p{1_{n-1},2}}} \\
    &&&&\quad\odot\sqrt{2\p{n-\range(n)}-1}\\
    \\
    \text{with }\xi&\colon&x&\longmapsto&
        \operatorname{arctan2}\big(x_{:n-1},\\
        &&&&\quad\concat\p{\flip\p{\cumsum\p{\flip\p{x_{1:}}\owedge2}_{1:}\owedge\frac12},x_{n}}\big) \\
  \end{array}}\label{eq-sphrinv-impl}\end{equation}

\hypertarget{implementation-example-2}{%
\paragraph{Implementation example}\label{implementation-example-2}}

\begin{Shaded}
\begin{Highlighting}[]
\ImportTok{import}\NormalTok{ numpy }\ImportTok{as}\NormalTok{ np}

\KeywordTok{def}\NormalTok{ reals\_to\_sphere(x):}
\NormalTok{    n }\OperatorTok{=}\NormalTok{ x.size}
\NormalTok{    ksi }\OperatorTok{=}\NormalTok{ np.tanh(x}\OperatorTok{/}\DecValTok{2}\OperatorTok{/}\NormalTok{np.sqrt(}\DecValTok{2}\OperatorTok{*}\NormalTok{np.arange(n, }\DecValTok{0}\NormalTok{, }\OperatorTok{{-}}\DecValTok{1}\NormalTok{)}\OperatorTok{{-}}\DecValTok{1}\NormalTok{))}\OperatorTok{*}\NormalTok{np.pi}\OperatorTok{/}\DecValTok{2}
\NormalTok{    ksi[}\OperatorTok{{-}}\DecValTok{1}\NormalTok{] }\OperatorTok{*=} \DecValTok{2}
\NormalTok{    zeta }\OperatorTok{=}\NormalTok{ np.exp(np.cumsum(np.log(np.cos(ksi[:}\OperatorTok{{-}}\DecValTok{1}\NormalTok{]))))}
    \ControlFlowTok{return}\NormalTok{ (}
\NormalTok{        np.concatenate((np.sin(ksi), (}\DecValTok{1}\NormalTok{,)))}
        \OperatorTok{*}\NormalTok{ np.concatenate(((}\DecValTok{1}\NormalTok{,), zeta, (zeta[}\OperatorTok{{-}}\DecValTok{1}\NormalTok{] }\OperatorTok{*}\NormalTok{ np.cos(ksi[}\OperatorTok{{-}}\DecValTok{1}\NormalTok{]),)))}
\NormalTok{    )}

\KeywordTok{def}\NormalTok{ sphere\_to\_reals(x):}
\NormalTok{    n }\OperatorTok{=}\NormalTok{ x.size }\OperatorTok{{-}} \DecValTok{1}
\NormalTok{    ksi }\OperatorTok{=}\NormalTok{ np.arctan2(}
\NormalTok{        x[:}\OperatorTok{{-}}\DecValTok{1}\NormalTok{],}
\NormalTok{        np.concatenate(}
\NormalTok{            (}
\NormalTok{                np.sqrt(np.cumsum(x[:}\DecValTok{0}\NormalTok{:}\OperatorTok{{-}}\DecValTok{1}\NormalTok{] }\OperatorTok{**} \DecValTok{2}\NormalTok{)[:}\DecValTok{0}\NormalTok{:}\OperatorTok{{-}}\DecValTok{1}\NormalTok{]),}
\NormalTok{                (x[}\OperatorTok{{-}}\DecValTok{1}\NormalTok{],),}
\NormalTok{            )}
\NormalTok{        ),}
\NormalTok{    )}
\NormalTok{    ksi[}\OperatorTok{{-}}\DecValTok{1}\NormalTok{] }\OperatorTok{/=} \DecValTok{2}
    \ControlFlowTok{return} \DecValTok{2} \OperatorTok{*}\NormalTok{ np.arctanh(ksi}\OperatorTok{/}\NormalTok{(np.pi}\OperatorTok{/}\DecValTok{2}\NormalTok{)) }\OperatorTok{*}\NormalTok{ np.sqrt(}\DecValTok{2}\OperatorTok{*}\NormalTok{np.arange(n, }\DecValTok{0}\NormalTok{, }\OperatorTok{{-}}\DecValTok{1}\NormalTok{)}\OperatorTok{{-}}\DecValTok{1}\NormalTok{)}
\end{Highlighting}
\end{Shaded}

\hypertarget{sec-half-sphr}{%
\subsubsection{\texorpdfstring{Half sphere
\(\mathbf{HS}_n\)}{Half sphere \textbackslash mathbf\{HS\}\_n}}\label{sec-half-sphr}}

Let be the unit half \(n\)-sphere of dimension \(n\) (as part of
\(\reels^{n+1}\)):

\[\hsphere n = \braces{x\in\reels^{n}\times\reelsp: \sum_i x_i^2=1} \subset \reels^{n+1}\]

This parametrization is needed, for example, when we want to
parameterize a normal vector of a hyperplane. A vector and its opposite
describe the same hyperplane, so we look for a vector restricted to the
half \(n\)-sphere.

\hypertarget{proposed-parametrization}{%
\paragraph{Proposed parametrization}\label{proposed-parametrization}}

The proposed parametrization is built like the one of the \(n\)-sphere,
but using \(\ouv{-\frac\pi2}{\frac\pi2}^n\), with the function \(h_n\):

\begin{equation}\protect\hypertarget{eq-hsphr}{}{\begin{array}{ccrcl}
  h_n &\colon& \ouv{-\frac\pi2}{\frac\pi2}^{n}
        &\longrightarrow& \hsphere n\\
        && x &\longmapsto&
        \concat\p{\sin\p x,1}\\
        &&&&\quad\odot\concat\p{1,\cumprod\p{\cos\p x}} \\
        \\
        h_n^{-1} &\colon& \hsphere n &\longrightarrow& \ouv{-\frac\pi2}{\frac\pi2}^{n}\\
        && x &\longmapsto&
        \operatorname{arctan2}\big(x_{:n-1},\\
        &&&&\quad\flip\p{\cumsum\p{\flip\p{x_{1:}}\owedge2}}\owedge\frac12\big) \\
      \end{array}
  }\label{eq-hsphr}\end{equation}

As for the \(n\)-sphere, (see Section~\ref{sec-sphere} for details), the
proposed approach is therefore to transform
\(\ouv{-\frac\pi2}{\frac\pi2}^{n}\) into \(\reels^n\), and apply a
scaling factor on the dimensions to obtain the variance of a logistic
law.

The transformation of \(\ouv{-\frac\pi2}{\frac\pi2}^{n}\) is
\(\repar{\reels^n}{\ouv{-\frac\pi2}{\frac\pi2}}\) introduced earlier.

Thus, if \(y\sim\mathcal U_{\msphere n}\), and
\(x = \repar{\ouv{-\frac\pi2}{\frac\pi2}}{\reels^n}\p{h_n^{-1}(y)}\),
for \(k: 0\le k<n\), we have \(\mathbb{E}\p{x_k}=0\), and
\(\operatorname{Var}\p{x_k} \approx \frac{\pi^2}{3\p{2(n-k)-1}}\). By
introducing \(z = \sqrt{2\p{n-\range(n)}-1}\odot x\), we have
\(\mathbb{E}\p{z_k}=0\) and
\(\operatorname{Var}\p{z_k}\approx\frac{\pi^2}3\), which is the variance
of a logistic distribution.

We will therefore consider the transformation:

\begin{equation}\protect\hypertarget{eq-hsphr-trsf-reals}{}{\begin{array}{ccrcl}
        \repar{\reels^n}{\hsphere n} &\colon& \reels^n& \longrightarrow & \hsphere n\\
        && x & \longmapsto & h_n\p{\repar{\reels^n}{\ouv{-\frac\pi2}{\frac\pi2}}\p{x\oslash\sqrt{2\p{n-\range(n)}-1}}} \\
        \\
        \repar{\hsphere n}{\reels^n} &\colon& \hsphere n & \longrightarrow & \reels^n\\
        && x & \longmapsto & \sqrt{2\p{n-\range(n)}-1}\odot \repar{\ouv{-\frac\pi2}{\frac\pi2}}{\reels^n}\p{h_n^{-1}(x)} \\
      \end{array}}\label{eq-hsphr-trsf-reals}\end{equation}

\begin{figure}

{\centering \includegraphics{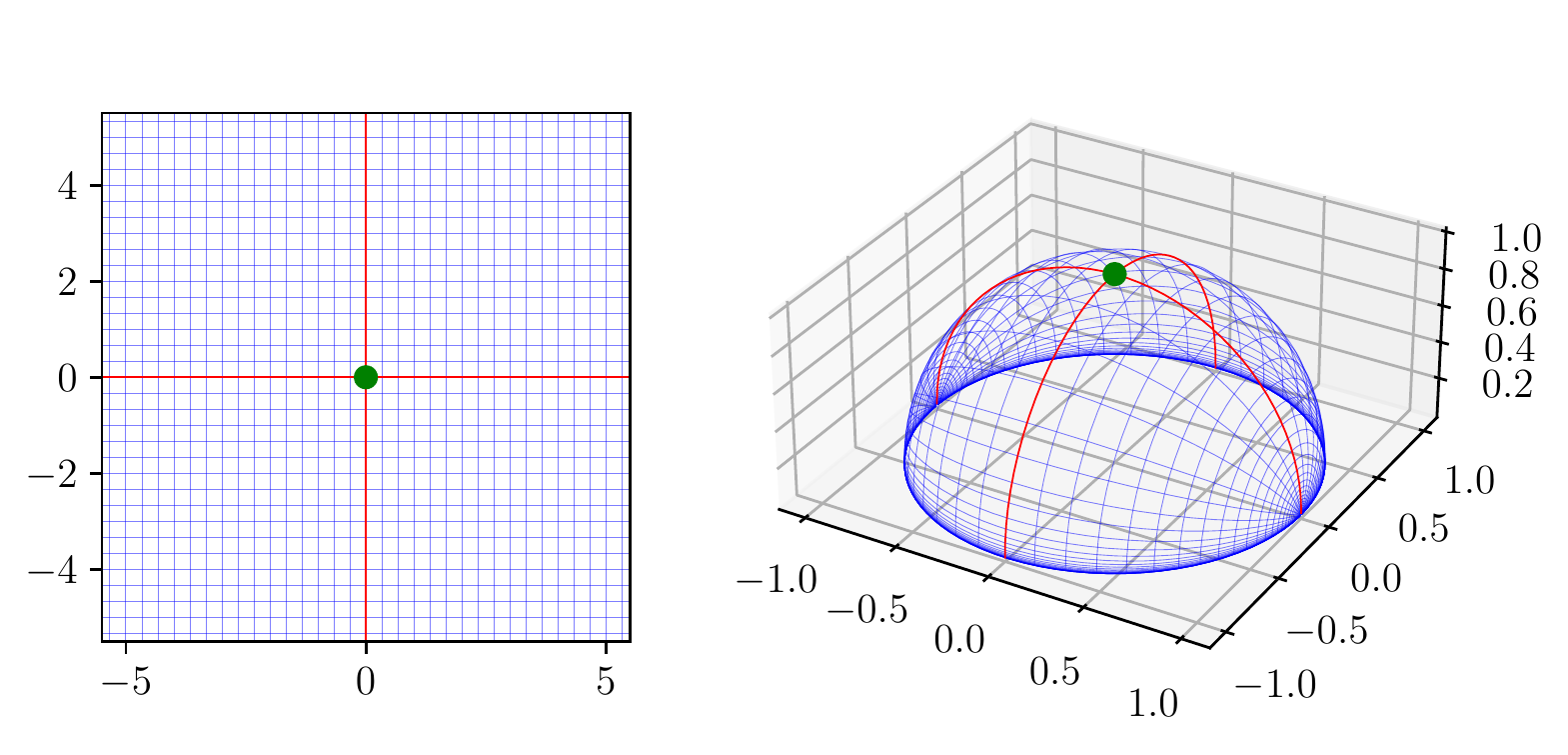}

}

\caption{\label{fig-hsphr}Representation of
\(\repar{\reels^2}{\hsphere 2}\). Left: a grid in \(\reels^2\). Right:
image of the grid by the transformation
\(\repar{\reels^2}{\hsphere 2}\).}

\end{figure}

For \(n=2\), a representation of \(\repar{\reels^n}{\hsphere n}\) is
shown Figure~\ref{fig-hsphr}.

\hypertarget{properties-4}{%
\paragraph{Properties}\label{properties-4}}

\begin{itemize}
\item
  \(\repar{\reels^n}{\hsphere n}(0) = \p{0,\cdots,0,1}\),
\item
  If \(x\sim\Logistic_n\), then \(\repar{\reels^n}{\hsphere n}(x)\)
  follows approximately \(\mathcal U_{\hsphere n}\),
\item
  If \(x\sim\mathcal U_{\hsphere n}\), then
  \(\repar{\hsphere n}{\reels^n}(x)\) follows approximately
  \(\Logistic_n\).
\end{itemize}

\hypertarget{implementation-details-3}{%
\paragraph{Implementation details}\label{implementation-details-3}}

The parametrization in \(\ouv{-\frac\pi2}{\frac\pi2}\) is done in a
symmetrical interval. It is therefore advisable to use the variant of
the implementation using the hyperbolic tangeant.

The parametrization of \(\reels^n\) in
\(\ouv{-\frac\pi2}{\frac\pi2}^{n}\) described in
Equation~\ref{eq-hsphr-trsf-reals} must be done with the hyperbolic
tangent to ensure good stability around \(0\) (see implementation
details for \(\ouv{-a}a\) in Section~\ref{sec-expit}).

Moreover, since the term \(\cumprod\p{\cos x}\) only involves positive
values, and since the cumulative product of small elements can induce
numerical errors, it should be easily computed as
\(\exp\p{\cumsum\p{\log\p{\cos x}}}\). We thus obtain the following
formulation of the transformation:

\begin{equation}\protect\hypertarget{eq-hsphr-impl}{}{\begin{array}{ccrcl}
        \repar{\reels^n}{\hsphere n} &\colon& \reels^n
        &\longrightarrow& \hsphere n\\
        && x &\longmapsto&
        \concat\p{\sin\p{\xi\p x},1}\\
        &&&&\quad\odot\concat\p{1,\exp\p{\cumsum\p{\log\p{\cos\p{\xi\p x}}}}}\\
        \\
        \text{with }\xi &\colon& x &\longmapsto&
        \frac\pi2\tanh\p{\frac x2\oslash\sqrt{2\p{n-\range(n)}-1}}
      \end{array}}\label{eq-hsphr-impl}\end{equation}

The inverse parametrization is written as:

\begin{equation}\protect\hypertarget{eq-hsphrinv-impl}{}{\begin{array}{ccrcl}
    \repar{\hsphere n}{\reels^n} &\colon& \hsphere n& \longrightarrow& \reels^n
    \\
    &&x&\longmapsto& 2\operatorname{arctanh}\p{\frac{\xi\p x}{\frac\pi2}}\odot\sqrt{2\p{n-\range(n)}-1}\\
    \\
    \text{with }\xi&\colon&x&\longmapsto&
        \operatorname{arctan2}\p{x_{:n-1}, \flip\p{\cumsum\p{\flip\p{x_{1:}}\owedge2}\owedge\frac12}} \\
  \end{array}}\label{eq-hsphrinv-impl}\end{equation}

\hypertarget{implementation-example-3}{%
\paragraph{Implementation example}\label{implementation-example-3}}

\begin{Shaded}
\begin{Highlighting}[]
\ImportTok{import}\NormalTok{ numpy }\ImportTok{as}\NormalTok{ np}

\KeywordTok{def}\NormalTok{ reals\_to\_half\_sphere(x):}
\NormalTok{    n }\OperatorTok{=}\NormalTok{ x.size}
\NormalTok{    ksi }\OperatorTok{=}\NormalTok{ np.pi}\OperatorTok{/}\DecValTok{2} \OperatorTok{*}\NormalTok{ np.tanh(x}\OperatorTok{/}\DecValTok{2}\OperatorTok{/}\NormalTok{np.sqrt(}\DecValTok{2}\OperatorTok{*}\NormalTok{np.arange(n, }\DecValTok{0}\NormalTok{, }\OperatorTok{{-}}\DecValTok{1}\NormalTok{)}\OperatorTok{{-}}\DecValTok{1}\NormalTok{))}
    \ControlFlowTok{return}\NormalTok{ (}
\NormalTok{        np.concatenate((np.sin(ksi), (}\DecValTok{1}\NormalTok{,)))}
        \OperatorTok{*}\NormalTok{ np.concatenate(((}\DecValTok{1}\NormalTok{,), np.exp(np.cumsum(np.log(np.cos(ksi))))))}
\NormalTok{    )}

\KeywordTok{def}\NormalTok{ half\_sphere\_to\_reals(x):}
\NormalTok{    n }\OperatorTok{=}\NormalTok{ x.size }\OperatorTok{{-}} \DecValTok{1}
\NormalTok{    ksi }\OperatorTok{=}\NormalTok{ np.arctan2(x[:}\OperatorTok{{-}}\DecValTok{1}\NormalTok{], np.sqrt(np.cumsum(x[:}\DecValTok{0}\NormalTok{:}\OperatorTok{{-}}\DecValTok{1}\NormalTok{] }\OperatorTok{**} \DecValTok{2}\NormalTok{)[::}\OperatorTok{{-}}\DecValTok{1}\NormalTok{]))}
    \ControlFlowTok{return} \DecValTok{2} \OperatorTok{*}\NormalTok{ np.arctanh(}\DecValTok{2}\OperatorTok{*}\NormalTok{ksi}\OperatorTok{/}\NormalTok{np.pi) }\OperatorTok{*}\NormalTok{ np.sqrt(}\DecValTok{2}\OperatorTok{*}\NormalTok{np.arange(n, }\DecValTok{0}\NormalTok{, }\OperatorTok{{-}}\DecValTok{1}\NormalTok{)}\OperatorTok{{-}}\DecValTok{1}\NormalTok{)}
\end{Highlighting}
\end{Shaded}

\hypertarget{sec-ball}{%
\subsubsection{\texorpdfstring{Ball
\(\mathbf B_n\)}{Ball \textbackslash mathbf B\_n}}\label{sec-ball}}

The unit \(n\)-ball is defined by:

\[\ball n = \braces{x\in\reels^n: \sum_i x_i^2 \le 1}\]

It is not possible to obtain a bijection with \(\reels^n\) because of
the boundary points. We will therefore choose to introduce a
parametrization with the \(n\)-open ball:

\[\oball n = \braces{x\in\reels^n: \sum_i x_i^2<1}\]

\hypertarget{construction-of-the-parametrization}{%
\paragraph{Construction of the
parametrization}\label{construction-of-the-parametrization}}

There are two naive ways to construct a parametrization with
\(\reels^n\):

\begin{itemize}
\item
  Starting from the parametrization of \(\sphere n\), indeed,
  \(\forall x\in\sphere n,\; x_{:n}\in\ball n\) and
  \(\forall x\in\ball n,\; \concat\p{x,1-\sum_ix_i}\in\sphere n\). But
  the whole sphere \(\sphere n\) is not parametrizable and we introduce
  uncontrolled effects on \(\sphere n\setminus\msphere n\).
\item
  By transforming the \(\ouv{-1}1^n\) into \(\oball n\) with the
  application \(x\mapsto \frac{{\|x\|}_\infty}{{\|x\|}_2}x\). However,
  as the ratio between the volume of \(\ball n\) and \(\ouv{-1}1^n\)
  tends very quickly to \(0\) when \(n\) increases, this transformation
  induces very strong deformation and does not make parametrization
  possible to obtain a uniformity on the ball by applying a
  transformation on each of the coordinates.
\end{itemize}

We will use an alternative approach, the general idea is that if
\(x\sim\mathcal N\p{0, I_n}\), posing \(y = {\|x\|}_2^2\), and
\(z = \frac x{\sqrt y}\), then \(y\sim\chi^2_n\),
\(z\sim\mathcal U_{\sphere n}\), and \(y\) and \(z\) are independent.

Now we know that if \(u\sim\mathcal U_{\ouv01}\) and
\(y\sim\mathcal U_{\sphere n}\), then
\(u^{\frac1n}y\sim\mathcal U_{\oball n}\). It is thus enough to
transform \(z\sim\chi^2_n\) into \(u\sim\mathcal U_{\ouv01}\), which can
be done by means of the cumulative distribution function of the
\(\chi^2_n\) probability distribution noted \(F_{\chi^2_n}\).

We would like to introduce the function:

\[\begin{array}{ccrcl}
    h_n &\colon& \reels^n &\longrightarrow&\oball n\\
    &&x&\longmapsto&
   \p{F_{\chi^2_n}\p{{\|x\|}_2^2}}^{\frac 1n} \frac x{{\|x\|}_2} \\
    \\
    h_n^{-1} &\colon& \oball n&\longrightarrow&\reels^n\\
    &&x&\longmapsto&
    F_{\chi^2_n}^{-1}\p{{\|x\|}_2^n} \frac x{{\|x\|}_2}
  \end{array}\]

Thus, if \(x\sim\mathcal N\p{0,I_n}\), then
\(h_n(x)\sim\mathcal U_{\oball n}\). This property implies a similar
importance of all components.

The writing of \(h_n\) requires the cumulative distribution function of
\({\chi^2_n}\), but this function and its reciprocal are not easily
computable so as to be automatically differentiable if \(n>2\). For
\(n=2\) we will use the property \(\chi^2_2 = \mathcal{E}\p{\frac12}\)
where \(\mathcal E(\theta)\) is the exponential probability distribution
of intensity \(\theta\). For \(n>2\), we know that if \(y\sim\chi^2_n\),
then \(y^{\frac13}\) is approximated by the law
\(\mathcal N\p{n^{\frac13}\p{1-\frac2{9n}},\frac2{9n^{\frac13}}}\)
(\protect\hyperlink{ref-wilson1931distribution}{Wilson and Hilferty
1931}). This approximation does not respect the support of the
distribution (which is essential to define a bijection). We choose to
use the transformation (which tends to identity when \(n\) is large),
\(t\mapsto \frac14\logexpmu\p{4t}\) for mapping \(\reelsp\) to
\(\reels\) with an identity asymptote when \(t\to+\infty\), the factor
\(4\) is choosen empirically to reduce the approximation error. Then we
consider
\(\frac14\logexpmu\p{4y^{\frac13}}\appsim\mathcal N\p{n^{\frac13}\p{1-\frac2{9n}},\frac2{9n^{\frac13}}}\).
To sum up, we will approximate \(F_{\chi^2_n}\) by \(m_n\) defined as:

\[\begin{array}{ccrcl}
    m_n&\colon&\reelsp&\rightarrow&\ouv01\\
    &&x&\longmapsto&
    \begin{cases}
      1-\exp\p{-\frac x2} & \text{if $n=2$}\\
      \Phi\p{\frac{\frac14\logexpmu\p{4y^{\frac13}}-n^{\frac13}\p{1-\frac2{9n}}}{\sqrt{\frac2{9n^{\frac13}}}}}
      & \text{if $n\ge3$}
    \end{cases}
    \\
    \\
    m_n^{-1}&\colon&\ouv01&\rightarrow&\reelsp\\
    &&x&\longmapsto&
    \begin{cases}
      -2\log\p{1-x} & \text{if $n=2$}\\
      \p{\frac14\logupexp\p{4\p{n^{\frac13}\p{1-\frac2{9n}}+\Phi^{-1}(x)\sqrt{\frac2{9n^{\frac13}}}}}}^{\frac13}
      & \text{if $n\ge3$}
    \end{cases}
  \end{array}\]

where \(\Phi\) is the distribution function of \(\mathcal N(0,1)\).

And we will use the function \(\tilde h_n\) as a approximation of
\(h_n\):

\[\begin{array}{ccrcl}
    \tilde h_n &\colon& \reels^n &\longrightarrow&\oball n\\
    &&x&\longmapsto&
   \p{m_n\p{{\|x\|}_2^2}}^{\frac 1n} \frac x{{\|x\|}_2}
    \\
    \tilde h_n^{-1} &\colon& \oball n&\longrightarrow&\reels^n\\
    &&x&\longmapsto&
    m_n^{-1}\p{{\|x\|}_2^n} \frac x{{\|x\|}_2}
  \end{array}\]

And if \(x\sim\mathcal N\p{0,I_n}\),
\(\tilde h_n\p{x}\appsim\mathcal U_{\oball n}\) (for \(n=2\) where this
result is exact). But for stability reasons, the light tails of the
normal distribution are problematic, and like the other
parametrizations, we want a parametrization which transforms the
distribution \(\Logistic_n\) of \(\reels_n\) into the uniform
distribution of the target space. It is therefore sufficient to consider
the transformation from \(\reels^n\) to \(\reels^n\):

\[\begin{array}{ccrcl}
    g_n&\colon&\reels^n&\longrightarrow&\reels^n\\
    &&x&\longmapsto& \p{\Phi^{-1}\p{\expit\p{x_i}}}_{i: 0\le i<n}\\
    \\
    g_n^{-1}&\colon&\reels^n&\longrightarrow&\reels^n\\
    &&x&\longmapsto& \p{\logit\p{\Phi\p{x_i}}}_{i: 0\le i<n}\\
  \end{array}\]

If \(x\sim\Logistic\), then \(g(x)\sim\mathcal N(0,1)\). Thus we obtain
the proposed parametrization:

\begin{equation}\protect\hypertarget{eq-ball}{}{\begin{array}{ccrcl}
    \repar{\reels^n}{\oball n} &\colon& \reels^n &\longrightarrow&\oball n\\
    &&x&\longmapsto&\tilde h_n\p{g_n(x)} \\
    \\
    \repar{\oball n}{\reels^n} &\colon& \oball n &\longrightarrow&\reels^n\\
    &&x&\longmapsto&g_n^{-1}\p{\tilde h_n^{-1}(x)}
  \end{array}}\label{eq-ball}\end{equation}

\begin{figure}

{\centering \includegraphics{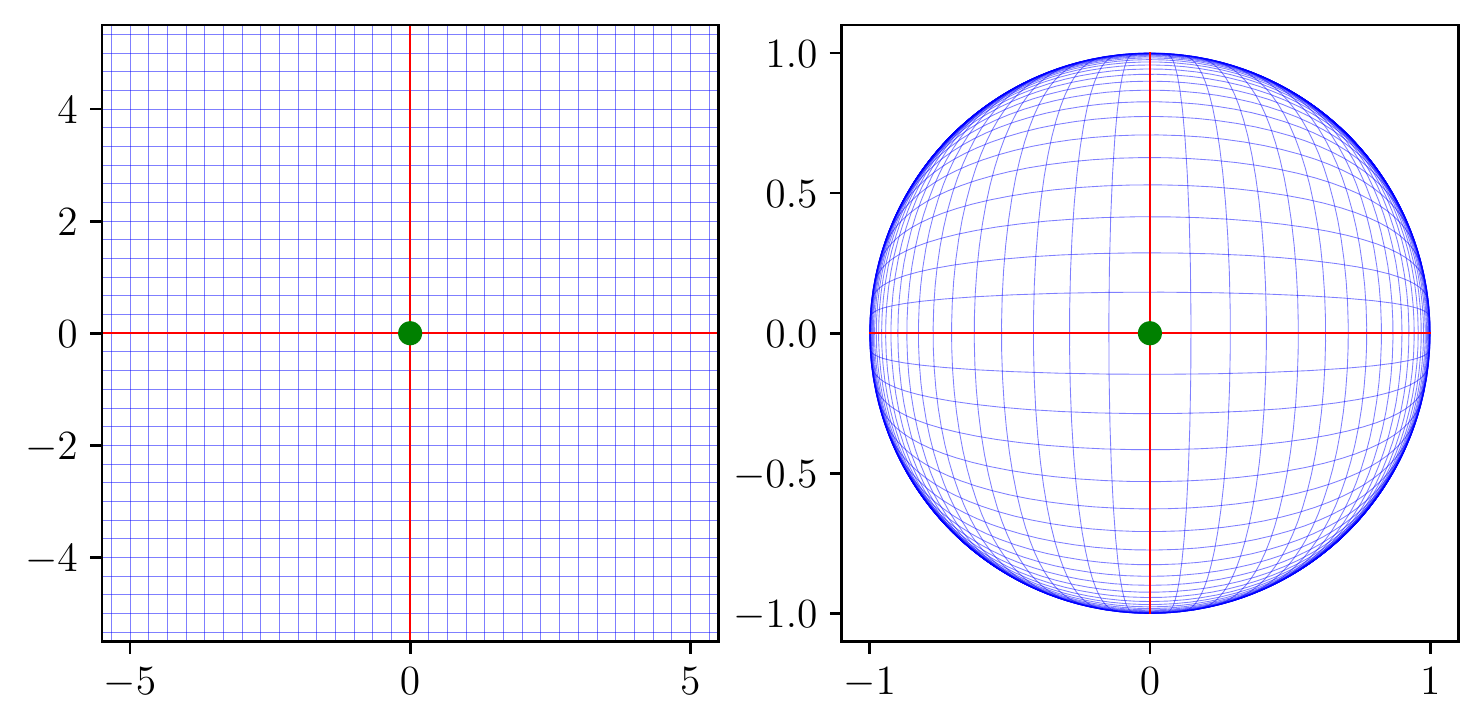}

}

\caption{\label{fig-ball}Representation of
\(\repar{\reels^2}{\ball 2}\). Left: a grid in \(\reels^2\). Right:
image of the grid by the transformation \(\repar{\reels^2}{\ball 2}\).}

\end{figure}

For \(n=2\), a representation of \(\repar{\reels^n}{\oball n}\) is shown
Figure~\ref{fig-ball}.

Note that \(\repar{\reels^n}{\oball n}(0)\) is not defined with the
previous formula, but we can note that the parametrization is extendable
by continuity in \(0\) by \(\repar{\reels^n}{\oball n}(0)=0\) and that
this extension is differentiable.

\hypertarget{properties-5}{%
\paragraph{Properties}\label{properties-5}}

\begin{itemize}
\item
  If \(x\sim\Logistic n\), then
  \(\repar{\reels^n}{\oball n}(x)\appsim\mathcal U_{\oball n}\) (for
  \(n=2\) this result is exact),
\item
  If \(x\sim\mathcal U_{\oball n}\), then
  \(\repar{\oball n}{\reels^n}(x)\appsim\Logistic_n\) (for \(n=2\) this
  result is exact),
\item
  \(\repar{\reels^n}{\oball n}(0) = 0\),
\item
  For \(x\in\reels^n\) and \(y = \repar{\reels^n}{\oball n}(x)\), we
  have
  \(\forall i: 0\le i<n,\, \operatorname{sign}(x_k)=\operatorname{sign}(y_k)\).
\item
  For \(x\in\reels^n\) and \(y = \repar{\reels^n}{\oball n}(x)\), we
  have
  \(\forall i,j: 0\le i,j<n,\, x_i\le x_j \Longleftrightarrow y_i\le y_j\).
\end{itemize}

\hypertarget{implementation-details-4}{%
\paragraph{Implementation details}\label{implementation-details-4}}

First of all, the function \(\Phi\) and the function \(\Phi^{-1}\) are
expressed in terms of the functions \(\erf\) and \(\erfinv\) introduced
in all the calculation libraries (and which are continuously
automatically differentiable in these libraries). We will therefore use:

\[\begin{array}{ccrcl}
    \Phi&\colon&\reels&\rightarrow&\ouv01\\
    &&x&\longmapsto&\frac12\p{1+\erf\p{\frac x{\sqrt 2}}} \\
    \Phi^{-1}&\colon&\ouv01&\rightarrow\reels\\
    &&x&\longmapsto&\sqrt 2\erfinv\p{2x-1}
  \end{array}\]

In the particular case of \(\p{\Phi^{-1}\circ\expit}\) and
\(\p{\logit\circ\,\Phi}\), we note the use of \(\ouv01\) as a
intermediary, whereas the functions \(\erf\) and \(\erfinv\) are
intended to pass by \(\ouv{-1}1\). It is thus appropriate to use the
identity \(\forall x\in\reels,\, \expit(x)=\frac{1+\tanh\p{\frac x2}}2\)
and to carry out the calculation \(\p{\Phi^{-1}\circ\expit}\) in one
time, in the same way for \(\p{\logit\circ\,\Phi}\). We will therefore
use:

\[\begin{array}{ccrcl}
    \Phi^{-1}\circ\expit&\colon&\reels&\rightarrow&\reels\\
    &&x&\longmapsto&\sqrt 2\erfinv\p{\tanh\p{\frac x2}} \\
    \\
    \logit\circ\,\Phi&\colon&\reels&\rightarrow&\reels\\
    &&x&\longmapsto&2\operatorname{arctanh}\p{\erf\p{\frac x{\sqrt2}}} \\
  \end{array}\]

Moreover, we do not calculate \(m_n\), but \(\log m_n\), allowing the
power operation to be a simple product, and a more precise calculation
by using \(\log\Phi\) which is implemented by the calculation libraries
under the name of \(\operatorname{log\_ndtr}\).

We thus obtain the following form to implement the transformation (using
the function \(\p{\Phi^{-1}\circ\expit}\) just defined):

\begin{equation}\protect\hypertarget{eq-ball-impl}{}{\begin{array}{ccrcl}
    \repar{\reels^n}{\oball n} &\colon& \reels^n &\longrightarrow&\oball n\\
    &&x&\longmapsto&\begin{cases}\tilde h_n\p{g_n(x)}&\text{if $x\neq0$}\\0
    &\text{if $x=0$}\end{cases} \\
    \\
    \text{with }g_n&\colon&x&\longmapsto& \p{\p{\Phi^{-1}\circ\expit}\p{x_i}}_{i: 0\le i<n}\\
    \\
    \text{with }\tilde h_n &\colon&x&\longmapsto&
    \exp\p{\frac1n\p{\log m_n\p{{\|x\|}_2^2}}-\frac12\log\p{{\|x\|}_2^2}} x\\
    \\
    \text{with }\log m_n&\colon&x&\longmapsto&
    \begin{cases}
      \operatorname{log1p}\p{-\exp\p{-\frac x2}} & \text{if $n=2$}\\
      \p{\log\Phi}\p{\frac{\frac14\logexpmu\p{4y^{\frac13}}-n^{\frac13}\p{1-\frac2{9n}}}{\sqrt{\frac2{9n^{\frac13}}}}}
      & \text{if $n\ge3$} \end{cases} \\
    \text{with }\log\Phi&\colon&x&\longmapsto&\operatorname{log\_ndtr}(x)\\
    \\
  \end{array}}\label{eq-ball-impl}\end{equation}

And in a similar way, we obtain the implemented form of the reciprocal
of the transformation (using the functions \(\Phi^{-1}\) and
\(\p{\logit\circ\,\Phi}\) just defined):

\begin{equation}\protect\hypertarget{eq-ballinv-impl}{}{\begin{array}{ccrcl}
    \repar{\oball n}{\reels^n} &\colon& \oball n &\longrightarrow&\reels^n\\
    &&x&\longmapsto&\begin{cases}g_n^{-1}\p{\tilde h_n^{-1}(x)}&\text{if
    $x\neq0$}\\0&\text{if $x=0$}\end{cases}\\
    \\
    \text{with }g_n^{-1}&\colon&x&\longmapsto& \p{\p{\logit\circ\,\Phi}\p{x_i}}_{i: 0\le i<n}\\
    \\
    \text{with }\tilde h_n^{-1} &\colon&x&\longmapsto&
    \sqrt{m_n^{-1}\p{{\|x\|}_2^n}} \frac x{{\|x\|}_2} \\
    \\
    \text{with }m_n^{-1}&\colon&x&\longmapsto&
    \begin{cases}
      -2\log\p{1-x} & \text{if $n=2$}\\
      \p{\frac14\logupexp\p{4\p{n^{\frac13}\p{1-\frac2{9n}}+\Phi^{-1}(x)\sqrt{\frac2{9n^{\frac13}}}}}}^3
      & \text{if $n\ge3$}
    \end{cases}
  \end{array}}\label{eq-ballinv-impl}\end{equation}

\hypertarget{implementation-example-4}{%
\paragraph{Implementation example}\label{implementation-example-4}}

This implementation uses \(\logupexp\) and \(\logexpmu\) functions
defined in implementation example of Section~\ref{sec-softplus}.

\begin{Shaded}
\begin{Highlighting}[]
\ImportTok{import}\NormalTok{ numpy }\ImportTok{as}\NormalTok{ np}
\ImportTok{import}\NormalTok{ scipy.special}

\KeywordTok{def}\NormalTok{ reals\_to\_ball(x):}
\NormalTok{    n }\OperatorTok{=}\NormalTok{ x.size}
\NormalTok{    g }\OperatorTok{=}\NormalTok{ np.sqrt(}\DecValTok{2}\NormalTok{) }\OperatorTok{*}\NormalTok{ scipy.special.erfinv(np.tanh(x }\OperatorTok{/} \DecValTok{2}\NormalTok{))}
\NormalTok{    normsq\_g }\OperatorTok{=}\NormalTok{ (g}\OperatorTok{**}\DecValTok{2}\NormalTok{).}\BuiltInTok{sum}\NormalTok{()}
\NormalTok{    normsq\_g\_rep\_01 }\OperatorTok{=}\NormalTok{ normsq\_g }\OperatorTok{+}\NormalTok{ (normsq\_g }\OperatorTok{==} \DecValTok{0}\NormalTok{) }\OperatorTok{*} \FloatTok{1.0}
    \ControlFlowTok{if}\NormalTok{ n }\OperatorTok{==} \DecValTok{2}\NormalTok{:}
\NormalTok{        log\_m }\OperatorTok{=}\NormalTok{ np.log(}\OperatorTok{{-}}\NormalTok{np.expm1(}\OperatorTok{{-}}\NormalTok{normsq\_g\_rep\_01 }\OperatorTok{/} \DecValTok{2}\NormalTok{))}
    \ControlFlowTok{else}\NormalTok{:}
\NormalTok{        log\_m }\OperatorTok{=}\NormalTok{ scipy.special.log\_ndtr(}
\NormalTok{            (}
                \DecValTok{1} \OperatorTok{/} \DecValTok{4} \OperatorTok{*}\NormalTok{ logexpm1(}\DecValTok{4} \OperatorTok{*}\NormalTok{ normsq\_g\_rep\_01 }\OperatorTok{**}\NormalTok{ (}\DecValTok{1} \OperatorTok{/} \DecValTok{3}\NormalTok{))}
                \OperatorTok{{-}}\NormalTok{ n }\OperatorTok{**}\NormalTok{ (}\DecValTok{1} \OperatorTok{/} \DecValTok{3}\NormalTok{) }\OperatorTok{*}\NormalTok{ (}\DecValTok{1} \OperatorTok{{-}} \DecValTok{2} \OperatorTok{/}\NormalTok{ (}\DecValTok{9} \OperatorTok{*}\NormalTok{ n))}
\NormalTok{            )}
            \OperatorTok{/}\NormalTok{ np.sqrt(}\DecValTok{2} \OperatorTok{/}\NormalTok{ (}\DecValTok{9} \OperatorTok{*}\NormalTok{ n }\OperatorTok{**}\NormalTok{ (}\DecValTok{1} \OperatorTok{/} \DecValTok{3}\NormalTok{)))}
\NormalTok{        )}
    \ControlFlowTok{return}\NormalTok{ np.exp(log\_m }\OperatorTok{/}\NormalTok{ n }\OperatorTok{{-}} \FloatTok{0.5} \OperatorTok{*}\NormalTok{ np.log(normsq\_g\_rep\_01)) }\OperatorTok{*}\NormalTok{ g}

\KeywordTok{def}\NormalTok{ ball\_to\_reals(x):}
\NormalTok{    n }\OperatorTok{=}\NormalTok{ x.size}
\NormalTok{    normsq\_x }\OperatorTok{=}\NormalTok{ (x}\OperatorTok{**}\DecValTok{2}\NormalTok{).}\BuiltInTok{sum}\NormalTok{()}
\NormalTok{    normsq\_x\_rep\_01 }\OperatorTok{=}\NormalTok{ normsq\_x }\OperatorTok{+}\NormalTok{ (normsq\_x }\OperatorTok{==} \DecValTok{0}\NormalTok{) }\OperatorTok{*} \FloatTok{1.0}
    \ControlFlowTok{if}\NormalTok{ n }\OperatorTok{==} \DecValTok{2}\NormalTok{:}
\NormalTok{        m\_inv }\OperatorTok{=} \OperatorTok{{-}}\DecValTok{2} \OperatorTok{*}\NormalTok{ np.log1p(}\OperatorTok{{-}}\NormalTok{normsq\_x\_rep\_01)}
    \ControlFlowTok{else}\NormalTok{:}
\NormalTok{        phi\_inv }\OperatorTok{=}\NormalTok{ np.sqrt(}\DecValTok{2}\NormalTok{) }\OperatorTok{*}\NormalTok{ scipy.special.erfinv(}\DecValTok{2}\OperatorTok{*}\NormalTok{normsq\_x\_rep\_01}\OperatorTok{**}\NormalTok{(n}\OperatorTok{/}\DecValTok{2}\NormalTok{)}\OperatorTok{{-}}\DecValTok{1}\NormalTok{)}
\NormalTok{        m\_inv }\OperatorTok{=}\NormalTok{ (}
            \FloatTok{0.25} \OperatorTok{*}\NormalTok{ log1pexp(}
                \DecValTok{4} \OperatorTok{*}\NormalTok{ (}
\NormalTok{                    n }\OperatorTok{**}\NormalTok{ (}\DecValTok{1} \OperatorTok{/} \DecValTok{3}\NormalTok{) }\OperatorTok{*}\NormalTok{ (}\DecValTok{1} \OperatorTok{{-}} \DecValTok{2} \OperatorTok{/}\NormalTok{ (}\DecValTok{9} \OperatorTok{*}\NormalTok{ n))}
                    \OperatorTok{+}\NormalTok{ phi\_inv }\OperatorTok{*}\NormalTok{ np.sqrt(}\DecValTok{2} \OperatorTok{/}\NormalTok{ (}\DecValTok{9} \OperatorTok{*}\NormalTok{ n }\OperatorTok{**}\NormalTok{ (}\DecValTok{1} \OperatorTok{/} \DecValTok{3}\NormalTok{)))}
\NormalTok{                )}
\NormalTok{            )}
\NormalTok{        ) }\OperatorTok{**} \DecValTok{3}
\NormalTok{    h\_inv }\OperatorTok{=}\NormalTok{ (m\_inv}\OperatorTok{**}\FloatTok{0.5} \OperatorTok{/}\NormalTok{ normsq\_x\_rep\_01}\OperatorTok{**}\FloatTok{0.5}\NormalTok{) }\OperatorTok{*}\NormalTok{ x}
    \ControlFlowTok{return} \DecValTok{2} \OperatorTok{*}\NormalTok{ np.arctanh(scipy.special.erf(h\_inv }\OperatorTok{/}\NormalTok{ np.sqrt(}\DecValTok{2}\NormalTok{)))}
\end{Highlighting}
\end{Shaded}

\hypertarget{parametrization-of-matrices}{%
\subsection{Parametrization of
matrices}\label{parametrization-of-matrices}}

\hypertarget{sec-mat-diag}{%
\subsubsection{Diagonal matrices}\label{sec-mat-diag}}

The set of diagonal matrices of \(\reels^{n\times n}\) is defined by:

\[\mathsf D^n = \braces{M\in\reels^{n\times n}: \forall i,j,\; i=j\vee M_{ij}=0}\]

The parametrization used will be:

\begin{equation}\protect\hypertarget{eq-diag}{}{\begin{array}{ccrcl}
    \repar{\mathsf D^n}{\reels^n}&\colon&\mathsf D^n&\longrightarrow&\reels^n \\
    &&M&\longmapsto&\operatorname{diag}\p M = \p{M_{ii}}_{i:0\le i<n} \\
    \\
    \repar{\reels^n}{\mathsf D^n}&\colon&\reels^n&\longrightarrow&\mathsf D^n \\
    &&x&\longmapsto&\operatorname{undiag}\p x \\
    &&&&=
    \begin{bmatrix}
      x_0  \\
      & x_1 \\
      && \ddots \\
      &&& x_{n-1}
    \end{bmatrix}
  \end{array}}\label{eq-diag}\end{equation}

These two operations \(\operatorname{diag}\) and
\(\operatorname{undiag}\) are often present in a single polymorphic
function in most calculation software under the name
\(\operatorname{diag}\). It is therefore not useful to provide a
reference implementation.

\hypertarget{sec-mat-sym}{%
\subsubsection{Symmetric matrices}\label{sec-mat-sym}}

The set of symmetric matrices of \(\reels^{n\times n}\) is defined by:
\[\mathsf S^n = \braces{M\in\reels^{n\times n}: M=M^T}\]

The parametrization used will be:

\begin{equation}\protect\hypertarget{eq-sym}{}{\begin{array}{ccrcl}
    \repar{\mathsf S^n}{\reels^{n(n+1)/2}}&\colon&\mathsf
    S^n&\longrightarrow&\reels^{n(n+1)/2} \\
    &&M&\longmapsto&
    \concat\p{\p{M_{k,:k+1}}_{k:0\le k<n}} \\
    &&&&= \concat\p{M_{0,:1},M_{1,:2},M_{2,:3},\ldots,M_{:,n-1}}\\
    &&&&= \p{
      M_{0,0},
      M_{1,0}, M_{1,1},
      M_{2,0},\ldots M_{2,2},
      \ldots,
      M_{n-1,0},\ldots M_{n-1,n-1},
    }
  \end{array}}\label{eq-sym}\end{equation}

The reciprocal parametrization will be:
\begin{equation}\protect\hypertarget{eq-syminv}{}{\begin{array}{ccrcl}
    \repar{\reels^{n(n+1)/2}}{\mathsf S^n}&\colon&\reels^{n(n+1)/2}&\longrightarrow&\mathsf S^n \\
    &&x&\longmapsto&
    \begin{bmatrix}
      x_0&&&&\!\!\!\!\!\!\!\!\text{(sym)}\\
      x_1 & x_2\\
      x_3 & \cdots & x_5 \\
      \vdots & & & \ddots\\
      x_{n(n-1)/2} & \cdots & \cdots & \cdots & x_{n(n+1)/2-1}
    \end{bmatrix}
  \end{array}}\label{eq-syminv}\end{equation}

\hypertarget{implementation-example-5}{%
\paragraph{Implementation example}\label{implementation-example-5}}

\begin{Shaded}
\begin{Highlighting}[]
\ImportTok{import}\NormalTok{ numpy }\ImportTok{as}\NormalTok{ np}

\KeywordTok{def}\NormalTok{ sym\_matrix\_to\_reals(x):}
    \ControlFlowTok{assert}\NormalTok{ x.shape[}\DecValTok{0}\NormalTok{] }\OperatorTok{==}\NormalTok{ x.shape[}\DecValTok{1}\NormalTok{]}
\NormalTok{    n }\OperatorTok{=}\NormalTok{ x.shape[}\DecValTok{0}\NormalTok{]}
    \ControlFlowTok{return}\NormalTok{ x[np.tril\_indices(n)]}

\KeywordTok{def}\NormalTok{ reals\_to\_sym\_matrix(x, scale}\OperatorTok{=}\FloatTok{1.0}\NormalTok{):}
\NormalTok{    n }\OperatorTok{=} \BuiltInTok{int}\NormalTok{((}\DecValTok{8} \OperatorTok{*}\NormalTok{ x.size }\OperatorTok{+} \DecValTok{1}\NormalTok{) }\OperatorTok{**} \FloatTok{0.5} \OperatorTok{/} \DecValTok{2}\NormalTok{)}
    \ControlFlowTok{assert}\NormalTok{ (}
\NormalTok{        x.size }\OperatorTok{==}\NormalTok{ n }\OperatorTok{*}\NormalTok{ (n }\OperatorTok{+} \DecValTok{1}\NormalTok{) }\OperatorTok{//} \DecValTok{2}
\NormalTok{    ), }\SpecialStringTok{f"Incorect size. It does not exist n such as n*(n+1)/2==}\SpecialCharTok{\{}\NormalTok{x}\SpecialCharTok{.}\NormalTok{size}\SpecialCharTok{\}}\SpecialStringTok{"}

\NormalTok{    y }\OperatorTok{=}\NormalTok{ np.zeros((n, n))}
\NormalTok{    indices }\OperatorTok{=}\NormalTok{ np.tril\_indices(n)}
\NormalTok{    y[indices] }\OperatorTok{=}\NormalTok{ x}
\NormalTok{    y\_transposed }\OperatorTok{=}\NormalTok{ y.T.copy()}
\NormalTok{    y\_transposed[(np.arange(n),) }\OperatorTok{*} \DecValTok{2}\NormalTok{] }\OperatorTok{=} \DecValTok{0}
\NormalTok{    y }\OperatorTok{+=}\NormalTok{ y\_transposed}
    \ControlFlowTok{return}\NormalTok{ y}
\end{Highlighting}
\end{Shaded}

\hypertarget{diagonal-positive-definite-matrices}{%
\subsubsection{Diagonal positive definite
matrices}\label{diagonal-positive-definite-matrices}}

The set of diagonal positive definite matrices of \(\reels^{n\times n}\)
is defined by:

\[\mathsf D^n_{++} = \braces{M\in\reels^{n\times n}: \p{\forall i M_{ii}>0}
  \wedge \p{\forall i,j,\; i=j\vee M_{ij}=0}}\]

The proposed parametrization will be the composition of the
parametrization of \(\reelsp^n\) and the parametrization of
\(\mathsf D^n\).

For \(s\in\reelsp^n\) a vector of scaling parameters of the diagonal
will be:

\[\begin{aligned}
  \reparp{s}{\mathsf D^n_{++}}{\reels^n}
  &=
  \reparp{s}{\reelsp^n}{\reels^n}\circ\repar{\mathsf D^n}{\reels^n}
  \\
  \reparp{s}{\reels^n}{\mathsf D^n_{++}}
  &=
  \repar{\reels^n}{\mathsf D^n}\circ\reparp{s}{\reels^n}{\reelsp^n}\end{aligned}\]

that is:

\begin{equation}\protect\hypertarget{eq-diag-pd}{}{\begin{array}{ccrcl}
    \reparp{s}{\mathsf D^n_{++}}{\reels^n}&\colon&\mathsf D^n_{++}&\longrightarrow&\reels^n \\
    &&M&\longmapsto&\reparp{s}{\reelsp^n}{\reels^n}\p{\operatorname{diag}\p M} =
    \p{\reparp{s_i}\reelsp\reels\p{M_{ii}}}_{i:0\le i<n} \\
    \\
    \reparp{s}{\reels^n}{\mathsf
    D^n_{++}}&\colon&\reels^n&\longrightarrow&\mathsf D^n_{++} \\
    &&x&\longmapsto&\operatorname{undiag}\p{\reparp{s}{\reels^n}{\reelsp^n}\p x}\\
    &&&&=
    \begin{bmatrix}
      \reparp{s_0}\reels\reelsp\p{x_0}  \\
      & \reparp{s_1}\reels\reelsp\p{x_1} \\
      && \ddots \\
      &&& \reparp{s_{n-1}}\reels\reelsp\p{x_{n-1}}
    \end{bmatrix}
  \end{array}}\label{eq-diag-pd}\end{equation}

\hypertarget{choice-of-the-vector-of-scaling-parameters-of-the-diagonal}{%
\paragraph{Choice of the vector of scaling parameters of the
diagonal}\label{choice-of-the-vector-of-scaling-parameters-of-the-diagonal}}

We will often choose the scaling parameters of the diagonal as the
expected order of magnitude of the values of the diagonal. Choosing
parameters that are too small can lead to very large steps in iterative
algorithms using the gradient and to very negative values in real space,
implying very small magnitudes without the optimum being reached. The
choice of scale too large leads the parametrization of the diagonal to
behave like an exponnential and can lead to unstable algorithms.

\hypertarget{implementation-example-6}{%
\paragraph{Implementation example}\label{implementation-example-6}}

In this implementation the scaling parameter vector can be provided as a
vector or as a scalar (and in this case is applied to the whole
diagonal). Functions \(\operatorname{softplus}\) and
\(\operatorname{softplusinv}\) defined in the implementation example in
Section~\ref{sec-softplus} are used.

\begin{Shaded}
\begin{Highlighting}[]
\ImportTok{import}\NormalTok{ numpy }\ImportTok{as}\NormalTok{ np}

\KeywordTok{def}\NormalTok{ reals\_to\_diag\_pd\_matrix(x, scale}\OperatorTok{=}\FloatTok{1.0}\NormalTok{):}
    \ControlFlowTok{return}\NormalTok{ np.diag(softplus(x, scale}\OperatorTok{=}\NormalTok{scale))}

\KeywordTok{def}\NormalTok{ diag\_pd\_matrix\_to\_reals(x, scale}\OperatorTok{=}\FloatTok{1.0}\NormalTok{):}
    \ControlFlowTok{return}\NormalTok{ softplusinv(np.diag(x), scale}\OperatorTok{=}\NormalTok{scale)}
\end{Highlighting}
\end{Shaded}

\hypertarget{sec-mat-spd}{%
\subsubsection{Symmetric positive definite matrices}\label{sec-mat-spd}}

The set of symmetric positive definite matrices of
\(\reels^{n\times n}\) is defined by:
\[\mathsf S^n_{++} = \braces{M\in\reels^{n\times n}: M=M^T \wedge \forall v\in\reels^n,\, v^TMv>0}\]

\hypertarget{construction-of-the-parametrization-1}{%
\paragraph{Construction of the
parametrization}\label{construction-of-the-parametrization-1}}

We will construct the parametrization starting from \(\mathsf S^n_{++}\)
to \(\reels^k\) (with \(k=\frac{n(n+1)}2\)).

Considering a matrix \(M\in\mathsf S^n_{++}\) which has a diagonal of
order of magnitude \(s\in\reelsp^n\). By posing
\(M'=D_s^{-\frac12}MD_s^{-\frac12}\), with
\(D_s=\operatorname{undiag}(s)\), we obtain a positive definite
symmetric matrix which has a diagonal of order of magnitude \(1\).

A symmetric positive definite matrix \(M\) has a unique decomposition of
the form \(M'=LL^T\) with

\begin{itemize}
\item
  \(L\) a lower triangular matrix,
\item
  \(\operatorname{diag}(L)\in\reelsp^n\).
\end{itemize}

This decomposition is the Cholesky factorization. The constraint of
positivity of the diagonal is essential to obtain the uniqueness of the
Cholesky factorization. Thus, the Cholesky factorization defines a
bijection between \(\mathsf S^n_{++}\) and the space of lower triangular
matrices with positive diagonal.

Consider the line \(i\) of the matrix \(L\), since \(L\) is lower
triangular, only the first part of the line \(i\) up to the diagonal is
useful, it is \(l_i = L_{i,:i+1}\in\reels^{i+1}\). Using the matrix
product, we find that \({\|l_i\|}^2 = l_i^Tl_i = M'_{ii}\). So the order
of magnitude of \({\|l_i\|}^2\) is \(1\). From this, we deduce that the
elements (in absolute value) of \(l_i\) have an order of magnitude of
\(\frac1{\sqrt{i+1}}\).

So we introduce:

\[L' = \operatorname{undiag}\p{
    \p{\sqrt{i+1}}_{i:0\le i<n}
  }L\]

Introducing \(l'_i\) in the same way as \(l_i\) with
\(l'_i = L'_{i,:i+1}\in\reels^{i+1}\), we have \(l'_i = \sqrt{i+1}l_i\).
Thus the elements (in absolute value) of \(l'_i\) have a magnitude of
\(1\).

So we obtain a matrix \(L'\) with positive diagonal and all the elements
having in absolute value an order of magnitude \(1\). It is therefore
sufficient to use \(\logexpmu=\reparp{1}{\reelsp}{\reels}\) for the
diagonal, and the identity for the extra-diagonal terms.

We obtain the element of \(\reels^{n(n+1)/2}\):

\[\concat\p{
    \logupexp\p{\operatorname{diag}\p{L'}},
    \concat\p{
      \p{L'_{i,:i}}_{i:0\le i<n}
    }
  }\]

The parametrization defined in this way is thus:

\begin{equation}\protect\hypertarget{eq-spd}{}{\begin{array}{ccrcl}
    \reparp s{\mathsf S^n_{++}}{\reels^{n(n+1)/2}}&\colon&
    \mathsf S^n_{++}&\longrightarrow&\reels^{n(n+1)/2} \\
    &&M&\longmapsto&g\p{\operatorname{undiag}\p{
      \p{\sqrt{i+1}}_{i:0\le i<n}
    }\operatorname{cholesky}\p{D_s^{-\frac12}MD_s^{-\frac12}}} \\
    \\
    \text{with g}&\colon&L&\longmapsto&
    \concat\p{
      \logexpmu\p{\operatorname{diag}\p L},
      \concat\p{
        \p{L_{i,:i}}_{i:0\le i<n}
      }
    } \\
    \\
    \text{with }&&D_s&=&\operatorname{undiag}(s)\\
  \end{array}}\label{eq-spd}\end{equation}

We deduce the reciprocal:

\[\begin{array}{ccrcl}
    \reparp s{\reels^{n(n+1)/2}}{\mathsf S^n_{++}}
    &\colon&\reels^{n(n+1)/2}&\rightarrow&\mathsf S^n_{++} \\
    &&x&\longmapsto&
    D_s^{\frac12}L(x)L(x)^TD_s^{\frac12} \\
    \\
    \text{with }L&\colon&x&\longmapsto&
    \operatorname{undiag}\p{\p{\frac1{\sqrt{i+1}}}_{i:0\le i<n}}
    \p{
      \operatorname{diag}\p{\logupexp\p{x_{:n}}} + g\p{x_{n:}}}\\
    \\
    \text{with }g&\colon&y&\longmapsto&
      \begin{bmatrix}
        0 \\
        y_0 & 0 \\
        y_{1} & y_{2} & 0 \\
        y_{3} & \cdots & y_{5} & 0 \\
        \vdots & & & \ddots & \ddots \\
        y_{n(n-1)/2-n+1} & \cdots & \cdots & \cdots & y_{n(n-1)/2-1} & 0
      \end{bmatrix} \\
      \\
    \text{with }&&D_s&=&\operatorname{undiag}(s)\\
  \end{array}\]

\hypertarget{implementation-details-5}{%
\paragraph{Implementation details}\label{implementation-details-5}}

When implementing, it is useful not to write products by diagonal
matrices as matrix products, but to write them as vector product
operations with broadcasting to reduce the number of operations.

\hypertarget{implementation-example-7}{%
\paragraph{Implementation example}\label{implementation-example-7}}

\begin{Shaded}
\begin{Highlighting}[]
\ImportTok{import}\NormalTok{ numpy }\ImportTok{as}\NormalTok{ np}

\KeywordTok{def}\NormalTok{ spd\_matrix\_to\_reals(x, scale}\OperatorTok{=}\FloatTok{1.0}\NormalTok{):}
    \ControlFlowTok{assert} \BuiltInTok{len}\NormalTok{(x.shape) }\OperatorTok{==} \DecValTok{2} \KeywordTok{and}\NormalTok{ x.shape[}\DecValTok{0}\NormalTok{] }\OperatorTok{==}\NormalTok{ x.shape[}\DecValTok{1}\NormalTok{]}
\NormalTok{    n }\OperatorTok{=}\NormalTok{ x.shape[}\OperatorTok{{-}}\DecValTok{1}\NormalTok{]}
    \ControlFlowTok{if} \BuiltInTok{hasattr}\NormalTok{(scale, }\StringTok{"shape"}\NormalTok{):}
        \ControlFlowTok{assert} \BuiltInTok{len}\NormalTok{(scale.shape) }\OperatorTok{==} \DecValTok{0} \KeywordTok{or}\NormalTok{ (}
            \BuiltInTok{len}\NormalTok{(scale.shape) }\OperatorTok{==} \DecValTok{1} \KeywordTok{and}\NormalTok{ scale.shape[}\DecValTok{0}\NormalTok{] }\OperatorTok{==}\NormalTok{ n}
\NormalTok{        ), }\StringTok{"Non broacastable shapes"}

    \ControlFlowTok{if} \BuiltInTok{hasattr}\NormalTok{(scale, }\StringTok{"shape"}\NormalTok{) }\KeywordTok{and} \BuiltInTok{len}\NormalTok{(scale.shape) }\OperatorTok{==} \DecValTok{1}\NormalTok{:}
\NormalTok{        sqrt\_scale }\OperatorTok{=}\NormalTok{ np.sqrt(scale)}
\NormalTok{        x\_rescaled }\OperatorTok{=}\NormalTok{ x }\OperatorTok{/}\NormalTok{ sqrt\_scale[:, }\VariableTok{None}\NormalTok{] }\OperatorTok{/}\NormalTok{ sqrt\_scale[}\VariableTok{None}\NormalTok{, :]}
    \ControlFlowTok{else}\NormalTok{:}
\NormalTok{        x\_rescaled }\OperatorTok{=}\NormalTok{ x }\OperatorTok{/}\NormalTok{ scale}

\NormalTok{    y }\OperatorTok{=}\NormalTok{ np.linalg.cholesky(x\_rescaled)}
\NormalTok{    y }\OperatorTok{*=}\NormalTok{ np.sqrt(np.arange(}\DecValTok{1}\NormalTok{, n }\OperatorTok{+} \DecValTok{1}\NormalTok{))[:, }\VariableTok{None}\NormalTok{]}
\NormalTok{    diag\_values }\OperatorTok{=}\NormalTok{ y[(np.arange(n),) }\OperatorTok{*} \DecValTok{2}\NormalTok{]}
\NormalTok{    tril\_values }\OperatorTok{=}\NormalTok{ y[np.tril\_indices(n, }\OperatorTok{{-}}\DecValTok{1}\NormalTok{)]}
    \ControlFlowTok{return}\NormalTok{ np.concatenate((logexpm1(diag\_values), tril\_values))}

\KeywordTok{def}\NormalTok{ reals\_to\_spd\_matrix(x, scale}\OperatorTok{=}\FloatTok{1.0}\NormalTok{):}
\NormalTok{    n }\OperatorTok{=} \BuiltInTok{int}\NormalTok{((}\DecValTok{8} \OperatorTok{*}\NormalTok{ x.size }\OperatorTok{+} \DecValTok{1}\NormalTok{) }\OperatorTok{**} \FloatTok{0.5} \OperatorTok{/} \DecValTok{2}\NormalTok{)}
    \ControlFlowTok{assert}\NormalTok{ (}
\NormalTok{        x.size }\OperatorTok{==}\NormalTok{ n }\OperatorTok{*}\NormalTok{ (n }\OperatorTok{+} \DecValTok{1}\NormalTok{) }\OperatorTok{//} \DecValTok{2}
\NormalTok{    ), }\SpecialStringTok{f"Incorect size. It does not exist n such as n*(n+1)/2==}\SpecialCharTok{\{}\NormalTok{x}\SpecialCharTok{.}\NormalTok{size}\SpecialCharTok{\}}\SpecialStringTok{"}
    \ControlFlowTok{if} \BuiltInTok{hasattr}\NormalTok{(scale, }\StringTok{"shape"}\NormalTok{):}
        \ControlFlowTok{assert} \BuiltInTok{len}\NormalTok{(scale.shape) }\OperatorTok{==} \DecValTok{0} \KeywordTok{or}\NormalTok{ (}
            \BuiltInTok{len}\NormalTok{(scale.shape) }\OperatorTok{==} \DecValTok{1} \KeywordTok{and}\NormalTok{ scale.shape[}\DecValTok{0}\NormalTok{] }\OperatorTok{==}\NormalTok{ n}
\NormalTok{        ), }\StringTok{"Non broacastable shapes"}
\NormalTok{    y }\OperatorTok{=}\NormalTok{ np.zeros((n, n))}
\NormalTok{    y[(np.arange(n),) }\OperatorTok{*} \DecValTok{2}\NormalTok{] }\OperatorTok{=}\NormalTok{ log1pexp(x[:n])}
\NormalTok{    y[np.tril\_indices(n, }\OperatorTok{{-}}\DecValTok{1}\NormalTok{)] }\OperatorTok{=}\NormalTok{ x[n:]}

\NormalTok{    y }\OperatorTok{/=}\NormalTok{ np.sqrt(np.arange(}\DecValTok{1}\NormalTok{, n }\OperatorTok{+} \DecValTok{1}\NormalTok{))[:, }\VariableTok{None}\NormalTok{]}
\NormalTok{    z\_rescaled }\OperatorTok{=}\NormalTok{ y }\OperatorTok{@}\NormalTok{ y.T}
    \ControlFlowTok{if} \BuiltInTok{hasattr}\NormalTok{(scale, }\StringTok{"shape"}\NormalTok{) }\KeywordTok{and} \BuiltInTok{len}\NormalTok{(scale.shape) }\OperatorTok{==} \DecValTok{1}\NormalTok{:}
\NormalTok{        sqrt\_scale }\OperatorTok{=}\NormalTok{ np.sqrt(scale)}
\NormalTok{        z }\OperatorTok{=}\NormalTok{ z\_rescaled }\OperatorTok{*}\NormalTok{ sqrt\_scale[:, }\VariableTok{None}\NormalTok{] }\OperatorTok{*}\NormalTok{ sqrt\_scale[}\VariableTok{None}\NormalTok{, :]}
    \ControlFlowTok{else}\NormalTok{:}
\NormalTok{        z }\OperatorTok{=}\NormalTok{ z\_rescaled }\OperatorTok{*}\NormalTok{ scale}
    \ControlFlowTok{return}\NormalTok{ z}
\end{Highlighting}
\end{Shaded}

\hypertarget{correlation-matrices}{%
\subsubsection{Correlation matrices}\label{correlation-matrices}}

A correlation matrix is a symmetric positive definite matrix with
constant diagonal equal to 1.

The set of correlation matrices of \(\reels^{n\times n}\) is defined by:
\[\mathsf C^n_{++} = \braces{M\in\reels^{n\times n}: M=M^T \wedge
  \p{\forall i,\,M_{ii}=1} \wedge \p{\forall v\in\reels^n,\, v^TMv>0}}\]

\hypertarget{construction-of-the-parametrization-2}{%
\paragraph{Construction of the
parametrization}\label{construction-of-the-parametrization-2}}

We will construct the parametrization of \(\mathsf C^n_{++}\) to
\(\reels^k\) (with \(k=\frac{n(n-1)}2\)), starting with the process as
the parametrization of \(\mathsf S^n_{++}\).

Considering a matrix \(M\in\mathsf C^n_{++}\), it is a positive definite
matrix, so it admits a unique decomposition of the form \(M'=LL^T\) with

\begin{itemize}
\item
  \(L\) a lower triangular matrix,
\item
  \(\operatorname{diag}(L)\in\reelsp^n\).
\end{itemize}

This decomposition is the Cholesky factorization (the constraint of
positivity of the diagonal is essential to obtain the uniqueness of the
Cholesky factorization). In this case, the Cholesky factorization does
not define a bijection between \(\mathsf C^n_{++}\) and the space of
lower triangular matrices with positive diagonal, it is only injective.
It is only possible to reach the lower triangular matrices of positive
diagonal \(L'\) such that \(L'L'^T\) has a diagonal of \(1\).

Consider the line \(i\) of a matrix \(L'\) lower triangular of positive
diagonal, only the first part of the line \(i\) up to the diagonal is
useful, it is \(l'_i = L'_{i,:i+1}\in\reels^{i+1}\). Using the matrix
product, we find that \({\|l'_i\|}^2 = {l'_i}^Tl'_i = (L'{L'}^T)_{ii}\).
So the matrix \(L'{L'}^T\) has a unit diagonal if and only if
\(\forall i,\, {\|l'_i\|}^2=1\).

We thus find that the Cholesky factorization defines a bijection between
the space of correlation matrices and the space of triangular matrices
of positive diagonal \(L'\) and whose rows are elements of the unit
spheres (formally \(\forall i,\, l'_i\in\sphere i\)).

Note that imposing the diagonal to be positive and
\(\forall i,\, l'_i\in\sphere i\) is equivalent to imposing that
\(\forall i,\, l'i\in\hsphere i\), where \(\hsphere i\) is the half
\(i\)-dimentional sphere definied in Section~\ref{sec-half-sphr}.

We know \(\hsphere0=\braces{1}\), and for \(i>0\) we have a
parametrization of \(\hsphere i\) in \(\reels^i\) so we obtain a
parametrization of the triangular matrix respecting the two constraints.
We obtain then a parametrization of the correlation matrices.

\begin{equation}\protect\hypertarget{eq-corr}{}{\begin{array}{ccrcl}
    \repar{\mathsf C^n_{++}}{\reels^{n(n-1)/2}}&\colon&
    \mathsf C^n_{++}&\longrightarrow&\reels^{n(n-1)/2} \\
    &&M&\longmapsto&g\p{\operatorname{cholesky}\p{M}} \\
    \\
    \text{with }g&\colon&L&\longmapsto&
    \concat\p{
      \p{
        \repar{\hsphere i}{\reels^i}\p{L_{i,:i+1}}
      }_{i:1\le i<n}
    }
  \end{array}}\label{eq-corr}\end{equation}

And we obtain the inverse parametrization by:

\begin{equation}\protect\hypertarget{eq-corr}{}{\begin{array}{ccrcl}
    \repar{\reels^{n(n-1)/2}}{\mathsf C^n_{++}}&\colon&
    \reels^{n(n-1)/2}&\longrightarrow&\mathsf C^n_{++} \\
    &&x&\longmapsto&L(x)L(x)^T \\
    \\
    \text{with }&\colon&L&\longmapsto&
    \left[
    \begin{array}{l}
      \boxed{\quad1\quad}\\
      \boxed{\;\repar{\reels^1}{\hsphere1}\p{x_0}\;}\\
      \boxed{\quad\repar{\reels^2}{\hsphere2}\p{x_{1:3}}\quad}\\
      \boxed{\quad\quad\repar{\reels^3}{\hsphere3}\p{x_{3:6}}\quad\quad}\\
      \quad\quad\quad\quad\vdots\\
      \boxed{\quad\quad\repar{\reels^{n-1}}{\hsphere{n-1}}\p{x_{(n-1)(n-2)/2:}}\quad\quad}\\
    \end{array}
    \right]
  \end{array}}\label{eq-corr}\end{equation}

\hypertarget{implementation-example-8}{%
\paragraph{Implementation example}\label{implementation-example-8}}

This implementation uses implementations of parametrization of
half-sphere defined in Section~\ref{sec-half-sphr}.

\begin{Shaded}
\begin{Highlighting}[]
\ImportTok{import}\NormalTok{ numpy }\ImportTok{as}\NormalTok{ np}

\KeywordTok{def}\NormalTok{ corr\_matrix\_to\_reals(x):}
    \ControlFlowTok{assert} \BuiltInTok{len}\NormalTok{(x.shape) }\OperatorTok{==} \DecValTok{2} \KeywordTok{and}\NormalTok{ x.shape[}\DecValTok{0}\NormalTok{] }\OperatorTok{==}\NormalTok{ x.shape[}\DecValTok{1}\NormalTok{]}
\NormalTok{    n }\OperatorTok{=}\NormalTok{ x.shape[}\OperatorTok{{-}}\DecValTok{1}\NormalTok{]}
    \ControlFlowTok{assert}\NormalTok{ n }\OperatorTok{\textgreater{}} \DecValTok{1}

\NormalTok{    y }\OperatorTok{=}\NormalTok{ np.linalg.cholesky(x)}
    \ControlFlowTok{return}\NormalTok{ np.concatenate(}
\NormalTok{        [half\_sphere\_to\_reals(y[i, :i}\OperatorTok{+}\DecValTok{1}\NormalTok{]) }\ControlFlowTok{for}\NormalTok{ i }\KeywordTok{in} \BuiltInTok{range}\NormalTok{(}\DecValTok{1}\NormalTok{, n)],}
\NormalTok{        axis}\OperatorTok{={-}}\DecValTok{1}\NormalTok{,}
\NormalTok{    )}

\KeywordTok{def}\NormalTok{ reals\_to\_corr\_matrix(x):}
\NormalTok{    n }\OperatorTok{=} \BuiltInTok{int}\NormalTok{((}\DecValTok{8} \OperatorTok{*}\NormalTok{ x.size }\OperatorTok{+} \DecValTok{1}\NormalTok{) }\OperatorTok{**} \FloatTok{0.5} \OperatorTok{/} \DecValTok{2} \OperatorTok{+} \DecValTok{1}\NormalTok{)}
    \ControlFlowTok{assert}\NormalTok{ n }\OperatorTok{*}\NormalTok{ (n }\OperatorTok{{-}} \DecValTok{1}\NormalTok{) }\OperatorTok{//} \DecValTok{2} \OperatorTok{==}\NormalTok{ x.size}
\NormalTok{    y }\OperatorTok{=}\NormalTok{ np.zeros(x.shape[:}\OperatorTok{{-}}\DecValTok{1}\NormalTok{] }\OperatorTok{+}\NormalTok{ (n, n))}
\NormalTok{    y[}\DecValTok{0}\NormalTok{, }\DecValTok{0}\NormalTok{] }\OperatorTok{=} \FloatTok{1.0}
    \ControlFlowTok{for}\NormalTok{ i }\KeywordTok{in} \BuiltInTok{range}\NormalTok{(}\DecValTok{1}\NormalTok{, n):}
\NormalTok{        y[i, :i}\OperatorTok{+}\DecValTok{1}\NormalTok{] }\OperatorTok{=}\NormalTok{ reals\_to\_half\_sphere(x[(i}\OperatorTok{*}\NormalTok{(i}\OperatorTok{{-}}\DecValTok{1}\NormalTok{)}\OperatorTok{//}\DecValTok{2}\NormalTok{) : (i}\OperatorTok{+}\DecValTok{1}\NormalTok{)}\OperatorTok{*}\NormalTok{i}\OperatorTok{//}\DecValTok{2}\NormalTok{])}
\NormalTok{    z }\OperatorTok{=}\NormalTok{ y }\OperatorTok{@}\NormalTok{ y.T}
    \ControlFlowTok{return}\NormalTok{ z}
\end{Highlighting}
\end{Shaded}

\hypertarget{python-package-parametrization_cookbook}{%
\section{\texorpdfstring{Python package
\texttt{parametrization\_cookbook}}{Python package parametrization\_cookbook}}\label{python-package-parametrization_cookbook}}

For easy handling of all the parameterizations introduced in this
cookbook, a Python package is provided. It is called
\texttt{parametrization\_cookbook}, and it can be installed from
\texttt{PyPI}:

\begin{Shaded}
\begin{Highlighting}[]
\ExtensionTok{pip}\NormalTok{ install parametrization\_cookbook}
\end{Highlighting}
\end{Shaded}

The complete documentation is available online on the following page:
\href{https://jbleger.gitlab.io/parametrization-cookbook/}{\texttt{https://jbleger.gitlab.io/parametrization-cookbook/}}.

The upper package \texttt{parametrization\_cookbook} may not be used
directly as it has no effects. Only sub-modules whould be loaded, they
are divided into two categories, high-level modules allowing to describe
a parametrization and to compose them, and low-level modules allowing to
manipulate the elementary functions introduced in this cookbook.

Each module type comes with three implementations, one using only
\texttt{numpy} and \texttt{scipy} (\protect\hyperlink{ref-numpy}{Harris
et al. 2020}; \protect\hyperlink{ref-scipy}{Virtanen et al. 2020}), one
using JAX (\protect\hyperlink{ref-jax}{Bradbury et al. 2018}), and one
using PyTorch (\protect\hyperlink{ref-pytorch}{Paszke et al. 2019}). Of
course, to use JAX, JAX must be installed, and to use PyTorch, PyTorch
must be installed.

It is usually necessary to load only one sub-module for the use of
parametrizations.

\hypertarget{high-level-modules-an-easy-way-to-define-and-use-a-parametrization}{%
\subsection{High-level modules: an easy way to define and use a
parametrization}\label{high-level-modules-an-easy-way-to-define-and-use-a-parametrization}}

Three high-level modules are provided:

\begin{itemize}
\item
  \texttt{parametrization\_cookbook.numpy}: for numpy and scipy
  implementation. This implementation does not support automatic
  differentiation.
\item
  \texttt{parametrization\_cookbook.jax}: for JAX implementation. This
  implementation does support automatic differentiation. All
  computations are done with JAX primitives using LAX-backend or
  directly with LAX-backend. All functions obtained after the definition
  of a parametrization are pure, compiled with JIT, and usable in
  user-defined JIT-compiled functions.
\item
  \texttt{parametrization\_cookbook.torch}: for PyTorch implementation.
  This implementation does support automatic differentiation. All
  computation are done with PyTorch primitives, using tensor with the
  same device than the provided device.
\end{itemize}

Each module provides classes to describe parametrizations or to
concatenate parametrizations. All classes are detailed in the following
sub-sections. Each module exposes the same API.

A first example, with the numpy implementation, to manipulate the
parametrization of \(\reelsp\):

\begin{Shaded}
\begin{Highlighting}[]
\ImportTok{import}\NormalTok{ numpy }\ImportTok{as}\NormalTok{ np}
\ImportTok{import}\NormalTok{ parametrization\_cookbook.numpy }\ImportTok{as}\NormalTok{ pc}

\NormalTok{parametrization }\OperatorTok{=}\NormalTok{ pc.RealPositive()}
\end{Highlighting}
\end{Shaded}

We can now use the parametrization from \(\reelsp\) to \(\reels\):

\begin{Shaded}
\begin{Highlighting}[]
\NormalTok{x }\OperatorTok{=}\NormalTok{ parametrization.params\_to\_reals1d(}\FloatTok{0.3}\NormalTok{)}
\NormalTok{x}
\end{Highlighting}
\end{Shaded}

\begin{tcolorbox}[boxrule=0pt, enhanced, borderline west={2pt}{0pt}{code-block-stdout-light}, interior hidden, frame hidden, breakable, sharp corners, grow to left by=-1em]

\begin{verbatim}
array([-1.05022561])
\end{verbatim}

\end{tcolorbox}

And we can use the parametrization from \(\reels\) to \(\reelsp\):

\begin{Shaded}
\begin{Highlighting}[]
\NormalTok{parametrization.reals1d\_to\_params(x)}
\end{Highlighting}
\end{Shaded}

\begin{tcolorbox}[boxrule=0pt, enhanced, borderline west={2pt}{0pt}{code-block-stdout-light}, interior hidden, frame hidden, breakable, sharp corners, grow to left by=-1em]

\begin{verbatim}
0.30000000000000004
\end{verbatim}

\end{tcolorbox}

\hypertarget{common-api-of-all-parametrization-classes}{%
\subsubsection{Common API of all parametrization
classes}\label{common-api-of-all-parametrization-classes}}

Parametrization instances are non-mutable objects, to change the
parametrization (\emph{e.g.} to change the bounds or the dimension) a
new parametrization instance must be defined. This behavior is not a
restriction for common usages. Therefore, for a defined parametrization
instance, bounds methods are pure functions, which is a very important
property for use with JAX.

Considering a parametrization of set \(E\), defined with the appropriate
class, we obtain a parametrization instance with the following
attributes:

\begin{itemize}
\item
  method \texttt{params\_to\_reals1d}. This method is the bijective
  mapping \(E\to\reels^k\), the value of \(k\) is automatically defined
  at the definition of the parametrization.

  This function takes as argument a value of \(E\), and returns a value
  of \(\reels^k\).

  For JAX module, this bound method is pure, JIT-compiled, and usable in
  JIT-compiled user-defined functions.
\item
  method \texttt{reals1d\_to\_params}. This method is the bijective
  mapping \(\reels^k\to E\), the value of \(k\) is automatically defined
  at the definition of the parametrization. This is the reciprocal
  function of the previous one.

  This function takes as argument a value of \(\reels^k\), and returns a
  value of \(E\).

  For JAX module, this bound method is pure, JIT-compiled, and usable in
  JIT-compiled user-defined functions.
\item
  property \texttt{size}. This property gives the value of \(k\), the
  dimension of the \(\reels^k\)-field used for the bijective mapping
  with \(E\). This value is computed at the definition of the
  parametrization.

  For JAX module, this property is pure and usable in JIT-compiled
  user-defined functions.
\end{itemize}

\hypertarget{parametrization-of-scalars-1}{%
\subsubsection{Parametrization of
scalars}\label{parametrization-of-scalars-1}}

We use these parametrizations when we are constrained using scalars, or
when we are using vectors or matrices (or n-dimensional array) where
each element is a constrained scalar.

\hypertarget{classes}{%
\paragraph{Classes}\label{classes}}

The following classes are provided for the scalar parametrization:

\begin{itemize}
\tightlist
\item
  \texttt{Real}: for parametrization of \(\reels\). This parametrization
  is not useful directly, but can be used to rescale reals and to define
  parametrization of unconstrained vectors and matrices.
\item
  \texttt{RealPositive}: for parametrization of \(\reelsp\).
\item
  \texttt{RealNegative}: for parametrization of \(\mathbb R_-^*\).
\item
  \texttt{RealLowerBounded}: for parametrization of
  \(\ouv a{+\infty}=\braces{x\in\reels: x>a}\) for a given \(a\).
\item
  \texttt{RealUpperBounded}: for parametrization of
  \(\ouv {-\infty}a=\braces{x\in\reels: x<a}\) for a given \(a\).
\item
  \texttt{RealBounded01}: for parametrization of \(\ouv01\).
\item
  \texttt{RealBounded}: for parametrization of \(\ouv ab\) for given
  \(a\) and \(b\).
\end{itemize}

Example, for the parametrization of \(\ouv0{12}\):

\begin{Shaded}
\begin{Highlighting}[]
\ImportTok{import}\NormalTok{ jax.numpy }\ImportTok{as}\NormalTok{ jnp}
\ImportTok{import}\NormalTok{ parametrization\_cookbook.jax }\ImportTok{as}\NormalTok{ pc}

\NormalTok{parametrization }\OperatorTok{=}\NormalTok{ pc.RealBounded(bound\_lower}\OperatorTok{=}\DecValTok{0}\NormalTok{, bound\_upper}\OperatorTok{=}\DecValTok{12}\NormalTok{)}
\BuiltInTok{print}\NormalTok{(}\SpecialStringTok{f"parametrization.size: }\SpecialCharTok{\{}\NormalTok{parametrization}\SpecialCharTok{.}\NormalTok{size}\SpecialCharTok{\}}\SpecialStringTok{"}\NormalTok{)}

\NormalTok{x\_real }\OperatorTok{=}\NormalTok{ jnp.array([}\OperatorTok{{-}}\FloatTok{1.2}\NormalTok{])}
\NormalTok{y }\OperatorTok{=}\NormalTok{ parametrization.reals1d\_to\_params(x\_real)}
\NormalTok{x\_real\_back }\OperatorTok{=}\NormalTok{ parametrization.params\_to\_reals1d(y)}
\BuiltInTok{print}\NormalTok{(}\SpecialStringTok{f"x\_real: }\SpecialCharTok{\{}\NormalTok{x\_real}\SpecialCharTok{\}}\SpecialStringTok{"}\NormalTok{)}
\BuiltInTok{print}\NormalTok{(}\SpecialStringTok{f"y: }\SpecialCharTok{\{}\NormalTok{y}\SpecialCharTok{\}}\SpecialStringTok{"}\NormalTok{)}
\BuiltInTok{print}\NormalTok{(}\SpecialStringTok{f"x\_real\_back: }\SpecialCharTok{\{}\NormalTok{x\_real\_back}\SpecialCharTok{\}}\SpecialStringTok{"}\NormalTok{)}
\end{Highlighting}
\end{Shaded}

\begin{tcolorbox}[boxrule=0pt, enhanced, borderline west={2pt}{0pt}{code-block-stdout-light}, interior hidden, frame hidden, breakable, sharp corners, grow to left by=-1em]

\begin{verbatim}
parametrization.size: 1
x_real: [-1.2]
y: 2.777702569961548
x_real_back: [-1.1999999]
\end{verbatim}

\end{tcolorbox}

\hypertarget{shaped-scalars-to-build-vectors-matrices-n-dimensional-arrays}{%
\paragraph{Shaped scalars to build vectors, matrices, n-dimensional
arrays}\label{shaped-scalars-to-build-vectors-matrices-n-dimensional-arrays}}

It is possible to handle vectors, matrices or n-dimensional arrays of
scalars, and all classes supports the \texttt{shape} argument in the
class initialization.

For example, it is possible to build the parametrization from
\({\ouv01}^{3\times3}\) (the set of \(3\times3\)-matrices where values
are between \(0\) and \(1\)):

\begin{Shaded}
\begin{Highlighting}[]
\ImportTok{import}\NormalTok{ jax.numpy }\ImportTok{as}\NormalTok{ jnp}
\ImportTok{import}\NormalTok{ parametrization\_cookbook.jax }\ImportTok{as}\NormalTok{ pc}

\NormalTok{parametrization }\OperatorTok{=}\NormalTok{ pc.RealBounded01(shape}\OperatorTok{=}\NormalTok{(}\DecValTok{3}\NormalTok{,}\DecValTok{3}\NormalTok{))}

\NormalTok{x\_real }\OperatorTok{=}\NormalTok{ jnp.linspace(}\OperatorTok{{-}}\DecValTok{4}\NormalTok{, }\DecValTok{4}\NormalTok{, parametrization.size)}
\NormalTok{y }\OperatorTok{=}\NormalTok{ parametrization.reals1d\_to\_params(x\_real)}
\NormalTok{x\_real\_back }\OperatorTok{=}\NormalTok{ parametrization.params\_to\_reals1d(y)}
\BuiltInTok{print}\NormalTok{(}\SpecialStringTok{f"parametrization.size: }\SpecialCharTok{\{}\NormalTok{parametrization}\SpecialCharTok{.}\NormalTok{size}\SpecialCharTok{\}}\SpecialStringTok{"}\NormalTok{)}
\BuiltInTok{print}\NormalTok{(}\SpecialStringTok{f"x\_real:}\CharTok{\textbackslash{}n}\SpecialCharTok{\{}\NormalTok{x\_real}\SpecialCharTok{\}}\SpecialStringTok{"}\NormalTok{)}
\BuiltInTok{print}\NormalTok{(}\SpecialStringTok{f"y:}\CharTok{\textbackslash{}n}\SpecialCharTok{\{}\NormalTok{y}\SpecialCharTok{\}}\SpecialStringTok{"}\NormalTok{)}
\BuiltInTok{print}\NormalTok{(}\SpecialStringTok{f"x\_real\_back:}\CharTok{\textbackslash{}n}\SpecialCharTok{\{}\NormalTok{x\_real\_back}\SpecialCharTok{\}}\SpecialStringTok{"}\NormalTok{)}
\end{Highlighting}
\end{Shaded}

\begin{tcolorbox}[boxrule=0pt, enhanced, borderline west={2pt}{0pt}{code-block-stdout-light}, interior hidden, frame hidden, breakable, sharp corners, grow to left by=-1em]

\begin{verbatim}
parametrization.size: 9
x_real:
[-4. -3. -2. -1.  0.  1.  2.  3.  4.]
y:
[[0.01798621 0.04742587 0.11920292]
 [0.26894143 0.5        0.7310586 ]
 [0.880797   0.95257413 0.98201376]]
x_real_back:
[-4.        -3.        -2.        -1.         0.         1.
  1.9999995  3.0000002  3.9999983]
\end{verbatim}

\end{tcolorbox}

\hypertarget{parametrization-of-vectors-1}{%
\subsubsection{Parametrization of
vectors}\label{parametrization-of-vectors-1}}

We use these parametrizations when we are using constrained vector, or
matrices where each row is a constrained vector, or n-dimensional array
where each slice w.r.t. the last dimension is a constrained vector.

\hypertarget{classes-1}{%
\paragraph{Classes}\label{classes-1}}

The following classes are provided for vector parametrization:

\begin{itemize}
\item
  \texttt{VectorSimplex}: for parametrization of
  \(\osimplex n = \braces{x\in\reels_+^{n+1}: \sum_i x_i=1}\), the unit
  \(n\)-simplex. A vector of \(n\)-simplex is a vector with \(n+1\)
  coordinates.
\item
  \texttt{VectorSphere}: for parametrization of
  \(\sphere{n,r} = \braces{x\in\reels^{n+1}: \sum_i x_i^2 = r^2}\), the
  \(n\)-sphere with a radius \(r\). By default the radius is 1. A vector
  of \(n\)-sphere is a vector with \(n+1\) coordinates.
\item
  \texttt{VectorHalfSphere}: for parametrization of
  \(\hsphere{n,r} = \braces{x\in\reels^{n+1}: x_n>0\wedge\sum_i x_i^2 = r^2}\),
  the half \(n\)-sphere with a radius \(r\). By default the radius is 1.
  A vector of half \(n\)-sphere is a vector with \(n+1\) coordinates.
\item
  \texttt{VectorBall}: for parametrization of
  \(\oball{n,r} = \braces{x\in\reels^n: x_i^2<r}\), the \(n\)-ball with
  a radius \(r\). By default the radius is 1. A vector of \(n\)-ball is
  a vector with \(n\) coordinates.
\end{itemize}

Example, for the parametrization of \(\simplex3\):

\begin{Shaded}
\begin{Highlighting}[]
\ImportTok{import}\NormalTok{ jax.numpy }\ImportTok{as}\NormalTok{ jnp}
\ImportTok{import}\NormalTok{ parametrization\_cookbook.jax }\ImportTok{as}\NormalTok{ pc}

\NormalTok{parametrization }\OperatorTok{=}\NormalTok{ pc.VectorSimplex(dim}\OperatorTok{=}\DecValTok{3}\NormalTok{)}
\BuiltInTok{print}\NormalTok{(}\SpecialStringTok{f"parametrization.size: }\SpecialCharTok{\{}\NormalTok{parametrization}\SpecialCharTok{.}\NormalTok{size}\SpecialCharTok{\}}\SpecialStringTok{"}\NormalTok{)}

\NormalTok{x\_real }\OperatorTok{=}\NormalTok{ jnp.array([}\OperatorTok{{-}}\FloatTok{0.5}\NormalTok{, }\FloatTok{0.5}\NormalTok{, }\DecValTok{1}\NormalTok{])}
\NormalTok{y }\OperatorTok{=}\NormalTok{ parametrization.reals1d\_to\_params(x\_real)}
\NormalTok{x\_real\_back }\OperatorTok{=}\NormalTok{ parametrization.params\_to\_reals1d(y)}
\BuiltInTok{print}\NormalTok{(}\SpecialStringTok{f"x\_real: }\SpecialCharTok{\{}\NormalTok{x\_real}\SpecialCharTok{\}}\SpecialStringTok{"}\NormalTok{)}
\BuiltInTok{print}\NormalTok{(}\SpecialStringTok{f"y: }\SpecialCharTok{\{}\NormalTok{y}\SpecialCharTok{\}}\SpecialStringTok{"}\NormalTok{)}
\BuiltInTok{print}\NormalTok{(}\SpecialStringTok{f"x\_real\_back: }\SpecialCharTok{\{}\NormalTok{x\_real\_back}\SpecialCharTok{\}}\SpecialStringTok{"}\NormalTok{)}
\end{Highlighting}
\end{Shaded}

\begin{tcolorbox}[boxrule=0pt, enhanced, borderline west={2pt}{0pt}{code-block-stdout-light}, interior hidden, frame hidden, breakable, sharp corners, grow to left by=-1em]

\begin{verbatim}
parametrization.size: 3
\end{verbatim}

\end{tcolorbox}

\begin{tcolorbox}[boxrule=0pt, enhanced, borderline west={2pt}{0pt}{code-block-stdout-light}, interior hidden, frame hidden, breakable, sharp corners, grow to left by=-1em]

\begin{verbatim}
x_real: [-0.5  0.5  1. ]
y: [0.14617212 0.32919896 0.38353443 0.14109443]
x_real_back: [-0.49999988  0.49999988  1.        ]
\end{verbatim}

\end{tcolorbox}

\hypertarget{shaped-vectors-to-build-vectors-n-dimensional-arrays}{%
\paragraph{Shaped vectors to build vectors, n-dimensional
arrays}\label{shaped-vectors-to-build-vectors-n-dimensional-arrays}}

It is possible to handle matrices where each row is a constrained vector
or more generally \(n\)-dimensional arrays where each slice w.r.t. the
last dimension is a constrained vector with the \texttt{shape} argument
in class initialization.

For example, it is possible to build the parametrization from
\({\osimplex 2}^5\) (set of matrices of size \(5\times3\) where each row
is in \(\osimplex 2\)):

\begin{Shaded}
\begin{Highlighting}[]
\ImportTok{import}\NormalTok{ jax.numpy }\ImportTok{as}\NormalTok{ jnp}
\ImportTok{import}\NormalTok{ parametrization\_cookbook.jax }\ImportTok{as}\NormalTok{ pc}

\NormalTok{parametrization }\OperatorTok{=}\NormalTok{ pc.VectorSimplex(dim}\OperatorTok{=}\DecValTok{2}\NormalTok{, shape}\OperatorTok{=}\NormalTok{(}\DecValTok{5}\NormalTok{,))}

\NormalTok{x\_real }\OperatorTok{=}\NormalTok{ jnp.linspace(}\OperatorTok{{-}}\DecValTok{4}\NormalTok{, }\DecValTok{4}\NormalTok{, parametrization.size)}
\NormalTok{y }\OperatorTok{=}\NormalTok{ parametrization.reals1d\_to\_params(x\_real)}
\NormalTok{x\_real\_back }\OperatorTok{=}\NormalTok{ parametrization.params\_to\_reals1d(y)}
\BuiltInTok{print}\NormalTok{(}\SpecialStringTok{f"parametrization.size: }\SpecialCharTok{\{}\NormalTok{parametrization}\SpecialCharTok{.}\NormalTok{size}\SpecialCharTok{\}}\SpecialStringTok{"}\NormalTok{)}
\BuiltInTok{print}\NormalTok{(}\SpecialStringTok{f"x\_real:}\CharTok{\textbackslash{}n}\SpecialCharTok{\{}\NormalTok{x\_real}\SpecialCharTok{\}}\SpecialStringTok{"}\NormalTok{)}
\BuiltInTok{print}\NormalTok{(}\SpecialStringTok{f"y:}\CharTok{\textbackslash{}n}\SpecialCharTok{\{}\NormalTok{y}\SpecialCharTok{\}}\SpecialStringTok{"}\NormalTok{)}
\BuiltInTok{print}\NormalTok{(}\SpecialStringTok{f"x\_real\_back:}\CharTok{\textbackslash{}n}\SpecialCharTok{\{}\NormalTok{x\_real\_back}\SpecialCharTok{\}}\SpecialStringTok{"}\NormalTok{)}
\end{Highlighting}
\end{Shaded}

\begin{tcolorbox}[boxrule=0pt, enhanced, borderline west={2pt}{0pt}{code-block-stdout-light}, interior hidden, frame hidden, breakable, sharp corners, grow to left by=-1em]

\begin{verbatim}
parametrization.size: 10
x_real:
[-4.         -3.1111112  -2.2222223  -1.3333331  -0.44444454  0.44444466
  1.3333335   2.2222223   3.1111112   4.        ]
y:
[[0.00903391 0.04226595 0.9487002 ]
 [0.05014347 0.19814818 0.7517083 ]
 [0.21941198 0.475626   0.30496198]
 [0.54326326 0.41208047 0.04465634]
 [0.7934782  0.20280725 0.00371455]]
x_real_back:
[-4.0000005  -3.1111112  -2.2222226  -1.3333333  -0.4444449   0.44444472
  1.3333337   2.2222223   3.1111112   3.9999995 ]
\end{verbatim}

\end{tcolorbox}

\hypertarget{parametrization-of-matrices-1}{%
\subsubsection{Parametrization of
matrices}\label{parametrization-of-matrices-1}}

We use these parametrizations when we are using constrained matrices or
n-dimentional array where each slice w.r.t. the two last dimensions is a
constrained vector.

\hypertarget{classes-2}{%
\paragraph{Classes}\label{classes-2}}

The following classes are provided for matrix parametrization:

\begin{itemize}
\tightlist
\item
  \texttt{MatrixDiag}: for diagonal matrices.
\item
  \texttt{MatrixDiagPosDef}: for diagonal positive definite matrices.
\item
  \texttt{MatrixSym}: for symmetric matrices.
\item
  \texttt{MatrixSymDefPos}: for symmetric definite positive matrices.
\item
  \texttt{MatrixCorrelation}: for correlation matrices (\emph{i.e.}
  symmetric positive definite matrices with unit diagonal).
\end{itemize}

Example, for the parametrization a symmetric positive definite
\(3\times3\) matrix:

\begin{Shaded}
\begin{Highlighting}[]
\ImportTok{import}\NormalTok{ jax.numpy }\ImportTok{as}\NormalTok{ jnp}
\ImportTok{import}\NormalTok{ parametrization\_cookbook.jax }\ImportTok{as}\NormalTok{ pc}

\NormalTok{parametrization }\OperatorTok{=}\NormalTok{ pc.MatrixSymPosDef(dim}\OperatorTok{=}\DecValTok{3}\NormalTok{)}
\BuiltInTok{print}\NormalTok{(}\SpecialStringTok{f"parametrization.size: }\SpecialCharTok{\{}\NormalTok{parametrization}\SpecialCharTok{.}\NormalTok{size}\SpecialCharTok{\}}\SpecialStringTok{"}\NormalTok{)}

\NormalTok{x\_real }\OperatorTok{=}\NormalTok{ jnp.array([}\OperatorTok{{-}}\FloatTok{0.5}\NormalTok{, }\FloatTok{0.5}\NormalTok{, }\DecValTok{1}\NormalTok{, }\OperatorTok{{-}}\DecValTok{1}\NormalTok{, }\DecValTok{0}\NormalTok{, }\FloatTok{1.5}\NormalTok{])}
\NormalTok{y }\OperatorTok{=}\NormalTok{ parametrization.reals1d\_to\_params(x\_real)}
\NormalTok{x\_real\_back }\OperatorTok{=}\NormalTok{ parametrization.params\_to\_reals1d(y)}
\BuiltInTok{print}\NormalTok{(}\SpecialStringTok{f"x\_real:}\CharTok{\textbackslash{}n}\SpecialCharTok{\{}\NormalTok{x\_real}\SpecialCharTok{\}}\SpecialStringTok{"}\NormalTok{)}
\BuiltInTok{print}\NormalTok{(}\SpecialStringTok{f"y:}\CharTok{\textbackslash{}n}\SpecialCharTok{\{}\NormalTok{y}\SpecialCharTok{\}}\SpecialStringTok{"}\NormalTok{)}
\BuiltInTok{print}\NormalTok{(}\SpecialStringTok{f"min\_eigenvalue(y): }\SpecialCharTok{\{}\NormalTok{jnp}\SpecialCharTok{.}\NormalTok{linalg}\SpecialCharTok{.}\NormalTok{eigh(y)[}\DecValTok{0}\NormalTok{]}\SpecialCharTok{.}\BuiltInTok{min}\NormalTok{()}\SpecialCharTok{\}}\SpecialStringTok{"}\NormalTok{)}
\BuiltInTok{print}\NormalTok{(}\SpecialStringTok{f"x\_real\_back:}\CharTok{\textbackslash{}n}\SpecialCharTok{\{}\NormalTok{x\_real\_back}\SpecialCharTok{\}}\SpecialStringTok{"}\NormalTok{)}
\end{Highlighting}
\end{Shaded}

\begin{tcolorbox}[boxrule=0pt, enhanced, borderline west={2pt}{0pt}{code-block-stdout-light}, interior hidden, frame hidden, breakable, sharp corners, grow to left by=-1em]

\begin{verbatim}
parametrization.size: 6
\end{verbatim}

\end{tcolorbox}

\begin{tcolorbox}[boxrule=0pt, enhanced, borderline west={2pt}{0pt}{code-block-stdout-light}, interior hidden, frame hidden, breakable, sharp corners, grow to left by=-1em]

\begin{verbatim}
x_real:
[-0.5  0.5  1.  -1.   0.   1.5]
y:
[[ 0.22474898 -0.33522305  0.        ]
 [-0.33522305  0.9744129   0.5964979 ]
 [ 0.          0.5964979   1.3248855 ]]
\end{verbatim}

\end{tcolorbox}

\begin{tcolorbox}[boxrule=0pt, enhanced, borderline west={2pt}{0pt}{code-block-stdout-light}, interior hidden, frame hidden, breakable, sharp corners, grow to left by=-1em]

\begin{verbatim}
min_eigenvalue(y): 0.050460200756788254
x_real_back:
[-0.49999988  0.49999982  0.99999976 -1.          0.          1.5000001 ]
\end{verbatim}

\end{tcolorbox}

\hypertarget{shaped-matrices-to-build-n-dimensional-arrays}{%
\paragraph{Shaped matrices to build n-dimensional
arrays}\label{shaped-matrices-to-build-n-dimensional-arrays}}

It is possible to handle n-dimensional arrays where each slice w.r.t.
the two last dimentions (each matrix with the form
\texttt{M{[}i0,\ …,\ ik,\ :,\ :{]}}) is a constrained matrix with the
\texttt{shape} argument in class initialization.

For example, to build a vector of size 4 of symmetric definite positive
\(3\times3\)-matrices:

\begin{Shaded}
\begin{Highlighting}[]
\ImportTok{import}\NormalTok{ jax.numpy }\ImportTok{as}\NormalTok{ jnp}
\ImportTok{import}\NormalTok{ parametrization\_cookbook.jax }\ImportTok{as}\NormalTok{ pc}

\NormalTok{parametrization }\OperatorTok{=}\NormalTok{ pc.MatrixSymPosDef(dim}\OperatorTok{=}\DecValTok{3}\NormalTok{, shape}\OperatorTok{=}\DecValTok{4}\NormalTok{)}

\NormalTok{x\_real }\OperatorTok{=}\NormalTok{ jnp.linspace(}\OperatorTok{{-}}\DecValTok{3}\NormalTok{, }\DecValTok{3}\NormalTok{, parametrization.size)}
\NormalTok{y }\OperatorTok{=}\NormalTok{ parametrization.reals1d\_to\_params(x\_real)}
\NormalTok{x\_real\_back }\OperatorTok{=}\NormalTok{ parametrization.params\_to\_reals1d(y)}
\BuiltInTok{print}\NormalTok{(}\SpecialStringTok{f"parametrization.size: }\SpecialCharTok{\{}\NormalTok{parametrization}\SpecialCharTok{.}\NormalTok{size}\SpecialCharTok{\}}\SpecialStringTok{"}\NormalTok{)}
\BuiltInTok{print}\NormalTok{(}\SpecialStringTok{f"yreal.shape: }\SpecialCharTok{\{}\NormalTok{y}\SpecialCharTok{.}\NormalTok{shape}\SpecialCharTok{\}}\SpecialStringTok{"}\NormalTok{)}
\BuiltInTok{print}\NormalTok{(}\SpecialStringTok{f"min\_eigenvalue(y[0]): }\SpecialCharTok{\{}\NormalTok{jnp}\SpecialCharTok{.}\NormalTok{linalg}\SpecialCharTok{.}\NormalTok{eigh(y[}\DecValTok{0}\NormalTok{])[}\DecValTok{0}\NormalTok{]}\SpecialCharTok{.}\BuiltInTok{min}\NormalTok{()}\SpecialCharTok{\}}\SpecialStringTok{"}\NormalTok{)}
\BuiltInTok{print}\NormalTok{(}\SpecialStringTok{f"min\_eigenvalue(y[1]): }\SpecialCharTok{\{}\NormalTok{jnp}\SpecialCharTok{.}\NormalTok{linalg}\SpecialCharTok{.}\NormalTok{eigh(y[}\DecValTok{1}\NormalTok{])[}\DecValTok{0}\NormalTok{]}\SpecialCharTok{.}\BuiltInTok{min}\NormalTok{()}\SpecialCharTok{\}}\SpecialStringTok{"}\NormalTok{)}
\BuiltInTok{print}\NormalTok{(}\SpecialStringTok{f"min\_eigenvalue(y[2]): }\SpecialCharTok{\{}\NormalTok{jnp}\SpecialCharTok{.}\NormalTok{linalg}\SpecialCharTok{.}\NormalTok{eigh(y[}\DecValTok{2}\NormalTok{])[}\DecValTok{0}\NormalTok{]}\SpecialCharTok{.}\BuiltInTok{min}\NormalTok{()}\SpecialCharTok{\}}\SpecialStringTok{"}\NormalTok{)}
\BuiltInTok{print}\NormalTok{(}\SpecialStringTok{f"min\_eigenvalue(y[3]): }\SpecialCharTok{\{}\NormalTok{jnp}\SpecialCharTok{.}\NormalTok{linalg}\SpecialCharTok{.}\NormalTok{eigh(y[}\DecValTok{3}\NormalTok{])[}\DecValTok{0}\NormalTok{]}\SpecialCharTok{.}\BuiltInTok{min}\NormalTok{()}\SpecialCharTok{\}}\SpecialStringTok{"}\NormalTok{)}
\BuiltInTok{print}\NormalTok{(}\SpecialStringTok{f"max(abs(x\_real{-}x\_real\_back)): }\SpecialCharTok{\{}\NormalTok{jnp}\SpecialCharTok{.}\BuiltInTok{abs}\NormalTok{(jnp.}\BuiltInTok{max}\NormalTok{(x\_real}\OperatorTok{{-}}\NormalTok{x\_real\_back))}\SpecialCharTok{\}}\SpecialStringTok{"}\NormalTok{)}
\end{Highlighting}
\end{Shaded}

\begin{tcolorbox}[boxrule=0pt, enhanced, borderline west={2pt}{0pt}{code-block-stdout-light}, interior hidden, frame hidden, breakable, sharp corners, grow to left by=-1em]

\begin{verbatim}
parametrization.size: 24
yreal.shape: (4, 3, 3)
min_eigenvalue(y[0]): 3.703959361445186e-09
min_eigenvalue(y[1]): 0.0035347489174455404
min_eigenvalue(y[2]): 0.10191547125577927
min_eigenvalue(y[3]): 0.5219957828521729
max(abs(x_real-x_real_back)): 4.76837158203125e-07
\end{verbatim}

\end{tcolorbox}

\hypertarget{parametrization-of-cartesian-product}{%
\subsubsection{Parametrization of Cartesian
product}\label{parametrization-of-cartesian-product}}

When we have a parametrization of sets \(E_k\) for \(k:0\le k<K\),
defining a parametrization of \(E_0\times\cdots\times E_{K-1}\) is an
index-rewriting task. Two classes are provided to handle this:

\begin{itemize}
\tightlist
\item
  \texttt{Tuple}: which returns parameters as a Python \texttt{tuple}
  instance,
\item
  \texttt{NamedTuple}: which returns parameters as a Python
  \texttt{namedtuple} instance.
\end{itemize}

These two classes have exactly the same goal, the user choice is driven
by their preference for manipulating indexes or names for elementary
parameters.

For example, if we have two parameters \(\alpha\) and \(\beta\) with
constraints \(\alpha\in\ouv01\) and \(\beta\in\reelsp\), we can do:

\begin{itemize}
\item
  with \texttt{Tuple}

\begin{Shaded}
\begin{Highlighting}[]
\ImportTok{import}\NormalTok{ jax.numpy }\ImportTok{as}\NormalTok{ jnp}
\ImportTok{import}\NormalTok{ parametrization\_cookbook.jax }\ImportTok{as}\NormalTok{ pc}

\NormalTok{parametrization }\OperatorTok{=}\NormalTok{ pc.Tuple(}
\NormalTok{    pc.RealBounded01(),}
\NormalTok{    pc.RealPositive(),}
\NormalTok{)}
\BuiltInTok{print}\NormalTok{(}\SpecialStringTok{f"parametrization.size: }\SpecialCharTok{\{}\NormalTok{parametrization}\SpecialCharTok{.}\NormalTok{size}\SpecialCharTok{\}}\SpecialStringTok{"}\NormalTok{)}

\NormalTok{x\_real }\OperatorTok{=}\NormalTok{ jnp.array([}\OperatorTok{{-}}\FloatTok{0.5}\NormalTok{, }\FloatTok{0.5}\NormalTok{])}
\NormalTok{y }\OperatorTok{=}\NormalTok{ parametrization.reals1d\_to\_params(x\_real)}

\BuiltInTok{print}\NormalTok{(}\SpecialStringTok{f"y[0]: }\SpecialCharTok{\{}\NormalTok{y[}\DecValTok{0}\NormalTok{]}\SpecialCharTok{\}}\SpecialStringTok{"}\NormalTok{)}
\BuiltInTok{print}\NormalTok{(}\SpecialStringTok{f"y[1]: }\SpecialCharTok{\{}\NormalTok{y[}\DecValTok{1}\NormalTok{]}\SpecialCharTok{\}}\SpecialStringTok{"}\NormalTok{)}
\end{Highlighting}
\end{Shaded}

  \begin{tcolorbox}[boxrule=0pt, enhanced, borderline west={2pt}{0pt}{code-block-stdout-light}, interior hidden, frame hidden, breakable, sharp corners, grow to left by=-1em]

\begin{verbatim}
parametrization.size: 2
y[0]: 0.3775406777858734
y[1]: 0.9740769863128662
\end{verbatim}

  \end{tcolorbox}
\item
  with \texttt{NamedTuple}

\begin{Shaded}
\begin{Highlighting}[]
\ImportTok{import}\NormalTok{ jax.numpy }\ImportTok{as}\NormalTok{ jnp}
\ImportTok{import}\NormalTok{ parametrization\_cookbook.jax }\ImportTok{as}\NormalTok{ pc}

\NormalTok{parametrization }\OperatorTok{=}\NormalTok{ pc.NamedTuple(}
\NormalTok{    alpha}\OperatorTok{=}\NormalTok{pc.RealBounded01(),}
\NormalTok{    beta}\OperatorTok{=}\NormalTok{pc.RealPositive(),}
\NormalTok{)}
\BuiltInTok{print}\NormalTok{(}\SpecialStringTok{f"parametrization.size: }\SpecialCharTok{\{}\NormalTok{parametrization}\SpecialCharTok{.}\NormalTok{size}\SpecialCharTok{\}}\SpecialStringTok{"}\NormalTok{)}

\NormalTok{x\_real }\OperatorTok{=}\NormalTok{ jnp.array([}\OperatorTok{{-}}\FloatTok{0.5}\NormalTok{, }\FloatTok{0.5}\NormalTok{])}
\NormalTok{y }\OperatorTok{=}\NormalTok{ parametrization.reals1d\_to\_params(x\_real)}

\BuiltInTok{print}\NormalTok{(}\SpecialStringTok{f"y.alpha: }\SpecialCharTok{\{}\NormalTok{y}\SpecialCharTok{.}\NormalTok{alpha}\SpecialCharTok{\}}\SpecialStringTok{"}\NormalTok{)}
\BuiltInTok{print}\NormalTok{(}\SpecialStringTok{f"y.beta: }\SpecialCharTok{\{}\NormalTok{y}\SpecialCharTok{.}\NormalTok{beta}\SpecialCharTok{\}}\SpecialStringTok{"}\NormalTok{)}
\end{Highlighting}
\end{Shaded}

  \begin{tcolorbox}[boxrule=0pt, enhanced, borderline west={2pt}{0pt}{code-block-stdout-light}, interior hidden, frame hidden, breakable, sharp corners, grow to left by=-1em]

\begin{verbatim}
parametrization.size: 2
\end{verbatim}

  \end{tcolorbox}

  \begin{tcolorbox}[boxrule=0pt, enhanced, borderline west={2pt}{0pt}{code-block-stdout-light}, interior hidden, frame hidden, breakable, sharp corners, grow to left by=-1em]

\begin{verbatim}
y.alpha: 0.3775406777858734
y.beta: 0.9740769863128662
\end{verbatim}

  \end{tcolorbox}
\end{itemize}

\hypertarget{definition-of-custom-parametrization}{%
\subsubsection{Definition of custom
parametrization}\label{definition-of-custom-parametrization}}

For some specific cases, user of the package may want to introduce a
custom parametrization, and use it as other parametrizations (in
particular with \texttt{Tuple} or \texttt{NamedTuple}).

User-defined parametrization from \(\reels^n\) to \(E\) must inherit
from \texttt{Param} virtual class, and must define:

\begin{itemize}
\item
  attribute \texttt{\_size}. This attribute, masked to the user, is the
  one used by the \texttt{size} property, and is used by \texttt{Tuple}
  and \texttt{NamedTuple} to build parametrization of Cartesian product.
  This attribute must be a positive integer containing the value \(n\),
  the dimension of \(\reels^n\). This attribute should be set in
  \texttt{\_\_init\_\_} method, if the size does not depend on arguments
  on class construction, the attribute can be set as class attribute.
\item
  method \texttt{reals1d\_to\_params}. This method maps an argument from
  \(\reels^n\) to \(E\). When using JAX, this method must be pure and
  must not use Python control flows depending on values (but Python
  control flows depending on shape are allowed).
\item
  method \texttt{params\_to\_reals1d}. This method maps a value from
  \(E\) to \(\reels^n\). When using JAX, this method must be pure and
  must not use Python control flow depending on values (but Python
  control flows depending on shape are allowed).
\end{itemize}

Furthermore, optionally,

\begin{itemize}
\item
  the user could use the method \texttt{\_check\_reals1d\_size} in
  \texttt{reals1d\_to\_params} to check the shape of the provided input
  (and to reshape scalars to 1-dimensional array of size 1).
\item
  the user could check the shape of input in
  \texttt{params\_to\_reals1d}.
\item
  the user could define \texttt{\_repr} attribute in
  \texttt{\_\_init\_\_}. This attribute contains the representation of
  the object used when the object is converted to string or when the
  object is displayed.
\item
  if JAX is used, the user could JIT-compile the methods
  \texttt{reals1d\_to\_params} and \texttt{params\_to\_reals1d}. As the
  object must be non-mutable, the JIT-compilation of the method is done
  with \texttt{self} as static argument.
\end{itemize}

\hypertarget{examples}{%
\paragraph{Examples}\label{examples}}

We will consider the parametrization from \(\reels\) to
\(\ouv{-\frac\pi2}{\frac\pi2}\) with \(\operatorname{arctan}\) function.

\emph{Note: this example is for illustration purpose. In general, there
is no good reason to prefer parametrization with
\(\operatorname{arctan}\) rather than parametrization with
\(\operatorname{arctanh}\) used before and used in \texttt{RealBounded}
class.}

\begin{itemize}
\item
  with PyTorch

\begin{Shaded}
\begin{Highlighting}[]
\ImportTok{import}\NormalTok{ torch}
\ImportTok{import}\NormalTok{ parametrization\_cookbook.torch }\ImportTok{as}\NormalTok{ pc}

\KeywordTok{class}\NormalTok{ MyAngle(pc.Param):}
    \KeywordTok{def} \FunctionTok{\_\_init\_\_}\NormalTok{(}\VariableTok{self}\NormalTok{):}
        \VariableTok{self}\NormalTok{.\_size }\OperatorTok{=} \DecValTok{1}
        \VariableTok{self}\NormalTok{.\_repr }\OperatorTok{=} \StringTok{"MyAngle()"}

    \KeywordTok{def}\NormalTok{ reals1d\_to\_params(}\VariableTok{self}\NormalTok{, x):}
\NormalTok{        x }\OperatorTok{=} \VariableTok{self}\NormalTok{.\_check\_reals1d\_size(x)}
\NormalTok{        y }\OperatorTok{=}\NormalTok{ torch.arctan(x)}
        \ControlFlowTok{return}\NormalTok{ y[}\DecValTok{0}\NormalTok{]}

    \KeywordTok{def}\NormalTok{ params\_to\_reals1d(}\VariableTok{self}\NormalTok{, x):}
        \ControlFlowTok{assert}\NormalTok{ x.shape }\OperatorTok{==}\NormalTok{ () }\CommentTok{\# must be a scalar}
\NormalTok{        y }\OperatorTok{=}\NormalTok{ torch.tan(x)}
        \ControlFlowTok{return}\NormalTok{ y.ravel() }\CommentTok{\# convert to vector 1{-}d}
\end{Highlighting}
\end{Shaded}
\item
  with JAX (with JIT-compilation for \texttt{reals1d\_to\_params} and
  \texttt{params\_to\_reals1d})

\begin{Shaded}
\begin{Highlighting}[]
\ImportTok{import}\NormalTok{ functools}
\ImportTok{import}\NormalTok{ jax}
\ImportTok{import}\NormalTok{ jax.numpy }\ImportTok{as}\NormalTok{ jnp}
\ImportTok{import}\NormalTok{ parametrization\_cookbook.jax }\ImportTok{as}\NormalTok{ pc}

\KeywordTok{class}\NormalTok{ MyAngle(pc.Param):}
    \KeywordTok{def} \FunctionTok{\_\_init\_\_}\NormalTok{(}\VariableTok{self}\NormalTok{):}
        \VariableTok{self}\NormalTok{.\_size }\OperatorTok{=} \DecValTok{1}
        \VariableTok{self}\NormalTok{.\_repr }\OperatorTok{=} \StringTok{"MyAngle()"}

    \AttributeTok{@functools.partial}\NormalTok{(jax.jit, static\_argnums}\OperatorTok{=}\DecValTok{0}\NormalTok{)}
    \KeywordTok{def}\NormalTok{ reals1d\_to\_params(}\VariableTok{self}\NormalTok{, x):}
\NormalTok{        x }\OperatorTok{=} \VariableTok{self}\NormalTok{.\_check\_reals1d\_size(x)}
\NormalTok{        y }\OperatorTok{=}\NormalTok{ jnp.arctan(x)}
        \ControlFlowTok{return}\NormalTok{ y[}\DecValTok{0}\NormalTok{]}

    \AttributeTok{@functools.partial}\NormalTok{(jax.jit, static\_argnums}\OperatorTok{=}\DecValTok{0}\NormalTok{)}
    \KeywordTok{def}\NormalTok{ params\_to\_reals1d(}\VariableTok{self}\NormalTok{, x):}
        \ControlFlowTok{assert}\NormalTok{ x.shape }\OperatorTok{==}\NormalTok{ () }\CommentTok{\# must be a scalar}
\NormalTok{        y }\OperatorTok{=}\NormalTok{ jnp.tan(x)}
        \ControlFlowTok{return}\NormalTok{ y.ravel() }\CommentTok{\# convert to vector 1{-}d}
\end{Highlighting}
\end{Shaded}
\end{itemize}

\hypertarget{low-level-module-access-to-elementary-functions}{%
\subsection{Low-level module: access to elementary
functions}\label{low-level-module-access-to-elementary-functions}}

Functions described in Section~\ref{sec-repars} are implemented in
low-level modules. Three low-level modules are provided:

\begin{itemize}
\item
  \texttt{parametrization\_cookbook.functions.numpy}: for numpy and
  scipy implementation. This implementation does not support automatic
  differentiation.
\item
  \texttt{parametrization\_cookbook.functions.jax}: for JAX
  implementation. This implementation does support automatic
  differentiation. All computations are done with JAX primitives using
  LAX-backend or directly with LAX-backend. All functions are pure,
  compiled with JIT, and usable in user-defined JIT-compiled functions.
\item
  \texttt{parametrization\_cookbook.functions.torch}: for PyTorch
  implementation. This implementation does support automatic
  differentiation. All computation are done with PyTorch primitive,
  using tensor with the same device than the provided device.
\end{itemize}

Each module provides the functions described in
Section~\ref{sec-repars}, in both directions \(\reels\to E\) and
\(E\to\reels\). These functions are used by high-level modules and are
directly accessible to users.

All functions supports vectorization, see
Section~\ref{sec-pkg-ll-vectorization} below.

\hypertarget{sec-pkg-ll-scalars}{%
\subsubsection{Parametrization of scalars}\label{sec-pkg-ll-scalars}}

The following functions are available:

\begin{itemize}
\tightlist
\item
  \texttt{softplus}, \texttt{softplusinv}, \texttt{log1pexp},
  \texttt{logexpm1}: for parametrization of \(\reelsp\), these functions
  are described in Section~\ref{sec-softplus}.
\item
  \texttt{expit}, \texttt{logit}, \texttt{arctanh}, \texttt{tanh}: for
  parametrization of \(\ouv01\) or \(\ouv{-1}1\), these functions are
  described in Section~\ref{sec-expit}. These functions are not
  implemented in the package, the implementation of scipy, JAX or
  PyTorch is used.
\end{itemize}

For usage of scalar parametrization to build vector, matrices or
n-dimensional arrays with constraints on elements, see
Section~\ref{sec-pkg-ll-vectorization} below.

Example, with the parametrization of \(\reelsp\) with PyTorch:

\begin{Shaded}
\begin{Highlighting}[]
\ImportTok{import}\NormalTok{ torch}
\ImportTok{import}\NormalTok{ parametrization\_cookbook.functions.torch }\ImportTok{as}\NormalTok{ pcf}

\NormalTok{y }\OperatorTok{=}\NormalTok{ torch.tensor(}\FloatTok{2.4}\NormalTok{)}
\NormalTok{x }\OperatorTok{=}\NormalTok{ pcf.softplusinv(y)}
\NormalTok{y2 }\OperatorTok{=}\NormalTok{ pcf.softplus(x)}

\BuiltInTok{print}\NormalTok{(}\SpecialStringTok{f"y: }\SpecialCharTok{\{}\NormalTok{y}\SpecialCharTok{\}}\SpecialStringTok{"}\NormalTok{)}
\BuiltInTok{print}\NormalTok{(}\SpecialStringTok{f"x: }\SpecialCharTok{\{}\NormalTok{x}\SpecialCharTok{\}}\SpecialStringTok{"}\NormalTok{)}
\BuiltInTok{print}\NormalTok{(}\SpecialStringTok{f"y2: }\SpecialCharTok{\{}\NormalTok{y2}\SpecialCharTok{\}}\SpecialStringTok{"}\NormalTok{)}
\end{Highlighting}
\end{Shaded}

\begin{tcolorbox}[boxrule=0pt, enhanced, borderline west={2pt}{0pt}{code-block-stdout-light}, interior hidden, frame hidden, breakable, sharp corners, grow to left by=-1em]

\begin{verbatim}
y: 2.4000000953674316
x: 2.3049001693725586
y2: 2.4000000953674316
\end{verbatim}

\end{tcolorbox}

\hypertarget{sec-pkg-ll-vectors}{%
\subsubsection{Parametrization of vectors}\label{sec-pkg-ll-vectors}}

The following functions are available:

\begin{itemize}
\tightlist
\item
  \texttt{reals\_to\_simplex}, \texttt{simplex\_to\_reals}: for
  parametrization of the unit simplex \(\simplex n\), these functions
  are described in Section~\ref{sec-simplex}.
\item
  \texttt{reals\_to\_sphere}, \texttt{sphere\_to\_reals}: for
  parametrization of the unit sphere \(\sphere n\), these functions are
  described in Section~\ref{sec-sphere}.
\item
  \texttt{reals\_to\_half\_sphere}, \texttt{half\_sphere\_to\_reals}:
  for parametrization of the unit half-sphere \(\hsphere n\), these
  functions are described in Section~\ref{sec-half-sphr}.
\item
  \texttt{reals\_to\_ball}, \texttt{ball\_to\_reals}: for
  parametrization of the unit ball \(\ball n\), these functions are
  described in Section~\ref{sec-ball}.
\end{itemize}

For usage of vector parametrization to build matrices or n-dimensional
arrays with constraints on the last dimension, see
Section~\ref{sec-pkg-ll-vectorization} below.

Example, with the parametrization of \(\simplex2\) with PyTorch:

\begin{Shaded}
\begin{Highlighting}[]
\ImportTok{import}\NormalTok{ torch}
\ImportTok{import}\NormalTok{ parametrization\_cookbook.functions.torch }\ImportTok{as}\NormalTok{ pcf}

\NormalTok{y }\OperatorTok{=}\NormalTok{ torch.tensor([}\FloatTok{.3}\NormalTok{, }\FloatTok{.5}\NormalTok{, }\FloatTok{.2}\NormalTok{])}
\NormalTok{x }\OperatorTok{=}\NormalTok{ pcf.simplex\_to\_reals(y)}
\NormalTok{y2 }\OperatorTok{=}\NormalTok{ pcf.reals\_to\_simplex(x)}

\BuiltInTok{print}\NormalTok{(}\SpecialStringTok{f"y: }\SpecialCharTok{\{}\NormalTok{y}\SpecialCharTok{\}}\SpecialStringTok{"}\NormalTok{)}
\BuiltInTok{print}\NormalTok{(}\SpecialStringTok{f"x: }\SpecialCharTok{\{}\NormalTok{x}\SpecialCharTok{\}}\SpecialStringTok{"}\NormalTok{)}
\BuiltInTok{print}\NormalTok{(}\SpecialStringTok{f"y2: }\SpecialCharTok{\{}\NormalTok{y2}\SpecialCharTok{\}}\SpecialStringTok{"}\NormalTok{)}
\end{Highlighting}
\end{Shaded}

\begin{tcolorbox}[boxrule=0pt, enhanced, borderline west={2pt}{0pt}{code-block-stdout-light}, interior hidden, frame hidden, breakable, sharp corners, grow to left by=-1em]

\begin{verbatim}
y: tensor([0.3000, 0.5000, 0.2000])
x: tensor([0.0400, 0.9163])
y2: tensor([0.3000, 0.5000, 0.2000])
\end{verbatim}

\end{tcolorbox}

\hypertarget{sec-pkg-ll-matrices}{%
\subsubsection{Parametrization of matrices}\label{sec-pkg-ll-matrices}}

The following functions are available:

\begin{itemize}
\tightlist
\item
  \texttt{reals\_to\_diag\_matrix}, \texttt{diag\_matrix\_to\_reals}:
  for parametrization of diagonal matrices. These functions are
  described in Section~\ref{sec-mat-diag}. Without vectorization, these
  functions can be substituted by \texttt{diag} from numpy, JAX or
  PyTorch.
\item
  \texttt{reals\_to\_sym\_matrix}, \texttt{sym\_matrix\_to\_reals}: for
  parametrization of symmetric matrices. These function are described in
  Section~\ref{sec-mat-sym}.
\item
  \texttt{reals\_to\_spd\_matrix}, \texttt{spd\_matrix\_to\_reals}: for
  parametrization of symmetric positive definite matrices. These
  function are described in Section~\ref{sec-mat-spd}.
\end{itemize}

For usage of matrix parametrizations to n-dimensional arrays with
constraints on the two last dimensions, see
Section~\ref{sec-pkg-ll-vectorization} below.

Example, with the parametrization of a symmetric definite positive
\(3\times3\) matrix with PyTorch:

\begin{Shaded}
\begin{Highlighting}[]
\ImportTok{import}\NormalTok{ torch}
\ImportTok{import}\NormalTok{ parametrization\_cookbook.functions.torch }\ImportTok{as}\NormalTok{ pcf}

\NormalTok{y }\OperatorTok{=}\NormalTok{ torch.tensor([[}\DecValTok{3}\NormalTok{,}\DecValTok{1}\NormalTok{,}\FloatTok{1.5}\NormalTok{],[}\DecValTok{1}\NormalTok{,}\FloatTok{2.5}\NormalTok{,}\OperatorTok{{-}}\DecValTok{1}\NormalTok{],[}\FloatTok{1.5}\NormalTok{,}\OperatorTok{{-}}\DecValTok{1}\NormalTok{,}\DecValTok{2}\NormalTok{]])}
\ControlFlowTok{assert}\NormalTok{ (y}\OperatorTok{==}\NormalTok{y.T).}\BuiltInTok{all}\NormalTok{()}
\ControlFlowTok{assert}\NormalTok{ torch.linalg.eigh(y)[}\DecValTok{0}\NormalTok{].}\BuiltInTok{min}\NormalTok{()}\OperatorTok{\textgreater{}}\DecValTok{0}
\NormalTok{x }\OperatorTok{=}\NormalTok{ pcf.spd\_matrix\_to\_reals(y)}
\NormalTok{y2 }\OperatorTok{=}\NormalTok{ pcf.reals\_to\_spd\_matrix(x)}

\BuiltInTok{print}\NormalTok{(}\SpecialStringTok{f"y:}\CharTok{\textbackslash{}n}\SpecialCharTok{\{}\NormalTok{y}\SpecialCharTok{\}}\SpecialStringTok{"}\NormalTok{)}
\BuiltInTok{print}\NormalTok{(}\SpecialStringTok{f"x:}\CharTok{\textbackslash{}n}\SpecialCharTok{\{}\NormalTok{x}\SpecialCharTok{\}}\SpecialStringTok{"}\NormalTok{)}
\BuiltInTok{print}\NormalTok{(}\SpecialStringTok{f"y2:}\CharTok{\textbackslash{}n}\SpecialCharTok{\{}\NormalTok{y2}\SpecialCharTok{\}}\SpecialStringTok{"}\NormalTok{)}
\end{Highlighting}
\end{Shaded}

\begin{tcolorbox}[boxrule=0pt, enhanced, borderline west={2pt}{0pt}{code-block-stdout-light}, interior hidden, frame hidden, breakable, sharp corners, grow to left by=-1em]

\begin{verbatim}
y:
tensor([[ 3.0000,  1.0000,  1.5000],
        [ 1.0000,  2.5000, -1.0000],
        [ 1.5000, -1.0000,  2.0000]])
x:
tensor([ 1.5373,  1.9485,  0.1972,  0.8165,  1.5000, -1.7650])
y2:
tensor([[ 3.0000,  1.0000,  1.5000],
        [ 1.0000,  2.5000, -1.0000],
        [ 1.5000, -1.0000,  2.0000]])
\end{verbatim}

\end{tcolorbox}

\hypertarget{sec-pkg-ll-vectorization}{%
\subsubsection{Vectorization}\label{sec-pkg-ll-vectorization}}

All functions in this module support vectorization. This allows
computational performance for computing many parametrizations at the
same time with arbitrary shape.

\hypertarget{scalar-functions}{%
\paragraph{Scalar functions}\label{scalar-functions}}

For scalar functions listed in Section~\ref{sec-pkg-ll-scalars},
applying the function on a n-dimensional array (vector, matrices, or
higher dimension) results in the n-dimensional array of the function
result.

Example:

\begin{Shaded}
\begin{Highlighting}[]
\ImportTok{import}\NormalTok{ torch}
\ImportTok{import}\NormalTok{ parametrization\_cookbook.functions.torch }\ImportTok{as}\NormalTok{ pcf}

\NormalTok{x }\OperatorTok{=}\NormalTok{ torch.tensor([[}\OperatorTok{{-}}\DecValTok{3}\NormalTok{, }\OperatorTok{{-}}\DecValTok{2}\NormalTok{], [}\DecValTok{2}\NormalTok{, }\DecValTok{1}\NormalTok{]])}
\NormalTok{y }\OperatorTok{=}\NormalTok{ pcf.softplus(x)}

\BuiltInTok{print}\NormalTok{(}\SpecialStringTok{f"x.shape: }\SpecialCharTok{\{}\NormalTok{x}\SpecialCharTok{.}\NormalTok{shape}\SpecialCharTok{\}}\SpecialStringTok{"}\NormalTok{)}
\BuiltInTok{print}\NormalTok{(}\SpecialStringTok{f"y.shape: }\SpecialCharTok{\{}\NormalTok{y}\SpecialCharTok{.}\NormalTok{shape}\SpecialCharTok{\}}\SpecialStringTok{"}\NormalTok{)}
\BuiltInTok{print}\NormalTok{(}\SpecialStringTok{f"y[1,0]: }\SpecialCharTok{\{}\NormalTok{y[}\DecValTok{1}\NormalTok{,}\DecValTok{0}\NormalTok{]}\SpecialCharTok{\}}\SpecialStringTok{"}\NormalTok{)}
\BuiltInTok{print}\NormalTok{(}\SpecialStringTok{f"softplus(x[1,0]): }\SpecialCharTok{\{}\NormalTok{pcf}\SpecialCharTok{.}\NormalTok{softplus(x[}\DecValTok{1}\NormalTok{,}\DecValTok{0}\NormalTok{])}\SpecialCharTok{\}}\SpecialStringTok{"}\NormalTok{)}
\end{Highlighting}
\end{Shaded}

\begin{tcolorbox}[boxrule=0pt, enhanced, borderline west={2pt}{0pt}{code-block-stdout-light}, interior hidden, frame hidden, breakable, sharp corners, grow to left by=-1em]

\begin{verbatim}
x.shape: torch.Size([2, 2])
y.shape: torch.Size([2, 2])
y[1,0]: 2.1269280910491943
softplus(x[1,0]): 2.1269280910491943
\end{verbatim}

\end{tcolorbox}

\hypertarget{vector-functions}{%
\paragraph{Vector functions}\label{vector-functions}}

For vectors functions listed in Section~\ref{sec-pkg-ll-vectors},
applying the function on a n-dimensional array (matrices, or higher
dimension) results in the n-dimensional array of the function result on
last dimensions.

\emph{E.g.}, for an input with shape \texttt{(n,)} if the output shape
is \texttt{(n+1,)}, then for an input with shape \texttt{(a,b,c,n)} the
output shape will be \texttt{(a,b,c,n+1)}.

Example:

\begin{Shaded}
\begin{Highlighting}[]
\ImportTok{import}\NormalTok{ torch}
\ImportTok{import}\NormalTok{ parametrization\_cookbook.functions.torch }\ImportTok{as}\NormalTok{ pcf}

\NormalTok{x }\OperatorTok{=}\NormalTok{ torch.tensor([[}\OperatorTok{{-}}\DecValTok{3}\NormalTok{, }\OperatorTok{{-}}\DecValTok{2}\NormalTok{], [}\DecValTok{2}\NormalTok{, }\DecValTok{1}\NormalTok{]])}
\NormalTok{y }\OperatorTok{=}\NormalTok{ pcf.reals\_to\_simplex(x)}

\BuiltInTok{print}\NormalTok{(}\SpecialStringTok{f"x.shape: }\SpecialCharTok{\{}\NormalTok{x}\SpecialCharTok{.}\NormalTok{shape}\SpecialCharTok{\}}\SpecialStringTok{"}\NormalTok{)}
\BuiltInTok{print}\NormalTok{(}\SpecialStringTok{f"y.shape: }\SpecialCharTok{\{}\NormalTok{y}\SpecialCharTok{.}\NormalTok{shape}\SpecialCharTok{\}}\SpecialStringTok{"}\NormalTok{)}
\BuiltInTok{print}\NormalTok{(}\SpecialStringTok{f"y[1,:]: }\SpecialCharTok{\{}\NormalTok{y[}\DecValTok{1}\NormalTok{,:]}\SpecialCharTok{\}}\SpecialStringTok{"}\NormalTok{)}
\BuiltInTok{print}\NormalTok{(}\SpecialStringTok{f"reals\_to\_simplex(x[1,:]): }\SpecialCharTok{\{}\NormalTok{pcf}\SpecialCharTok{.}\NormalTok{reals\_to\_simplex(x[}\DecValTok{1}\NormalTok{,:])}\SpecialCharTok{\}}\SpecialStringTok{"}\NormalTok{)}
\end{Highlighting}
\end{Shaded}

\begin{tcolorbox}[boxrule=0pt, enhanced, borderline west={2pt}{0pt}{code-block-stdout-light}, interior hidden, frame hidden, breakable, sharp corners, grow to left by=-1em]

\begin{verbatim}
x.shape: torch.Size([2, 2])
y.shape: torch.Size([2, 3])
y[1,:]: tensor([0.6547, 0.2524, 0.0929])
reals_to_simplex(x[1,:]): tensor([0.6547, 0.2524, 0.0929])
\end{verbatim}

\end{tcolorbox}

\hypertarget{matrices-functions}{%
\paragraph{Matrices functions}\label{matrices-functions}}

For matrices functions listed in Section~\ref{sec-pkg-ll-matrices},
applying the function on a n-dimensional array results in the
n-dimensional array of the function result on two last dimensions.

\emph{E.g.}, for a input with shape \texttt{(N,)} if the output shape is
\texttt{(n,n)} with \(N=\frac{n(n+1)}2\), then for a input with shape
\texttt{(a,b,c,N)} the output shape will be \texttt{(a,b,c,n,n)}.

Example:

\begin{Shaded}
\begin{Highlighting}[]
\ImportTok{import}\NormalTok{ torch}
\ImportTok{import}\NormalTok{ parametrization\_cookbook.functions.torch }\ImportTok{as}\NormalTok{ pcf}

\NormalTok{x }\OperatorTok{=}\NormalTok{ torch.linspace(}\OperatorTok{{-}}\DecValTok{1}\NormalTok{,}\DecValTok{1}\NormalTok{,}\DecValTok{66}\NormalTok{).reshape((}\DecValTok{11}\NormalTok{,}\DecValTok{6}\NormalTok{))}
\NormalTok{y }\OperatorTok{=}\NormalTok{ pcf.reals\_to\_spd\_matrix(x)}

\BuiltInTok{print}\NormalTok{(}\SpecialStringTok{f"x.shape: }\SpecialCharTok{\{}\NormalTok{x}\SpecialCharTok{.}\NormalTok{shape}\SpecialCharTok{\}}\SpecialStringTok{"}\NormalTok{)}
\BuiltInTok{print}\NormalTok{(}\SpecialStringTok{f"y.shape: }\SpecialCharTok{\{}\NormalTok{y}\SpecialCharTok{.}\NormalTok{shape}\SpecialCharTok{\}}\SpecialStringTok{"}\NormalTok{)}
\BuiltInTok{print}\NormalTok{(}\SpecialStringTok{f"y[7,:,:]:}\CharTok{\textbackslash{}n}\SpecialCharTok{\{}\NormalTok{y[}\DecValTok{7}\NormalTok{,:,:]}\SpecialCharTok{\}}\SpecialStringTok{"}\NormalTok{)}
\BuiltInTok{print}\NormalTok{(}\SpecialStringTok{f"reals\_to\_spd\_matrix(x[7,:]):}\CharTok{\textbackslash{}n}\SpecialCharTok{\{}\NormalTok{pcf}\SpecialCharTok{.}\NormalTok{reals\_to\_spd\_matrix(x[}\DecValTok{7}\NormalTok{,:])}\SpecialCharTok{\}}\SpecialStringTok{"}\NormalTok{)}
\end{Highlighting}
\end{Shaded}

\begin{tcolorbox}[boxrule=0pt, enhanced, borderline west={2pt}{0pt}{code-block-stdout-light}, interior hidden, frame hidden, breakable, sharp corners, grow to left by=-1em]

\begin{verbatim}
x.shape: torch.Size([11, 6])
y.shape: torch.Size([11, 3, 3])
y[7,:,:]:
tensor([[0.7224, 0.2312, 0.2038],
        [0.2312, 0.4504, 0.2233],
        [0.2038, 0.2233, 0.3853]])
reals_to_spd_matrix(x[7,:]):
tensor([[0.7224, 0.2312, 0.2038],
        [0.2312, 0.4504, 0.2233],
        [0.2038, 0.2233, 0.3853]])
\end{verbatim}

\end{tcolorbox}

\hypertarget{examples-1}{%
\section{Examples}\label{examples-1}}

\hypertarget{high-level-modules-inference-of-parameters-of-a-multivariate-student-distribution}{%
\subsection{High-level modules: inference of parameters of a
Multivariate Student
distribution}\label{high-level-modules-inference-of-parameters-of-a-multivariate-student-distribution}}

\hypertarget{introduction-1}{%
\subsubsection{Introduction}\label{introduction-1}}

With \(\mu\in\reels^p\), \(\Sigma\) a symmetric definite positive matrix
of size \(p\), and \(\nu\in\reelsp\), the multivariate Student
distribution with parameters \((\mu,\Sigma,\nu)\) is defined by the
following probability distribution function:

\[
   \begin{array}{rcl}
    \mathbb R^p & \longrightarrow & \mathbb R_+ \\
    x & \longmapsto &
    \frac{\Gamma\left(\frac{\nu+p}2\right)}{\Gamma\left(\frac\nu2\right)\nu^{\frac p2}\pi^{\frac p2}\left|\Sigma\right|^{\frac12}}{\left(1+\frac1\nu(x-\mu)^T\Sigma^{-1}(x-\mu)\right)}^{-\frac{\nu+p}2}
   \end{array}
\]

With \(Y\sim\mathcal N(0,\Sigma)\), \(Z\sim\chi^2_\nu\), then
\(X = \mu+\frac{Y}{\sqrt{Z/\nu}}\) follows a multivariate Student
distribution with parameters \((\mu,\Sigma,\nu)\).

The idea of this example is to introduce the inference with Maximum
Likelihood Estimator (MLE) of the parameters, handling the constraints
by parametrization with high level interface of the package, using
automatic differentiation to compute derivatives.

We introduce \(\theta\in\reels^k\) the mapping of \((\mu, \Sigma, \nu)\)
by our bijective parametrization. Using invariance property of the MLE,
the mapping of MLE of \(\theta\) is equivalent to the MLE of
\((\mu, \Sigma, \nu)\).

In a second time, when a MLE is obtained, with sufficient regularity
conditions (not detailed here), using asymptotic properties of MLE and
Slutsky's lemma we have:

\[
    \widehat I_{n,\widehat\theta}^{-\frac12}\p{\widehat\theta-\theta_0}
    \underset{n\to+\infty}\longrightarrow \mathcal N\p{0, I}
\]

where:

\begin{itemize}
\tightlist
\item
  \(\widehat I_{n,\widehat\theta} = - {\left.\frac{\operatorname{d}^2\ell\p{\theta, \ldots}}{\operatorname{d}\theta^2}\right|}_{\theta=\widehat\theta}\)
  is the estimated Fisher information matrix.
\item
  \(I\) is the identity matrix.
\item
  \(\ell\) is the log-likelihood of the whole sample.
\end{itemize}

Now we can move this result to our original parameter space:

\[
\frac{\widehat\nu-\nu_0}{\sqrt{\delta_\nu^TI_{n,\widehat\theta}^{-1}\delta_\nu}}\underset{n\to+\infty}\longrightarrow\mathcal N(0,1)
\]

with:

\begin{itemize}
\tightlist
\item
  \(\delta_\nu = {\left.\frac{\operatorname{d}\nu}{\operatorname{d}\theta}\right|}_{\theta=\widehat\theta}\)
\end{itemize}

Therefore we can obtain asymptotic confidence interval:

\[
\mathbb P\p{
\nu_0\in
\left[\widehat\nu \pm
   u_{1-\alpha/2}\sqrt{\delta_\nu^TI_{n,\widehat\theta}^{-1}\delta_\nu}\right]
} \underset{n\to+\infty}\longrightarrow 1-\alpha
\]

Note that \(\delta_\nu\) and \(I_{n,\widehat\theta}\) will be computed
with automatic differentiation.

The same method is applicable with any parameter or function of
parameter, \emph{e.g.} we can have a confidence interval on
\(\abs{\Sigma}\):

\[
\mathbb P\p{
\abs{\Sigma_0}\in
\left[\abs{\widehat\Sigma} \pm
   u_{1-\alpha/2}\sqrt{\delta_{\abs\Sigma}^TI_{n,\widehat\theta}^{-1}\delta_{\abs\Sigma}}\right]
} \underset{n\to+\infty}\longrightarrow 1-\alpha
\]

with:

\begin{itemize}
\tightlist
\item
  \(\delta_{\abs\Sigma} = {\left.\frac{\operatorname{d}\abs{\Sigma}}{\operatorname{d}\theta}\right|}_{\theta=\widehat\theta}\)
\end{itemize}

\hypertarget{with-jax}{%
\subsubsection{With JAX}\label{with-jax}}

\hypertarget{simulating-the-data}{%
\paragraph{Simulating the data}\label{simulating-the-data}}

First we generate simulated data to illustrate the method.

\begin{Shaded}
\begin{Highlighting}[]
\ImportTok{import}\NormalTok{ numpy }\ImportTok{as}\NormalTok{ np}
\ImportTok{import}\NormalTok{ scipy.stats}

\NormalTok{n }\OperatorTok{=} \DecValTok{1000}
\NormalTok{mu }\OperatorTok{=}\NormalTok{ np.arange(}\DecValTok{3}\NormalTok{)}
\NormalTok{Sigma }\OperatorTok{=}\NormalTok{ np.array([[}\DecValTok{2}\NormalTok{, }\DecValTok{1}\NormalTok{, }\DecValTok{1}\NormalTok{], [}\DecValTok{1}\NormalTok{, }\DecValTok{2}\NormalTok{, }\FloatTok{1.5}\NormalTok{], [}\DecValTok{1}\NormalTok{, }\FloatTok{1.5}\NormalTok{, }\DecValTok{2}\NormalTok{]])}
\NormalTok{df }\OperatorTok{=} \DecValTok{7}

\CommentTok{\# we use a seeded random state only for reproducibility}
\NormalTok{random\_state }\OperatorTok{=}\NormalTok{ np.random.RandomState(np.random.MT19937(np.random.SeedSequence(}\DecValTok{0}\NormalTok{)))}
\NormalTok{X }\OperatorTok{=}\NormalTok{ scipy.stats.multivariate\_t(loc}\OperatorTok{=}\NormalTok{mu, shape}\OperatorTok{=}\NormalTok{Sigma, df}\OperatorTok{=}\NormalTok{df).rvs(}
\NormalTok{    size}\OperatorTok{=}\NormalTok{n, random\_state}\OperatorTok{=}\NormalTok{random\_state}
\NormalTok{)}

\CommentTok{\# convert to JAX array}
\ImportTok{import}\NormalTok{ jax.numpy }\ImportTok{as}\NormalTok{ jnp}
\NormalTok{X }\OperatorTok{=}\NormalTok{ jnp.array(X)}
\end{Highlighting}
\end{Shaded}

\hypertarget{definition-of-the-parametrization}{%
\paragraph{Definition of the
parametrization}\label{definition-of-the-parametrization}}

Our parameter space is the Cartesian product of \(\mathbb R^3\) (for
\(\mu\)), the space of symmetric definite positive matrices of size 3
(for \(\Sigma\)) and \(\mathbb R_+^*\) (for the degree of freedom
\(\nu\)). To handle these constraints we define a parametrization
between this space and \(\mathbb R^k\) (the value of \(k\) will be
automatically computed).

\begin{Shaded}
\begin{Highlighting}[]
\ImportTok{import}\NormalTok{ parametrization\_cookbook.jax }\ImportTok{as}\NormalTok{ pc}

\NormalTok{parametrization }\OperatorTok{=}\NormalTok{ pc.NamedTuple(}
\NormalTok{     mu}\OperatorTok{=}\NormalTok{pc.Real(shape}\OperatorTok{=}\DecValTok{3}\NormalTok{),}
\NormalTok{     Sigma}\OperatorTok{=}\NormalTok{pc.MatrixSymPosDef(dim}\OperatorTok{=}\DecValTok{3}\NormalTok{),}
\NormalTok{     df}\OperatorTok{=}\NormalTok{pc.RealPositive()}
\NormalTok{)}
\end{Highlighting}
\end{Shaded}

We can retrieve the value of \(k\) with \texttt{parametrization.size}.

\begin{Shaded}
\begin{Highlighting}[]
\NormalTok{parametrization.size}
\end{Highlighting}
\end{Shaded}

\begin{tcolorbox}[boxrule=0pt, enhanced, borderline west={2pt}{0pt}{code-block-stdout-light}, interior hidden, frame hidden, breakable, sharp corners, grow to left by=-1em]

\begin{verbatim}
10
\end{verbatim}

\end{tcolorbox}

\hypertarget{definition-of-the-log-likelihood-and-gradients}{%
\paragraph{Definition of the log-likelihood and
gradients}\label{definition-of-the-log-likelihood-and-gradients}}

First we define the log-likelihood depending on our original parameters:

\begin{Shaded}
\begin{Highlighting}[]
\KeywordTok{def}\NormalTok{ original\_loglikelihood(mu, Sigma, df, X):}
\NormalTok{    n, p }\OperatorTok{=}\NormalTok{ X.shape}

\NormalTok{    eigvals, eigvect }\OperatorTok{=}\NormalTok{ jnp.linalg.eigh(Sigma)}
\NormalTok{    U }\OperatorTok{=}\NormalTok{ eigvect }\OperatorTok{*}\NormalTok{ (eigvals}\OperatorTok{**{-}}\FloatTok{0.5}\NormalTok{)}
\NormalTok{    logdet }\OperatorTok{=}\NormalTok{ jnp.log(eigvals).}\BuiltInTok{sum}\NormalTok{()}

    \ControlFlowTok{return}\NormalTok{ (}
\NormalTok{        jax.scipy.special.gammaln((df }\OperatorTok{+}\NormalTok{ p) }\OperatorTok{/} \DecValTok{2}\NormalTok{)}
        \OperatorTok{{-}}\NormalTok{ jax.scipy.special.gammaln(df }\OperatorTok{/} \DecValTok{2}\NormalTok{)}
        \OperatorTok{{-}}\NormalTok{ p }\OperatorTok{/} \DecValTok{2} \OperatorTok{*}\NormalTok{ jnp.log(df }\OperatorTok{*}\NormalTok{ jnp.pi)}
        \OperatorTok{{-}} \DecValTok{1} \OperatorTok{/} \DecValTok{2} \OperatorTok{*}\NormalTok{ logdet}
        \OperatorTok{{-}}\NormalTok{ (}
\NormalTok{            (df }\OperatorTok{+}\NormalTok{ p)}
            \OperatorTok{/} \DecValTok{2}
            \OperatorTok{*}\NormalTok{ jnp.log1p((((X }\OperatorTok{{-}}\NormalTok{ mu) }\OperatorTok{@}\NormalTok{ U) }\OperatorTok{**} \DecValTok{2}\NormalTok{).}\BuiltInTok{sum}\NormalTok{(axis}\OperatorTok{=}\DecValTok{1}\NormalTok{) }\OperatorTok{/}\NormalTok{ df)}
\NormalTok{        )}
\NormalTok{    ).}\BuiltInTok{sum}\NormalTok{()}
\end{Highlighting}
\end{Shaded}

And we define the log-likelihood of our parametrized model:

\begin{Shaded}
\begin{Highlighting}[]
\ImportTok{import}\NormalTok{ jax}

\AttributeTok{@jax.jit}
\KeywordTok{def}\NormalTok{ loglikelihood(theta, X):}
\NormalTok{    my\_params }\OperatorTok{=}\NormalTok{ parametrization.reals1d\_to\_params(theta)}
    \ControlFlowTok{return}\NormalTok{ original\_loglikelihood(my\_params.mu, my\_params.Sigma, my\_params.df, X)}
\end{Highlighting}
\end{Shaded}

This function was JIT-compiled, as this function is run many-times, this
is very interesting to reduce computation time.

We now define the gradient and hessian functions (with JIT-compilation):

\begin{Shaded}
\begin{Highlighting}[]
\NormalTok{grad\_loglikelihood }\OperatorTok{=}\NormalTok{ jax.jit(jax.grad(loglikelihood))}
\NormalTok{hessian\_loglikelihood }\OperatorTok{=}\NormalTok{ jax.jit(jax.jacfwd(jax.jacrev(loglikelihood)))}
\end{Highlighting}
\end{Shaded}

\hypertarget{optimization}{%
\paragraph{Optimization}\label{optimization}}

We can use any optimization algorithm. We choose here a gradient descend
(with step conditioning by the highest eigenvalue of the hessian)
followed by the Newton-Raphson method. The gradient method is choosen
for its robustness, and the second for its quick convergence starting
from a initial point close to the optimum. This is only given for
illustration purpose, in a real application case, using optimization
algorithm developed with JAX in the Python module \texttt{jaxopt}
(\protect\hyperlink{ref-jaxopt}{Blondel et al. 2021}) can be a better
choice.

We choose here to initialize randomly \(\theta\). We can also build a
plausible value of \(\theta\) with
\texttt{parametrization.params\_to\_reals1d}.

\begin{Shaded}
\begin{Highlighting}[]
\ImportTok{import}\NormalTok{ itertools}

\CommentTok{\# we use a seeded random state only for reproducibility}
\NormalTok{random\_state }\OperatorTok{=}\NormalTok{ np.random.RandomState(np.random.MT19937(np.random.SeedSequence(}\DecValTok{1}\NormalTok{)))}
\NormalTok{theta }\OperatorTok{=}\NormalTok{ random\_state.normal(size}\OperatorTok{=}\NormalTok{parametrization.size)}

\NormalTok{current\_likeli }\OperatorTok{=}\NormalTok{ loglikelihood(theta, X)}
\BuiltInTok{print}\NormalTok{(}\SpecialStringTok{f"current\_likli = }\SpecialCharTok{\{}\NormalTok{current\_likeli}\SpecialCharTok{\}}\SpecialStringTok{"}\NormalTok{)}

\ControlFlowTok{for}\NormalTok{ it\_grad }\KeywordTok{in}\NormalTok{ itertools.count():}
\NormalTok{    g }\OperatorTok{=}\NormalTok{ grad\_loglikelihood(theta, X)}
\NormalTok{    H }\OperatorTok{=}\NormalTok{ hessian\_loglikelihood(theta, X)}
\NormalTok{    sdp\_eigenvalues }\OperatorTok{=} \OperatorTok{{-}}\NormalTok{jnp.linalg.eigh(H)[}\DecValTok{0}\NormalTok{]}
\NormalTok{    lr }\OperatorTok{=} \DecValTok{1}\OperatorTok{/}\NormalTok{sdp\_eigenvalues.}\BuiltInTok{max}\NormalTok{()}
\NormalTok{    theta }\OperatorTok{+=}\NormalTok{ lr}\OperatorTok{*}\NormalTok{g}
\NormalTok{    current\_likeli, old\_likeli }\OperatorTok{=}\NormalTok{ loglikelihood(theta, X), current\_likeli}
    \ControlFlowTok{if}\NormalTok{ current\_likeli}\OperatorTok{{-}}\NormalTok{old\_likeli}\OperatorTok{\textless{}}\FloatTok{1e{-}2}\NormalTok{:}
        \ControlFlowTok{break}
\BuiltInTok{print}\NormalTok{(}\SpecialStringTok{f\textquotesingle{}it\_grad: }\SpecialCharTok{\{}\NormalTok{it\_grad}\SpecialCharTok{\}}\SpecialStringTok{, current\_likli: }\SpecialCharTok{\{}\NormalTok{current\_likeli}\SpecialCharTok{\}}\SpecialStringTok{\textquotesingle{}}\NormalTok{)}

\ControlFlowTok{for}\NormalTok{ it\_nr }\KeywordTok{in}\NormalTok{ itertools.count():}
\NormalTok{    g }\OperatorTok{=}\NormalTok{ grad\_loglikelihood(theta, X)}
\NormalTok{    H }\OperatorTok{=}\NormalTok{ hessian\_loglikelihood(theta, X)}
\NormalTok{    theta }\OperatorTok{+=} \OperatorTok{{-}}\BuiltInTok{min}\NormalTok{(}\DecValTok{1}\NormalTok{,}\FloatTok{.1}\OperatorTok{*}\DecValTok{2}\OperatorTok{**}\NormalTok{it\_nr)}\OperatorTok{*}\NormalTok{jnp.linalg.solve(H, g)}
\NormalTok{    current\_likeli, old\_likeli }\OperatorTok{=}\NormalTok{ loglikelihood(theta, X), current\_likeli}
    \ControlFlowTok{if}\NormalTok{ it\_nr}\OperatorTok{\textgreater{}}\DecValTok{3} \KeywordTok{and}\NormalTok{ current\_likeli}\OperatorTok{{-}}\NormalTok{old\_likeli}\OperatorTok{\textless{}}\FloatTok{1e{-}6}\NormalTok{:}
        \ControlFlowTok{break}
\BuiltInTok{print}\NormalTok{(}\SpecialStringTok{f\textquotesingle{}it\_nr: }\SpecialCharTok{\{}\NormalTok{it\_nr}\SpecialCharTok{\}}\SpecialStringTok{, current\_likli: }\SpecialCharTok{\{}\NormalTok{current\_likeli}\SpecialCharTok{\}}\SpecialStringTok{\textquotesingle{}}\NormalTok{)}
\BuiltInTok{print}\NormalTok{(}\SpecialStringTok{f\textquotesingle{}theta: }\SpecialCharTok{\{}\NormalTok{theta}\SpecialCharTok{\}}\SpecialStringTok{\textquotesingle{}}\NormalTok{)}
\end{Highlighting}
\end{Shaded}

\begin{tcolorbox}[boxrule=0pt, enhanced, borderline west={2pt}{0pt}{code-block-stdout-light}, interior hidden, frame hidden, breakable, sharp corners, grow to left by=-1em]

\begin{verbatim}
current_likli = -17553.708984375
\end{verbatim}

\end{tcolorbox}

\begin{tcolorbox}[boxrule=0pt, enhanced, borderline west={2pt}{0pt}{code-block-stdout-light}, interior hidden, frame hidden, breakable, sharp corners, grow to left by=-1em]

\begin{verbatim}
it_grad: 481, current_likli: -5160.08642578125
it_nr: 6, current_likli: -5156.76318359375
theta: [0.01115316 1.0506299  2.0488813  1.1360171  1.6467183  1.3518243
 1.0945351  1.2938223  1.5352948  7.492975  ]
\end{verbatim}

\end{tcolorbox}

\hypertarget{using-the-value}{%
\paragraph{Using the value}\label{using-the-value}}

It is easy to retrieve estimates
\((\widehat\mu,\widehat\Sigma,\widehat\nu)\):

\begin{Shaded}
\begin{Highlighting}[]
\NormalTok{my\_params }\OperatorTok{=}\NormalTok{ parametrization.reals1d\_to\_params(theta)}
\NormalTok{my\_params.mu}
\end{Highlighting}
\end{Shaded}

\begin{tcolorbox}[boxrule=0pt, enhanced, borderline west={2pt}{0pt}{code-block-stdout-light}, interior hidden, frame hidden, breakable, sharp corners, grow to left by=-1em]

\begin{verbatim}
DeviceArray([0.01115316, 1.0506299 , 2.0488813 ], dtype=float32)
\end{verbatim}

\end{tcolorbox}

\begin{Shaded}
\begin{Highlighting}[]
\NormalTok{my\_params.Sigma}
\end{Highlighting}
\end{Shaded}

\begin{tcolorbox}[boxrule=0pt, enhanced, borderline west={2pt}{0pt}{code-block-stdout-light}, interior hidden, frame hidden, breakable, sharp corners, grow to left by=-1em]

\begin{verbatim}
DeviceArray([[2.0007489, 1.09474  , 1.0565993],
             [1.09474  , 2.260526 , 1.720708 ],
             [1.0565993, 1.720708 , 2.1778986]], dtype=float32)
\end{verbatim}

\end{tcolorbox}

\begin{Shaded}
\begin{Highlighting}[]
\NormalTok{my\_params.df}
\end{Highlighting}
\end{Shaded}

\begin{tcolorbox}[boxrule=0pt, enhanced, borderline west={2pt}{0pt}{code-block-stdout-light}, interior hidden, frame hidden, breakable, sharp corners, grow to left by=-1em]

\begin{verbatim}
DeviceArray(7.493532, dtype=float32)
\end{verbatim}

\end{tcolorbox}

We can see we recover good estimate of the simulated parameters.

\hypertarget{building-confidence-interval}{%
\paragraph{Building confidence
interval}\label{building-confidence-interval}}

The first step is to compute the inverse of the estimated Fisher
information matrix \(\widehat I_{n,\widehat\theta}\):

\begin{Shaded}
\begin{Highlighting}[]
\NormalTok{FIM\_inv }\OperatorTok{=}\NormalTok{ jnp.linalg.inv(}\OperatorTok{{-}}\NormalTok{H)}
\end{Highlighting}
\end{Shaded}

And we can compute \(\delta_\nu\), then the confidence interval:

\begin{Shaded}
\begin{Highlighting}[]
\NormalTok{delta\_df }\OperatorTok{=}\NormalTok{ jax.grad(}
    \KeywordTok{lambda}\NormalTok{ theta: parametrization.reals1d\_to\_params(theta).df}
\NormalTok{)(theta)}
\NormalTok{df\_asymptotic\_variance }\OperatorTok{=}\NormalTok{ delta\_df }\OperatorTok{@}\NormalTok{ FIM\_inv }\OperatorTok{@}\NormalTok{ delta\_df}
\NormalTok{df\_confidence\_interval }\OperatorTok{=}\NormalTok{ (}
\NormalTok{    parametrization.reals1d\_to\_params(theta).df}
    \OperatorTok{+}\NormalTok{ (}
\NormalTok{        jnp.array([}\OperatorTok{{-}}\DecValTok{1}\NormalTok{, }\DecValTok{1}\NormalTok{])}
        \OperatorTok{*}\NormalTok{ scipy.stats.norm.ppf(}\FloatTok{0.975}\NormalTok{)}
        \OperatorTok{*}\NormalTok{ jnp.sqrt(df\_asymptotic\_variance)}
\NormalTok{    )}
\NormalTok{)}
\BuiltInTok{print}\NormalTok{(df\_confidence\_interval)}
\end{Highlighting}
\end{Shaded}

\begin{tcolorbox}[boxrule=0pt, enhanced, borderline west={2pt}{0pt}{code-block-stdout-light}, interior hidden, frame hidden, breakable, sharp corners, grow to left by=-1em]

\begin{verbatim}
[5.815369 9.171696]
\end{verbatim}

\end{tcolorbox}

The simulated value was \(7\).

For the confidence interval on \(\abs\Sigma\), we have:

\begin{Shaded}
\begin{Highlighting}[]
\NormalTok{delta\_det }\OperatorTok{=}\NormalTok{ jax.grad(}
    \KeywordTok{lambda}\NormalTok{ theta: jnp.linalg.det(parametrization.reals1d\_to\_params(theta).Sigma)}
\NormalTok{)(theta)}
\NormalTok{det\_asymptotic\_variance }\OperatorTok{=}\NormalTok{ delta\_det }\OperatorTok{@}\NormalTok{ FIM\_inv }\OperatorTok{@}\NormalTok{ delta\_det}
\NormalTok{det\_confidence\_interval }\OperatorTok{=}\NormalTok{ (}
\NormalTok{    jnp.linalg.det(parametrization.reals1d\_to\_params(theta).Sigma)}
    \OperatorTok{+}\NormalTok{ (}
\NormalTok{        jnp.array([}\OperatorTok{{-}}\DecValTok{1}\NormalTok{, }\DecValTok{1}\NormalTok{]) }
        \OperatorTok{*}\NormalTok{ scipy.stats.norm.ppf(}\FloatTok{0.975}\NormalTok{) }
        \OperatorTok{*}\NormalTok{ np.sqrt(det\_asymptotic\_variance)}
\NormalTok{    )}
\NormalTok{)}
\BuiltInTok{print}\NormalTok{(det\_confidence\_interval)}
\end{Highlighting}
\end{Shaded}

\begin{tcolorbox}[boxrule=0pt, enhanced, borderline west={2pt}{0pt}{code-block-stdout-light}, interior hidden, frame hidden, breakable, sharp corners, grow to left by=-1em]

\begin{verbatim}
[2.1325927 3.4136271]
\end{verbatim}

\end{tcolorbox}

The simulated value was \(2.5\).

\hypertarget{with-pytorch}{%
\subsubsection{With PyTorch}\label{with-pytorch}}

\hypertarget{simulating-the-data-1}{%
\paragraph{Simulating the data}\label{simulating-the-data-1}}

First we generate simulated data to illustrate the method.

\begin{Shaded}
\begin{Highlighting}[]
\ImportTok{import}\NormalTok{ numpy }\ImportTok{as}\NormalTok{ np}
\ImportTok{import}\NormalTok{ scipy.stats}

\NormalTok{n }\OperatorTok{=} \DecValTok{1000}
\NormalTok{mu }\OperatorTok{=}\NormalTok{ np.arange(}\DecValTok{3}\NormalTok{)}
\NormalTok{Sigma }\OperatorTok{=}\NormalTok{ np.array([[}\DecValTok{2}\NormalTok{, }\DecValTok{1}\NormalTok{, }\DecValTok{1}\NormalTok{], [}\DecValTok{1}\NormalTok{, }\DecValTok{2}\NormalTok{, }\FloatTok{1.5}\NormalTok{], [}\DecValTok{1}\NormalTok{, }\FloatTok{1.5}\NormalTok{, }\DecValTok{2}\NormalTok{]])}
\NormalTok{df }\OperatorTok{=} \DecValTok{7}

\CommentTok{\# we use a seeded random state only for reproducibility}
\NormalTok{random\_state }\OperatorTok{=}\NormalTok{ np.random.RandomState(np.random.MT19937(np.random.SeedSequence(}\DecValTok{0}\NormalTok{)))}
\NormalTok{X }\OperatorTok{=}\NormalTok{ scipy.stats.multivariate\_t(loc}\OperatorTok{=}\NormalTok{mu, shape}\OperatorTok{=}\NormalTok{Sigma, df}\OperatorTok{=}\NormalTok{df).rvs(}
\NormalTok{    size}\OperatorTok{=}\NormalTok{n, random\_state}\OperatorTok{=}\NormalTok{random\_state}
\NormalTok{)}

\CommentTok{\# convert to torch}
\ImportTok{import}\NormalTok{ torch}
\NormalTok{X }\OperatorTok{=}\NormalTok{ torch.tensor(X, dtype}\OperatorTok{=}\NormalTok{torch.float32)}
\end{Highlighting}
\end{Shaded}

\hypertarget{definition-of-the-parametrization-1}{%
\paragraph{Definition of the
parametrization}\label{definition-of-the-parametrization-1}}

Our parameter space is the Cartesian product of \(\mathbb R^3\) (for
\(\mu\)), the space of symmetric definite positive matrices of size 3
(for \(\Sigma\)) and \(\mathbb R_+^*\) (for the degree of freedom
\(\nu\)). To handle these constraints we define a parametrization
between this space and \(\mathbb R^k\) (the value of \(k\) will be
automatically computed).

\begin{Shaded}
\begin{Highlighting}[]
\ImportTok{import}\NormalTok{ parametrization\_cookbook.torch }\ImportTok{as}\NormalTok{ pc}

\NormalTok{parametrization }\OperatorTok{=}\NormalTok{ pc.NamedTuple(}
\NormalTok{     mu}\OperatorTok{=}\NormalTok{pc.Real(shape}\OperatorTok{=}\DecValTok{3}\NormalTok{),}
\NormalTok{     Sigma}\OperatorTok{=}\NormalTok{pc.MatrixSymPosDef(dim}\OperatorTok{=}\DecValTok{3}\NormalTok{),}
\NormalTok{     df}\OperatorTok{=}\NormalTok{pc.RealPositive()}
\NormalTok{)}
\end{Highlighting}
\end{Shaded}

We can retrieve the value of \(k\) with \texttt{parametrization.size}.

\begin{Shaded}
\begin{Highlighting}[]
\NormalTok{parametrization.size}
\end{Highlighting}
\end{Shaded}

\begin{tcolorbox}[boxrule=0pt, enhanced, borderline west={2pt}{0pt}{code-block-stdout-light}, interior hidden, frame hidden, breakable, sharp corners, grow to left by=-1em]

\begin{verbatim}
10
\end{verbatim}

\end{tcolorbox}

\hypertarget{definition-of-the-log-likelihood-and-gradients-1}{%
\paragraph{Definition of the log-likelihood and
gradients}\label{definition-of-the-log-likelihood-and-gradients-1}}

First we define the log-likelihood depending on our original parameters:

\begin{Shaded}
\begin{Highlighting}[]
\KeywordTok{def}\NormalTok{ original\_loglikelihood(mu, Sigma, df, X):}
\NormalTok{    n, p }\OperatorTok{=}\NormalTok{ X.shape}

\NormalTok{    eigvals, eigvect }\OperatorTok{=}\NormalTok{ torch.linalg.eigh(Sigma)}
\NormalTok{    U }\OperatorTok{=}\NormalTok{ eigvect }\OperatorTok{*}\NormalTok{ (eigvals}\OperatorTok{**{-}}\FloatTok{0.5}\NormalTok{)}
\NormalTok{    logdet }\OperatorTok{=}\NormalTok{ torch.log(eigvals).}\BuiltInTok{sum}\NormalTok{()}

    \ControlFlowTok{return}\NormalTok{ (}
\NormalTok{        torch.special.gammaln((df }\OperatorTok{+}\NormalTok{ p) }\OperatorTok{/} \DecValTok{2}\NormalTok{)}
        \OperatorTok{{-}}\NormalTok{ torch.special.gammaln(df }\OperatorTok{/} \DecValTok{2}\NormalTok{)}
        \OperatorTok{{-}}\NormalTok{ p }\OperatorTok{/} \DecValTok{2} \OperatorTok{*}\NormalTok{ torch.log(df }\OperatorTok{*}\NormalTok{ torch.pi)}
        \OperatorTok{{-}} \DecValTok{1} \OperatorTok{/} \DecValTok{2} \OperatorTok{*}\NormalTok{ logdet}
        \OperatorTok{{-}}\NormalTok{ (}
\NormalTok{            (df }\OperatorTok{+}\NormalTok{ p)}
            \OperatorTok{/} \DecValTok{2}
            \OperatorTok{*}\NormalTok{ torch.log1p((((X }\OperatorTok{{-}}\NormalTok{ mu) }\OperatorTok{@}\NormalTok{ U) }\OperatorTok{**} \DecValTok{2}\NormalTok{).}\BuiltInTok{sum}\NormalTok{(axis}\OperatorTok{=}\DecValTok{1}\NormalTok{) }\OperatorTok{/}\NormalTok{ df)}
\NormalTok{        )}
\NormalTok{    ).}\BuiltInTok{sum}\NormalTok{()}
\end{Highlighting}
\end{Shaded}

And we define the log-likelihood of our parametrized model:

\begin{Shaded}
\begin{Highlighting}[]
\KeywordTok{def}\NormalTok{ loglikelihood(theta, X):}
\NormalTok{    my\_params }\OperatorTok{=}\NormalTok{ parametrization.reals1d\_to\_params(theta)}
    \ControlFlowTok{return}\NormalTok{ original\_loglikelihood(my\_params.mu, my\_params.Sigma, my\_params.df, X)}
\end{Highlighting}
\end{Shaded}

\hypertarget{optimization-1}{%
\paragraph{Optimization}\label{optimization-1}}

We can use any optimization algorithm. We choose here a ADAM gradient.

We choose here to initialize randomly \(\theta\). We can also build a
plausible value of \(\theta\) with
\texttt{parametrization.params\_to\_reals1d}.

\begin{Shaded}
\begin{Highlighting}[]
\ImportTok{import}\NormalTok{ itertools}

\CommentTok{\# we use a seeded random state only for reproducibility}
\NormalTok{random\_state }\OperatorTok{=}\NormalTok{ np.random.RandomState(np.random.MT19937(np.random.SeedSequence(}\DecValTok{1}\NormalTok{)))}
\NormalTok{theta }\OperatorTok{=}\NormalTok{ random\_state.normal(size}\OperatorTok{=}\NormalTok{parametrization.size)}
\NormalTok{theta }\OperatorTok{=}\NormalTok{ torch.tensor(theta, dtype}\OperatorTok{=}\NormalTok{torch.float32, requires\_grad}\OperatorTok{=}\VariableTok{True}\NormalTok{)}

\BuiltInTok{print}\NormalTok{(}\SpecialStringTok{f"log{-}likelihood before: }\SpecialCharTok{\{}\NormalTok{loglikelihood(theta, X)}\SpecialCharTok{\}}\SpecialStringTok{"}\NormalTok{)}
\NormalTok{optimizer }\OperatorTok{=}\NormalTok{ torch.optim.Adam([theta], lr}\OperatorTok{=}\DecValTok{1}\OperatorTok{/}\NormalTok{n)}
\NormalTok{last\_losses }\OperatorTok{=}\NormalTok{ []}
\ControlFlowTok{for}\NormalTok{ it }\KeywordTok{in}\NormalTok{ itertools.count():}
\NormalTok{    optimizer.zero\_grad()}
\NormalTok{    loss }\OperatorTok{=} \OperatorTok{{-}}\NormalTok{loglikelihood(theta, X)}
\NormalTok{    new\_loss }\OperatorTok{=}\NormalTok{ loss.detach()}
\NormalTok{    last\_losses.append(new\_loss)}
    \ControlFlowTok{if} \BuiltInTok{len}\NormalTok{(last\_losses)}\OperatorTok{\textgreater{}}\DecValTok{5000}\NormalTok{:}
\NormalTok{        last\_losses.pop(}\DecValTok{0}\NormalTok{)}
        \ControlFlowTok{if}\NormalTok{ last\_losses[}\DecValTok{0}\NormalTok{]}\OperatorTok{{-}}\NormalTok{last\_losses[}\OperatorTok{{-}}\DecValTok{1}\NormalTok{]}\OperatorTok{\textless{}}\DecValTok{0}\NormalTok{:}
            \ControlFlowTok{break}
\NormalTok{    loss.backward()}
\NormalTok{    optimizer.step()}
\BuiltInTok{print}\NormalTok{(}\SpecialStringTok{f"it: }\SpecialCharTok{\{}\NormalTok{it}\SpecialCharTok{\}}\SpecialStringTok{"}\NormalTok{)}
\BuiltInTok{print}\NormalTok{(}\SpecialStringTok{f"log{-}likelihood after: }\SpecialCharTok{\{}\NormalTok{loglikelihood(theta, X)}\SpecialCharTok{\}}\SpecialStringTok{"}\NormalTok{)}
\BuiltInTok{print}\NormalTok{(}\SpecialStringTok{f"theta: }\SpecialCharTok{\{}\NormalTok{theta}\SpecialCharTok{\}}\SpecialStringTok{"}\NormalTok{)}
\end{Highlighting}
\end{Shaded}

\begin{tcolorbox}[boxrule=0pt, enhanced, borderline west={2pt}{0pt}{code-block-stdout-light}, interior hidden, frame hidden, breakable, sharp corners, grow to left by=-1em]

\begin{verbatim}
log-likelihood before: -17554.00390625
\end{verbatim}

\end{tcolorbox}

\begin{tcolorbox}[boxrule=0pt, enhanced, borderline west={2pt}{0pt}{code-block-stdout-light}, interior hidden, frame hidden, breakable, sharp corners, grow to left by=-1em]

\begin{verbatim}
it: 23814
log-likelihood after: -5156.763671875
theta: tensor([0.0112, 1.0506, 2.0489, 1.1360, 1.6467, 1.3518, 1.0945, 1.2938, 1.5353,
        7.4930], requires_grad=True)
\end{verbatim}

\end{tcolorbox}

\hypertarget{using-the-value-1}{%
\paragraph{Using the value}\label{using-the-value-1}}

It is easy to retrieve estimates
\((\widehat\mu,\widehat\Sigma,\widehat\nu)\):

\begin{Shaded}
\begin{Highlighting}[]
\NormalTok{my\_params }\OperatorTok{=}\NormalTok{ parametrization.reals1d\_to\_params(theta)}
\NormalTok{my\_params.mu}
\end{Highlighting}
\end{Shaded}

\begin{tcolorbox}[boxrule=0pt, enhanced, borderline west={2pt}{0pt}{code-block-stdout-light}, interior hidden, frame hidden, breakable, sharp corners, grow to left by=-1em]

\begin{verbatim}
tensor([0.0112, 1.0506, 2.0489], grad_fn=<AddBackward0>)
\end{verbatim}

\end{tcolorbox}

\begin{Shaded}
\begin{Highlighting}[]
\NormalTok{my\_params.Sigma}
\end{Highlighting}
\end{Shaded}

\begin{tcolorbox}[boxrule=0pt, enhanced, borderline west={2pt}{0pt}{code-block-stdout-light}, interior hidden, frame hidden, breakable, sharp corners, grow to left by=-1em]

\begin{verbatim}
tensor([[2.0007, 1.0947, 1.0566],
        [1.0947, 2.2605, 1.7207],
        [1.0566, 1.7207, 2.1779]], grad_fn=<MulBackward0>)
\end{verbatim}

\end{tcolorbox}

\begin{Shaded}
\begin{Highlighting}[]
\NormalTok{my\_params.df}
\end{Highlighting}
\end{Shaded}

\begin{tcolorbox}[boxrule=0pt, enhanced, borderline west={2pt}{0pt}{code-block-stdout-light}, interior hidden, frame hidden, breakable, sharp corners, grow to left by=-1em]

\begin{verbatim}
tensor(7.4935, grad_fn=<SelectBackward0>)
\end{verbatim}

\end{tcolorbox}

We can see we recover good estimate of the simulated parameters.

\hypertarget{building-confidence-interval-1}{%
\paragraph{Building confidence
interval}\label{building-confidence-interval-1}}

The first step is to compute the inverse of the estimated Fisher
information matrix \(\widehat I_{n,\widehat\theta}\):

\begin{Shaded}
\begin{Highlighting}[]
\NormalTok{FIM }\OperatorTok{=} \OperatorTok{{-}}\NormalTok{ torch.autograd.functional.hessian(}
    \KeywordTok{lambda}\NormalTok{ theta: loglikelihood(theta, X), }
\NormalTok{    theta,}
\NormalTok{)}
\NormalTok{FIM\_inv }\OperatorTok{=}\NormalTok{ torch.linalg.inv(FIM)}
\end{Highlighting}
\end{Shaded}

And we can compute \(\delta_\nu\), then the confidence interval:

\begin{Shaded}
\begin{Highlighting}[]
\NormalTok{theta.grad.zero\_()}
\NormalTok{est\_df }\OperatorTok{=}\NormalTok{ parametrization.reals1d\_to\_params(theta).df}
\NormalTok{est\_df.backward()}
\NormalTok{delta\_df }\OperatorTok{=}\NormalTok{ theta.grad.detach()}
\NormalTok{df\_asymptotic\_variance }\OperatorTok{=}\NormalTok{ delta\_df }\OperatorTok{@}\NormalTok{ FIM\_inv }\OperatorTok{@}\NormalTok{ delta\_df}
\NormalTok{df\_confidence\_interval }\OperatorTok{=}\NormalTok{ (}
\NormalTok{    est\_df.detach()}
    \OperatorTok{+}\NormalTok{ (}
\NormalTok{        torch.tensor([}\OperatorTok{{-}}\DecValTok{1}\NormalTok{, }\DecValTok{1}\NormalTok{]) }
        \OperatorTok{*}\NormalTok{ scipy.stats.norm.ppf(}\FloatTok{0.975}\NormalTok{) }
        \OperatorTok{*}\NormalTok{ torch.sqrt(df\_asymptotic\_variance)}
\NormalTok{    )}
\NormalTok{)}
\BuiltInTok{print}\NormalTok{(df\_confidence\_interval)}
\end{Highlighting}
\end{Shaded}

\begin{tcolorbox}[boxrule=0pt, enhanced, borderline west={2pt}{0pt}{code-block-stdout-light}, interior hidden, frame hidden, breakable, sharp corners, grow to left by=-1em]

\begin{verbatim}
tensor([5.8149, 9.1722])
\end{verbatim}

\end{tcolorbox}

The simulated value was \(7\).

For the confidence interval on \(\abs\Sigma\), we have:

\begin{Shaded}
\begin{Highlighting}[]
\NormalTok{theta.grad.zero\_()}
\NormalTok{est\_det }\OperatorTok{=}\NormalTok{ torch.linalg.det(parametrization.reals1d\_to\_params(theta).Sigma)}
\NormalTok{est\_det.backward()}
\NormalTok{delta\_det }\OperatorTok{=}\NormalTok{ theta.grad.detach()}
\NormalTok{det\_asymptotic\_variance }\OperatorTok{=}\NormalTok{ delta\_det }\OperatorTok{@}\NormalTok{ FIM\_inv }\OperatorTok{@}\NormalTok{ delta\_det}
\NormalTok{det\_confidence\_interval }\OperatorTok{=}\NormalTok{ (}
\NormalTok{    est\_det.detach()}
    \OperatorTok{+}\NormalTok{ (}
\NormalTok{        torch.tensor([}\OperatorTok{{-}}\DecValTok{1}\NormalTok{, }\DecValTok{1}\NormalTok{]) }
        \OperatorTok{*}\NormalTok{ scipy.stats.norm.ppf(}\FloatTok{0.975}\NormalTok{) }
        \OperatorTok{*}\NormalTok{ torch.sqrt(det\_asymptotic\_variance)}
\NormalTok{    )}
\NormalTok{)}
\BuiltInTok{print}\NormalTok{(det\_confidence\_interval)}
\end{Highlighting}
\end{Shaded}

\begin{tcolorbox}[boxrule=0pt, enhanced, borderline west={2pt}{0pt}{code-block-stdout-light}, interior hidden, frame hidden, breakable, sharp corners, grow to left by=-1em]

\begin{verbatim}
tensor([2.1326, 3.4136])
\end{verbatim}

\end{tcolorbox}

The simulated value was \(2.5\).

\hypertarget{low-level-modules-inference-of-parameters-of-a-gumbel-distribution}{%
\subsection{Low-level modules: inference of parameters of a Gumbel
distribution}\label{low-level-modules-inference-of-parameters-of-a-gumbel-distribution}}

\hypertarget{introduction-2}{%
\subsubsection{Introduction}\label{introduction-2}}

With \(\mu\in\reels^p\) and \(\beta\in\reelsp\), the Gumbel distribution
is defined with the probability distribution function:

\[
   \begin{array}{rcl}
    \reels & \longrightarrow & \mathbb R_+ \\
    x & \longmapsto &
    \frac{\exp\p{-\exp\p{-\frac{x-\mu}\beta}}\exp\p{-\frac{x-\mu}\beta}}\beta
   \end{array}
\]

The idea of this example is to introduce the inference with Maximum
Likelihood Estimator (MLE) of the parameters, handling the constraints
by parametrization with low-level interface of the package, using
automatic differentiation to compute derivatives.

We introduce \(\theta\in\reels^2\) the mapping of \((\mu, \beta)\) by
our bijective parametrization. Using invariance property of the MLE, the
mapping of MLE of \(\theta\) is equivalent to the MLE of
\((\mu, \beta)\).

In a second time, when a MLE is obtained, with sufficient regularity
conditions (not detailed here), using asymptotic properties of MLE and
Slutsky's lemma we have:

\[
    \widehat I_{n,\widehat\theta}^{-\frac12}\p{\widehat\theta-\theta_0}
    \underset{n\to+\infty}\longrightarrow \mathcal N\p{0, I}
\]

where:

\begin{itemize}
\tightlist
\item
  the estimated Fisher information matrix
  \(\widehat I_{n,\widehat\theta} = - {\left.\frac{\operatorname{d}^2\ell\p{\theta, \ldots}}{\operatorname{d}\theta^2}\right|}_{\theta=\widehat\theta}\).
\item
  \(I\) the identity matrix.
\item
  \(\ell\) is the log-likelihood of the whole sample.
\end{itemize}

Now we can move this result in our original parameter space:

\[
\frac{\widehat\beta-\beta_0}{\sqrt{\delta_\beta^TI_{n,\widehat\theta}^{-1}\delta_\beta}}\underset{n\to+\infty}\longrightarrow\mathcal N(0,1)
\]

with:

\begin{itemize}
\tightlist
\item
  \(\delta_\beta = {\left.\frac{\operatorname{d}\beta}{\operatorname{d}\theta}\right|}_{\theta=\widehat\theta}\)
\end{itemize}

Therefore we can obtain asymptotic confidence interval:

\[
\mathbb P\p{
\beta_0\in
\left[\widehat\beta \pm
   u_{1-\alpha/2}\sqrt{\delta_\beta^TI_{n,\widehat\theta}^{-1}\delta_\beta}\right]
} \underset{n\to+\infty}\longrightarrow 1-\alpha
\]

Note that \(\delta_\beta\) and \(I_{n,\widehat\theta}\) will be computed
with automatic differentiation.

\hypertarget{with-jax-1}{%
\subsubsection{With JAX}\label{with-jax-1}}

\hypertarget{simulating-the-data-2}{%
\paragraph{Simulating the data}\label{simulating-the-data-2}}

First we generate simulated data to illustrate the method.

\begin{Shaded}
\begin{Highlighting}[]
\ImportTok{import}\NormalTok{ numpy }\ImportTok{as}\NormalTok{ np}
\ImportTok{import}\NormalTok{ scipy.stats}

\NormalTok{n }\OperatorTok{=} \DecValTok{1000}
\NormalTok{mu0 }\OperatorTok{=} \DecValTok{5}
\NormalTok{beta0 }\OperatorTok{=} \DecValTok{2}

\CommentTok{\# we use a seeded random state only for reproducibilty}
\NormalTok{random\_state }\OperatorTok{=}\NormalTok{ np.random.RandomState(np.random.MT19937(np.random.SeedSequence(}\DecValTok{0}\NormalTok{)))}
\NormalTok{X }\OperatorTok{=}\NormalTok{ scipy.stats.gumbel\_r(loc}\OperatorTok{=}\NormalTok{mu0, scale}\OperatorTok{=}\NormalTok{beta0).rvs(}
\NormalTok{    size}\OperatorTok{=}\NormalTok{n, random\_state}\OperatorTok{=}\NormalTok{random\_state}
\NormalTok{)}

\CommentTok{\# convert to JAX array}
\ImportTok{import}\NormalTok{ jax.numpy }\ImportTok{as}\NormalTok{ jnp}
\NormalTok{X }\OperatorTok{=}\NormalTok{ jnp.array(X)}
\end{Highlighting}
\end{Shaded}

\hypertarget{definition-of-the-log-likelihood-and-gradients-2}{%
\paragraph{Definition of the log-likelihood and
gradients}\label{definition-of-the-log-likelihood-and-gradients-2}}

First we define the likelihood depending on our original parameters:

\begin{Shaded}
\begin{Highlighting}[]
\KeywordTok{def}\NormalTok{ original\_loglikelihood(mu, beta, X):}
\NormalTok{    logz }\OperatorTok{=} \OperatorTok{{-}}\NormalTok{(X}\OperatorTok{{-}}\NormalTok{mu)}\OperatorTok{/}\NormalTok{beta}
    \ControlFlowTok{return}\NormalTok{ (}\OperatorTok{{-}}\NormalTok{jnp.exp(logz)}\OperatorTok{+}\NormalTok{logz}\OperatorTok{{-}}\NormalTok{jnp.log(beta)).}\BuiltInTok{sum}\NormalTok{()}
\end{Highlighting}
\end{Shaded}

And we define the log-likelihood of our parametrized model by using
functions from the \texttt{parametrization\_cookbook.functions.jax}
module:

\begin{Shaded}
\begin{Highlighting}[]
\ImportTok{import}\NormalTok{ parametrization\_cookbook.functions.jax }\ImportTok{as}\NormalTok{ pcf}
\ImportTok{import}\NormalTok{ jax}

\AttributeTok{@jax.jit}
\KeywordTok{def}\NormalTok{ loglikelihood(theta, X):}
\NormalTok{    mu }\OperatorTok{=}\NormalTok{ theta[}\DecValTok{0}\NormalTok{]}
\NormalTok{    beta }\OperatorTok{=}\NormalTok{ pcf.softplus(theta[}\DecValTok{1}\NormalTok{])}
    \ControlFlowTok{return}\NormalTok{ original\_loglikelihood(mu, beta, X)}
\end{Highlighting}
\end{Shaded}

This function was JIT-compiled, as this function is run many-times, this
is very interesting to reduce computation time.

We now define the gradient and hessian functions (with JIT-compilation):

\begin{Shaded}
\begin{Highlighting}[]
\NormalTok{grad\_loglikelihood }\OperatorTok{=}\NormalTok{ jax.jit(jax.grad(loglikelihood))}
\NormalTok{hessian\_loglikelihood }\OperatorTok{=}\NormalTok{ jax.jit(jax.jacfwd(jax.jacrev(loglikelihood)))}
\end{Highlighting}
\end{Shaded}

\hypertarget{optimization-2}{%
\paragraph{Optimization}\label{optimization-2}}

We can use any optimization algorithm. We choose here a gradient descend
(with step conditioning by the highest eigenvalue of the hessian). This
is only given for illustration purpose, in a real application case,
using optimization algorithm developed with JAX in the Python module
\texttt{jaxopt} (\protect\hyperlink{ref-jaxopt}{Blondel et al. 2021})
can be a better choice.

We choose here to initialize randomly \(\theta\). We can also build a
plausible value of \(\theta\) with reciprocal functions.

\begin{Shaded}
\begin{Highlighting}[]
\ImportTok{import}\NormalTok{ itertools}

\CommentTok{\# we use a seeded random state only for reproducibility}
\NormalTok{random\_state }\OperatorTok{=}\NormalTok{ np.random.RandomState(np.random.MT19937(np.random.SeedSequence(}\DecValTok{1}\NormalTok{)))}
\NormalTok{theta }\OperatorTok{=}\NormalTok{ random\_state.normal(size}\OperatorTok{=}\DecValTok{2}\NormalTok{)}
\NormalTok{theta }\OperatorTok{=}\NormalTok{ jnp.array(theta)}

\NormalTok{current\_likeli }\OperatorTok{=}\NormalTok{ loglikelihood(theta, X)}
\BuiltInTok{print}\NormalTok{(}\SpecialStringTok{f"Log{-}likeli: }\SpecialCharTok{\{}\NormalTok{current\_likeli}\SpecialCharTok{\}}\SpecialStringTok{"}\NormalTok{)}

\ControlFlowTok{for}\NormalTok{ it }\KeywordTok{in}\NormalTok{ itertools.count():}
\NormalTok{    g }\OperatorTok{=}\NormalTok{ grad\_loglikelihood(theta, X)}
\NormalTok{    H }\OperatorTok{=}\NormalTok{ hessian\_loglikelihood(theta, X)}
\NormalTok{    sdp\_eigenvalues }\OperatorTok{=} \OperatorTok{{-}}\NormalTok{jnp.linalg.eigh(H)[}\DecValTok{0}\NormalTok{]}
\NormalTok{    lr }\OperatorTok{=} \DecValTok{1}\OperatorTok{/}\NormalTok{sdp\_eigenvalues.}\BuiltInTok{max}\NormalTok{()}
\NormalTok{    theta }\OperatorTok{+=}\NormalTok{ lr}\OperatorTok{*}\NormalTok{g}
\NormalTok{    current\_likeli, old\_likeli }\OperatorTok{=}\NormalTok{ loglikelihood(theta, X), current\_likeli}
    \ControlFlowTok{if}\NormalTok{ current\_likeli}\OperatorTok{{-}}\NormalTok{old\_likeli}\OperatorTok{\textless{}}\FloatTok{1e{-}6}\NormalTok{:}
        \ControlFlowTok{break}
\BuiltInTok{print}\NormalTok{(}\SpecialStringTok{f"it: }\SpecialCharTok{\{}\NormalTok{it}\SpecialCharTok{\}}\SpecialStringTok{, Log{-}likeli: }\SpecialCharTok{\{}\NormalTok{current\_likeli}\SpecialCharTok{\}}\SpecialStringTok{"}\NormalTok{)}
\BuiltInTok{print}\NormalTok{(}\SpecialStringTok{f"theta: }\SpecialCharTok{\{}\NormalTok{theta}\SpecialCharTok{\}}\SpecialStringTok{"}\NormalTok{)}
\end{Highlighting}
\end{Shaded}

\begin{tcolorbox}[boxrule=0pt, enhanced, borderline west={2pt}{0pt}{code-block-stdout-light}, interior hidden, frame hidden, breakable, sharp corners, grow to left by=-1em]

\begin{verbatim}
Log-likeli: -14577.125
\end{verbatim}

\end{tcolorbox}

\begin{tcolorbox}[boxrule=0pt, enhanced, borderline west={2pt}{0pt}{code-block-stdout-light}, interior hidden, frame hidden, breakable, sharp corners, grow to left by=-1em]

\begin{verbatim}
it: 13, Log-likeli: -2253.14013671875
theta: [4.9774423 1.8211547]
\end{verbatim}

\end{tcolorbox}

\hypertarget{using-the-value-2}{%
\paragraph{Using the value}\label{using-the-value-2}}

To retrieve the initial parameter, we must use \(\widehat\theta\):

\begin{Shaded}
\begin{Highlighting}[]
\NormalTok{theta[}\DecValTok{0}\NormalTok{] }\CommentTok{\# this is estimated of mu}
\end{Highlighting}
\end{Shaded}

\begin{tcolorbox}[boxrule=0pt, enhanced, borderline west={2pt}{0pt}{code-block-stdout-light}, interior hidden, frame hidden, breakable, sharp corners, grow to left by=-1em]

\begin{verbatim}
DeviceArray(4.9774423, dtype=float32)
\end{verbatim}

\end{tcolorbox}

\begin{Shaded}
\begin{Highlighting}[]
\NormalTok{pcf.softplus(theta[}\DecValTok{1}\NormalTok{]) }\CommentTok{\# this is estimated of beta}
\end{Highlighting}
\end{Shaded}

\begin{tcolorbox}[boxrule=0pt, enhanced, borderline west={2pt}{0pt}{code-block-stdout-light}, interior hidden, frame hidden, breakable, sharp corners, grow to left by=-1em]

\begin{verbatim}
DeviceArray(1.9711586, dtype=float32)
\end{verbatim}

\end{tcolorbox}

We can see we recover good estimate of the simulated parameters.

\hypertarget{building-confidence-interval-2}{%
\paragraph{Building confidence
interval}\label{building-confidence-interval-2}}

The first step is to compute the inverse of the estimated Fisher
information matrix \(\widehat I_{n,\widehat\theta}\):

\begin{Shaded}
\begin{Highlighting}[]
\NormalTok{FIM\_inv }\OperatorTok{=}\NormalTok{ jnp.linalg.inv(}\OperatorTok{{-}}\NormalTok{H)}
\end{Highlighting}
\end{Shaded}

And we can compute \(\delta_\beta\), then the confidence interval:

\begin{Shaded}
\begin{Highlighting}[]
\NormalTok{delta\_beta }\OperatorTok{=}\NormalTok{ jax.grad(}\KeywordTok{lambda}\NormalTok{ theta: pcf.softplus(theta[}\DecValTok{1}\NormalTok{]))(theta)}
\NormalTok{beta\_asymptotic\_variance }\OperatorTok{=}\NormalTok{ delta\_beta }\OperatorTok{@}\NormalTok{ FIM\_inv }\OperatorTok{@}\NormalTok{ delta\_beta}
\NormalTok{beta\_confidence\_interval }\OperatorTok{=}\NormalTok{ (}
\NormalTok{    pcf.softplus(theta[}\DecValTok{1}\NormalTok{])}
    \OperatorTok{+}\NormalTok{ (}
\NormalTok{        jnp.array([}\OperatorTok{{-}}\DecValTok{1}\NormalTok{, }\DecValTok{1}\NormalTok{])}
        \OperatorTok{*}\NormalTok{ scipy.stats.norm.ppf(}\FloatTok{0.975}\NormalTok{)}
        \OperatorTok{*}\NormalTok{ jnp.sqrt(beta\_asymptotic\_variance)}
\NormalTok{    )}
\NormalTok{)}
\BuiltInTok{print}\NormalTok{(beta\_confidence\_interval)}
\end{Highlighting}
\end{Shaded}

\begin{tcolorbox}[boxrule=0pt, enhanced, borderline west={2pt}{0pt}{code-block-stdout-light}, interior hidden, frame hidden, breakable, sharp corners, grow to left by=-1em]

\begin{verbatim}
[1.8763951 2.065922 ]
\end{verbatim}

\end{tcolorbox}

The simulated value was \(2\).

\hypertarget{with-pytorch-1}{%
\subsubsection{With PyTorch}\label{with-pytorch-1}}

\hypertarget{simulating-the-data-3}{%
\paragraph{Simulating the data}\label{simulating-the-data-3}}

First we generate simulated data to illustrate the method.

\begin{Shaded}
\begin{Highlighting}[]
\ImportTok{import}\NormalTok{ numpy }\ImportTok{as}\NormalTok{ np}
\ImportTok{import}\NormalTok{ scipy.stats}

\NormalTok{n }\OperatorTok{=} \DecValTok{1000}
\NormalTok{mu0 }\OperatorTok{=} \DecValTok{5}
\NormalTok{beta0 }\OperatorTok{=} \DecValTok{2}

\CommentTok{\# we use a seeded random state only for reproducibility}
\NormalTok{random\_state }\OperatorTok{=}\NormalTok{ np.random.RandomState(np.random.MT19937(np.random.SeedSequence(}\DecValTok{0}\NormalTok{)))}
\NormalTok{X }\OperatorTok{=}\NormalTok{ scipy.stats.gumbel\_r(loc}\OperatorTok{=}\NormalTok{mu0, scale}\OperatorTok{=}\NormalTok{beta0).rvs(}
\NormalTok{    size}\OperatorTok{=}\NormalTok{n, random\_state}\OperatorTok{=}\NormalTok{random\_state}
\NormalTok{)}

\CommentTok{\# convert to PyTorch tensor}
\ImportTok{import}\NormalTok{ torch}
\NormalTok{X }\OperatorTok{=}\NormalTok{ torch.tensor(X)}
\end{Highlighting}
\end{Shaded}

\hypertarget{definition-of-the-log-likelihood-and-gradients-3}{%
\paragraph{Definition of the log-likelihood and
gradients}\label{definition-of-the-log-likelihood-and-gradients-3}}

First we define the likelihood depending on our original parameters:

\begin{Shaded}
\begin{Highlighting}[]
\KeywordTok{def}\NormalTok{ original\_loglikelihood(mu, beta, X):}
\NormalTok{    logz }\OperatorTok{=} \OperatorTok{{-}}\NormalTok{(X}\OperatorTok{{-}}\NormalTok{mu)}\OperatorTok{/}\NormalTok{beta}
    \ControlFlowTok{return}\NormalTok{ (}\OperatorTok{{-}}\NormalTok{torch.exp(logz)}\OperatorTok{+}\NormalTok{logz}\OperatorTok{{-}}\NormalTok{torch.log(beta)).}\BuiltInTok{sum}\NormalTok{()}
\end{Highlighting}
\end{Shaded}

And we define the log-likelihood of our parametrized model by using
functions from the \texttt{parametrization\_cookbook.functions.torch}
module:

\begin{Shaded}
\begin{Highlighting}[]
\ImportTok{import}\NormalTok{ parametrization\_cookbook.functions.torch }\ImportTok{as}\NormalTok{ pcf}

\KeywordTok{def}\NormalTok{ loglikelihood(theta, X):}
\NormalTok{    mu }\OperatorTok{=}\NormalTok{ theta[}\DecValTok{0}\NormalTok{]}
\NormalTok{    beta }\OperatorTok{=}\NormalTok{ pcf.softplus(theta[}\DecValTok{1}\NormalTok{])}
    \ControlFlowTok{return}\NormalTok{ original\_loglikelihood(mu, beta, X)}
\end{Highlighting}
\end{Shaded}

\hypertarget{optimization-3}{%
\paragraph{Optimization}\label{optimization-3}}

We can use any optimization algorithm. We choose here a ADAM gradient.

We choose here to initialize randomly \(\theta\). We can also build a
plausible value of \(\theta\) with reciprocal functions.

\begin{Shaded}
\begin{Highlighting}[]
\ImportTok{import}\NormalTok{ itertools}

\CommentTok{\# we use a seeded random state only for reproducibility}
\NormalTok{random\_state }\OperatorTok{=}\NormalTok{ np.random.RandomState(np.random.MT19937(np.random.SeedSequence(}\DecValTok{1}\NormalTok{)))}
\NormalTok{theta }\OperatorTok{=}\NormalTok{ random\_state.normal(size}\OperatorTok{=}\DecValTok{2}\NormalTok{)}
\NormalTok{theta }\OperatorTok{=}\NormalTok{ torch.tensor(theta, dtype}\OperatorTok{=}\NormalTok{torch.float32, requires\_grad}\OperatorTok{=}\VariableTok{True}\NormalTok{)}

\NormalTok{current\_likeli }\OperatorTok{=}\NormalTok{ loglikelihood(theta, X)}
\BuiltInTok{print}\NormalTok{(}\SpecialStringTok{f"log{-}likelihood before: }\SpecialCharTok{\{}\NormalTok{loglikelihood(theta, X)}\SpecialCharTok{\}}\SpecialStringTok{"}\NormalTok{)}
\NormalTok{optimizer }\OperatorTok{=}\NormalTok{ torch.optim.Adam([theta], lr}\OperatorTok{=}\DecValTok{1}\OperatorTok{/}\NormalTok{n)}
\NormalTok{last\_losses }\OperatorTok{=}\NormalTok{ []}
\ControlFlowTok{for}\NormalTok{ it }\KeywordTok{in}\NormalTok{ itertools.count():}
\NormalTok{    optimizer.zero\_grad()}
\NormalTok{    loss }\OperatorTok{=} \OperatorTok{{-}}\NormalTok{loglikelihood(theta, X)}
\NormalTok{    new\_loss }\OperatorTok{=}\NormalTok{ loss.detach()}
\NormalTok{    last\_losses.append(new\_loss)}
    \ControlFlowTok{if} \BuiltInTok{len}\NormalTok{(last\_losses)}\OperatorTok{\textgreater{}}\DecValTok{5000}\NormalTok{:}
\NormalTok{        last\_losses.pop(}\DecValTok{0}\NormalTok{)}
        \ControlFlowTok{if}\NormalTok{ last\_losses[}\DecValTok{0}\NormalTok{]}\OperatorTok{{-}}\NormalTok{last\_losses[}\OperatorTok{{-}}\DecValTok{1}\NormalTok{]}\OperatorTok{\textless{}}\DecValTok{0}\NormalTok{:}
            \ControlFlowTok{break}
\NormalTok{    loss.backward()}
\NormalTok{    optimizer.step()}
\BuiltInTok{print}\NormalTok{(}\SpecialStringTok{f"it: }\SpecialCharTok{\{}\NormalTok{it}\SpecialCharTok{\}}\SpecialStringTok{"}\NormalTok{)}
\BuiltInTok{print}\NormalTok{(}\SpecialStringTok{f"log{-}likelihood after: }\SpecialCharTok{\{}\NormalTok{loglikelihood(theta, X)}\SpecialCharTok{\}}\SpecialStringTok{"}\NormalTok{)}
\BuiltInTok{print}\NormalTok{(}\SpecialStringTok{f"theta: }\SpecialCharTok{\{}\NormalTok{theta}\SpecialCharTok{\}}\SpecialStringTok{"}\NormalTok{)}
\end{Highlighting}
\end{Shaded}

\begin{tcolorbox}[boxrule=0pt, enhanced, borderline west={2pt}{0pt}{code-block-stdout-light}, interior hidden, frame hidden, breakable, sharp corners, grow to left by=-1em]

\begin{verbatim}
log-likelihood before: -14577.124868295035
\end{verbatim}

\end{tcolorbox}

\begin{tcolorbox}[boxrule=0pt, enhanced, borderline west={2pt}{0pt}{code-block-stdout-light}, interior hidden, frame hidden, breakable, sharp corners, grow to left by=-1em]

\begin{verbatim}
it: 15350
log-likelihood after: -2253.1400991199
theta: tensor([4.9778, 1.8214], requires_grad=True)
\end{verbatim}

\end{tcolorbox}

\hypertarget{using-the-value-3}{%
\paragraph{Using the value}\label{using-the-value-3}}

To retrieve the initial parameter, we must use \(\widehat\theta\):

\begin{Shaded}
\begin{Highlighting}[]
\NormalTok{theta[}\DecValTok{0}\NormalTok{] }\CommentTok{\# this is estimated of mu}
\end{Highlighting}
\end{Shaded}

\begin{tcolorbox}[boxrule=0pt, enhanced, borderline west={2pt}{0pt}{code-block-stdout-light}, interior hidden, frame hidden, breakable, sharp corners, grow to left by=-1em]

\begin{verbatim}
tensor(4.9778, grad_fn=<SelectBackward0>)
\end{verbatim}

\end{tcolorbox}

\begin{Shaded}
\begin{Highlighting}[]
\NormalTok{pcf.softplus(theta[}\DecValTok{1}\NormalTok{]) }\CommentTok{\# this is estimated of beta}
\end{Highlighting}
\end{Shaded}

\begin{tcolorbox}[boxrule=0pt, enhanced, borderline west={2pt}{0pt}{code-block-stdout-light}, interior hidden, frame hidden, breakable, sharp corners, grow to left by=-1em]

\begin{verbatim}
tensor(1.9714, grad_fn=<MulBackward0>)
\end{verbatim}

\end{tcolorbox}

We can see we recover good estimate of the simulated parameters.

\hypertarget{building-confidence-interval-3}{%
\paragraph{Building confidence
interval}\label{building-confidence-interval-3}}

The first step is to compute the inverse of the estimated Fisher
information matrix \(\widehat I_{n,\widehat\theta}\):

\begin{Shaded}
\begin{Highlighting}[]
\NormalTok{FIM }\OperatorTok{=} \OperatorTok{{-}}\NormalTok{ torch.autograd.functional.hessian(}
    \KeywordTok{lambda}\NormalTok{ theta: loglikelihood(theta, X),}
\NormalTok{    theta,}
\NormalTok{)}
\NormalTok{FIM\_inv }\OperatorTok{=}\NormalTok{ torch.linalg.inv(FIM)}
\end{Highlighting}
\end{Shaded}

And we can compute \(\delta_\beta\), then the confidence interval:

\begin{Shaded}
\begin{Highlighting}[]
\NormalTok{theta.grad.zero\_()}
\NormalTok{beta }\OperatorTok{=}\NormalTok{ pcf.softplus(theta[}\DecValTok{1}\NormalTok{])}
\NormalTok{beta.backward()}
\NormalTok{delta\_beta }\OperatorTok{=}\NormalTok{ theta.grad.detach()}
\NormalTok{beta\_asymptotic\_variance }\OperatorTok{=}\NormalTok{ delta\_beta }\OperatorTok{@}\NormalTok{ FIM\_inv }\OperatorTok{@}\NormalTok{ delta\_beta}
\NormalTok{beta\_confidence\_interval }\OperatorTok{=}\NormalTok{ (}
\NormalTok{    beta.detach()}
    \OperatorTok{+}\NormalTok{ (}
\NormalTok{        torch.tensor([}\OperatorTok{{-}}\DecValTok{1}\NormalTok{, }\DecValTok{1}\NormalTok{])}
        \OperatorTok{*}\NormalTok{ scipy.stats.norm.ppf(}\FloatTok{0.975}\NormalTok{)}
        \OperatorTok{*}\NormalTok{ torch.sqrt(beta\_asymptotic\_variance)}
\NormalTok{    )}
\NormalTok{)}
\BuiltInTok{print}\NormalTok{(beta\_confidence\_interval)}
\end{Highlighting}
\end{Shaded}

\begin{tcolorbox}[boxrule=0pt, enhanced, borderline west={2pt}{0pt}{code-block-stdout-light}, interior hidden, frame hidden, breakable, sharp corners, grow to left by=-1em]

\begin{verbatim}
tensor([1.8766, 2.0662])
\end{verbatim}

\end{tcolorbox}

The simulated value was \(2\).

\hypertarget{acknowledgements}{%
\section*{Acknowledgements}\label{acknowledgements}}
\addcontentsline{toc}{section}{Acknowledgements}

The authors are very grateful to Matthias Bussonnier, Thibaud Le
Graverend and Charlotte Baey for their feedbacks, their corrections of
the cookbook and for testing the package. Nevertheless, any remaining
errors remain the sole responsibility of the authors.

\hypertarget{references}{%
\section*{References}\label{references}}
\addcontentsline{toc}{section}{References}

\hypertarget{refs}{}
\begin{CSLReferences}{1}{0}
\leavevmode\vadjust pre{\hypertarget{ref-jaxopt}{}}%
Blondel, Mathieu, Quentin Berthet, Marco Cuturi, Roy Frostig, Stephan
Hoyer, Felipe Llinares-López, Fabian Pedregosa, and Jean-Philippe Vert.
2021. {``Efficient and Modular Implicit Differentiation.''} \emph{arXiv
Preprint arXiv:2105.15183}.

\leavevmode\vadjust pre{\hypertarget{ref-jax}{}}%
Bradbury, James, Roy Frostig, Peter Hawkins, Matthew James Johnson,
Chris Leary, Dougal Maclaurin, George Necula, et al. 2018. \emph{{JAX}:
Composable Transformations of {P}ython+{N}um{P}y Programs} (version
0.3.13). \url{http://github.com/google/jax}.

\leavevmode\vadjust pre{\hypertarget{ref-numpy}{}}%
Harris, Charles R., K. Jarrod Millman, Stéfan J van der Walt, Ralf
Gommers, Pauli Virtanen, David Cournapeau, Eric Wieser, et al. 2020.
{``Array Programming with {NumPy}.''} \emph{Nature} 585: 357--62.
\url{https://doi.org/10.1038/s41586-020-2649-2}.

\leavevmode\vadjust pre{\hypertarget{ref-pytorch}{}}%
Paszke, Adam, Sam Gross, Francisco Massa, Adam Lerer, James Bradbury,
Gregory Chanan, Trevor Killeen, et al. 2019. {``{PyTorch: An Imperative
Style, High-Performance Deep Learning Library}.''} In \emph{Advances in
Neural Information Processing Systems 32}, edited by H. Wallach, H.
Larochelle, A. Beygelzimer, F. d'Alché-Buc, E. Fox, and R. Garnett,
8024--35. Curran Associates, Inc.
\url{http://papers.neurips.cc/paper/9015-pytorch-an-imperative-style-high-performance-deep-learning-library.pdf}.

\leavevmode\vadjust pre{\hypertarget{ref-scipy}{}}%
Virtanen, Pauli, Ralf Gommers, Travis E. Oliphant, Matt Haberland, Tyler
Reddy, David Cournapeau, Evgeni Burovski, et al. 2020. {``{{SciPy} 1.0:
Fundamental Algorithms for Scientific Computing in Python}.''}
\emph{Nature Methods} 17: 261--72.
\url{https://doi.org/10.1038/s41592-019-0686-2}.

\leavevmode\vadjust pre{\hypertarget{ref-wilson1931distribution}{}}%
Wilson, Edwin B, and Margaret M Hilferty. 1931. {``The Distribution of
Chi-Square.''} \emph{Proceedings of the National Academy of Sciences} 17
(12): 684--88.

\end{CSLReferences}

\end{document}